\documentclass[preprint,3p,12pt]{elsarticle}
\usepackage{mathrsfs}
\usepackage{amsmath}
\usepackage{stmaryrd}
\usepackage{bbding}
\usepackage{dcolumn}
\usepackage{graphicx}
\usepackage{amsfonts}
\usepackage{amssymb}
\usepackage{psfrag}
\usepackage{wrapfig}
\usepackage{subfigure}
\usepackage{makeidx}
\usepackage{bm}
\usepackage{epsf}
\usepackage{epsfig}
\usepackage{setspace}
\usepackage{graphicx}
\usepackage{epstopdf}
\usepackage{psfrag}
\usepackage{subfigure}
\usepackage{booktabs}
\usepackage{color}
\usepackage{comment}

\begin{document}

\title{UGKWP for three-dimensional simulation of gas-particle fluidized bed}
	
\author[HKUST1]{Xiaojian Yang}
\ead{xyangbm@connect.ust.hk}

\author[HKUST2]{Yufeng Wei}
\ead{yweibe@connect.ust.hk}
	
\author[HKUST1]{Wei Shyy}
\ead{weishyy@ust.hk}
	
\author[HKUST1,HKUST2,HKUST3,HKUST4]{Kun Xu\corref{cor1}}
\ead{makxu@ust.hk}	
	
\address[HKUST1]{Department of Mechanical and Aerospace Engineering, Hong Kong University of Science and Technology, Clear Water Bay, Kowloon, Hong Kong, China}
\address[HKUST2]{Department of Mathematics, Hong Kong University of Science and Technology, Clear Water Bay, Kowloon, Hong Kong, China}
\address[HKUST3]{Shenzhen Research Institute, Hong Kong University of Science and Technology, Shenzhen, China}
\address[HKUST4]{Guangdong-Hong Kong-Macao Joint Laboratory for Data-Driven Fluid Mechanics and Engineering Applications, Hong Kong University of Science and Technology, Hong Kong, China}
\cortext[cor1]{Corresponding author}

\begin{abstract}
The gas-solid particle two-phase flow in a fluidized bed shows complex physics. Following our previous work, the multi-scale framework based on gas-kinetic scheme (GKS) and unified gas-kinetic wave-particle method (UGKWP) for the gas-particle system is firstly extended to the three-dimensional simulation of the fluidized bed. For the solid particle evolution, different from the widely-used Eulerian and Lagrangian approaches, the UGKWP unifies the wave (dense particle region) and discrete particle (dilute particle region) formulation seamlessly according to a continuous variation of particle cell's Kundsen number ($Kn$). The GKS-UGKWP for the coupled gas-particle evolution system can automatically become an Eulerian-Eulerian (EE) method in the high particle collision regime and Eulerian-Lagrangian (EL) formulation in the collisionless particle regime. In the transition regime, the UGKWP can achieve a smooth transition between the Eulerian and Lagrangian limiting formulation. More importantly, the weights of mass distributions from analytical wave and discrete particle are related to the local $Kn$ by $\exp(-1/Kn)$ for wave and $(1-\exp(-1/Kn))$ for discrete particle. As a result, the UGKWP provides an optimal modeling for capturing the particle phase in terms of physical accuracy and numerical efficiency.
In the numerical simulation, the UGKWP does not need any prior division of dilute/dense regions, which makes it suitable for the fluidized bed problem, where the dilute/transition/dense regions instantaneously coexist and are dynamically interconvertible. In this paper, based on the GKS-UGKWP formulation two lab-scale fluidization cases, i.e., one turbulent fluidized bed and one circulating fluidized bed, are simulated in 3D and the simulation results are compared with the experimental measurements. The typical heterogeneous flow features of the fluidized bed are well captured and the statistics are in good agreement with experiment data.
\end{abstract}

\begin{keyword}
	Unified gas-kinetic wave-particle method, gas-kinetic scheme, gas-particle flow, gas-solid fluidization
\end{keyword}

\maketitle

\section{Introduction}
Gas-solid particle fluidization system is widely used in the energy and chemical industry. The vigorous interaction between gas and solid particle
involves complex dynamics in the determination of mass and heat transfer \cite{Gasparticle-review-Falah-alobaid2021progress}. Generally, the fluidization occurs in a container with a large number of solid particles and gas flow blown from below. This two-phase system usually shows rich and complex physics, such as the particle transport and collision, the clustering and dispersion of large number of particles, and the coexistence of dilute, transition, and dense regions, etc. For such a complex system, the prediction through an analytical solution is almost impossible, and the experiment measurement is expensive and depends highly on the measuring devices. Therefore, computational fluid dynamics (CFD) becomes a powerful and indispensable way for studying and understanding dynamics in fluidization, and guides the design and optimization of fluidized beds, etc \cite{Gasparticle-review-van2008numerical, Gasparticle-review-zhongwenqi-zhong2016cfd, Gasparticle-review-Falah-alobaid2021progress}.

Many numerical methods have been constructed to simulate gas-solid particle fluidization. The Eulerian-Eulerian (EE) approach, also called two fluid model (TFM), is one of the important approaches used in fluidization engineering \cite{Gasparticle-review-WangJunwu2020continuum, Gasparticle-book-luhuilin2021computational}. In the EE approach, both gas and solid particle phases are described under the Eulerian framework.
The kinetic theory for granular flow (KTGF) is one representative method of TFM, in which the constitutive relationship, i.e., the stress tensor, can be derived based on the Chapman-Enskog asymptotic analysis \cite{Gasparticle-KTGF-lun1984kinetic, Gasparticle-book-gidaspow1994multiphase, Gasparticle-book-luhuilin2021computational}. In general, the underlying assumption in TFM is that the solid phase stays in a near equilibrium state. In other words, with the slight deviation from local Maxwellian distribution for the particle phase, the corresponding hydrodynamic evolution equations based on the macroscopic variables (density, velocity, and granular temperature) can be obtained. In reality, the solid particle phase can be in equilibrium or non-equilibrium state according to the Knudsen number ($Kn$), which is defined by the ratio of the particle mean free path over the characteristic length scale \cite{Gasparticle-momentmethod-Fox2013computational, Gasparticle-Knudsen-number-wang2019quantifying}. The solid phase stays in an equilibrium state at small $Kn$ number under the intensive inter-particle collisions. This is likely to occur in the dense particle region in the gas-particle fluidization system.
At large $Kn$ number, the particle keeps the non-equilibrium state and the particle free transport plays a key role in the evolution, such as  the dilute particle region. One of the outstanding non-equilibrium phenomenon is the particle trajectory crossing (PTC), where multiple particle velocities have to be captured at the same space location. The averaged single fluid velocity model in TFM makes it difficult
to give an accurate prediction of non-equilibrium physics \cite{Gasparticle-momentmethod-Fox2013computational}.
Another alternative approach is Eulerian-Lagrangian (EL) method, such as the computational fluid dynamic-discrete element method (CFD-DEM), where all solid particles are tracked explicitly in the evolution \cite{Gasparticle-DEM-tsuji1993discrete, Gasparticle-review-Ge2017discrete}.
To improve the computational efficiency with limited number of tractable particles, the numerical parcel concept for grouping many particles with the same property is employed in the coarse graining particle method (CGPM) \cite{Gasparticle-coarse-grained-DEM-sakai2014verification, Gasparticle-coarse-grained-EMMS-DPM-gewei-lu2016computer}, and multiphase particle-in-cell (MP-PIC) \cite{Gasparticle-PIC-rourke-2009model}, etc. EL approach is theoretically able to give an accurate prediction of solid phase evolution in all regimes, but the computation cost increases gigantically in the dense particular flow for tracking the tremendous amount of solid particles or parcels and simulating their inter-particle collisions \cite{Gasparticle-review-van2008numerical, Gasparticle-book-gidaspow2010computational}. At current stage, the implementation of EL approach for industrial fluidization system is almost infeasible computationally, and the TFM is still the mainstream method in engineering applications \cite{Gasparticle-subgridmodel-EMMS-DPM-lu2014emms}.
In addition, the hybrid method that couples EE and EL approaches in different regions is studied in hope of maintaining both accuracy and efficiency in the simulation. The coupling strategy between EE/EL approaches plays an important role in order to achieve a smooth transition  and give a reliable prediction \cite{Gasparticle-hybrid-TFM-DEM-wang-lvjunfu-2015numerical, Gasparticle-hybrid-dynamic-multiscale-method-chen-wangjunwu-2016multiscale}.
Besides, other methods used for gas-solid fluidization system have been explored in the community, such as direct numerical simulation (DNS) \cite{Gasparticle-review-Kuipers-deen2014reviewDNS, Gasparticle-DNS-DEM-DFM-comparison-lu-gewei-2017assessing, Gasparticle-DNS-immersed-boundary-luokun-luo2019improved}, method of moment (MOM) \cite{Gasparticle-momentmethod-Fox2013computational, Gasparticle-hybrid-MOM-KTGF-kong2017solution}, and material point method (MPM) \cite{Gasparticle-MPM-baumgarten2019general}, etc.

To capture the non-equilibrium physics of particular flow, a multiscale numerical method GKS-UGKWP for gas-particle two-phase system
has been proposed, where the gas-kinetic scheme (GKS) is used for gas phase and unified gas-kinetic wave-particle method (UGKWP) for solid particle phase with dynamic coupling between them \cite{WP-six-gas-particle-yang2021unified, WP-gasparticle-dense-yang2022unified}.
UGKWP is a wave-particle version of the unified gas-kinetic scheme (UGKS) for multiscale flow dynamics simulations \cite{UGKS-xu2010unified, UGKS-book-xu2014direct}. UGKS is a direct modeling method on the scale of cell size and time step and captures the flow physics according to the  cell's $Kn$ number. It has been used for flow simulation in all regimes from the free molecular flow to the continuum Navier-Stokes solution.
Based on the same methodology, UGKS is also successfully extended to other multiscale transport problems,
such as radiative heat transfer, plasma, particular flow, etc \cite{UGKS-book-framework-xu2021cambridge, UGKS-radiative-sun2015asymptotic, UGKS-gas-particle-liu2019unified}.
In UGKS, both macroscopic flow variables and microscopic distribution function with discrete particle velocity points are updated in a deterministic way.
Later, in order to improve the efficiency of UGKS, especially for the hypersonic flow computation,
a particle-based UGKS, i.e., the so-called unified gas-kinetic particle (UGKP) method, has been proposed by updating the distribution function through stochastic particle \cite{WP-first-liu2020unified, WP-second-zhu-unstructured-mesh-zhu2019unified}.
In UGKP, the particles are categorized as free transport particles and collisional particles.
The collisionless particle will be tracked in the whole time step; while the collisional one is only tracked before the first collision and
eliminated within the time step. At the beginning of next time step, all annihilated particles will be re-sampled from an equilibrium state
determined by the updated macroscopic flow variables within each control volume (cell).
Furthermore, depending on the cell's Knudsen number $Kn_c$, in the next time step a proportion of (1-$e^{-1/Kn_c}$) re-sampled particles in UGKP
from the equilibrium state will get collision and be eliminated again. More importantly, it is realized that the contribution from these re-sampled particles to the flux function in a finite volume scheme can be evaluated analytically through a wave or field-type representation.
Therefore, in UGKP only the free transport particles need to be sampled and tracked in the next whole time step.
The scheme with analytical formulation for the flux transport from those collisional particles is called unified gas-kinetic wave-particle (UGKWP) method. Tremendous reduction in computation cost and memory requirement is achieved in UGKWP for high speed flow computations, especially in the transition and near continuum flow regimes \cite{WP-first-liu2020unified, WP-second-zhu-unstructured-mesh-zhu2019unified, WP-four-liu2020unified, WP-sample-xu2021modeling}. UGKWP is intrinsically suitable for capturing the multiscale non-equilibrium solid particle transport in the gas-particle two phase flow. At a very small cell $Kn_c$, no particles will be sampled in UGKWP, and the hydrodynamic formulation for solid particle evolution will automatically emerged. As a result, GKS-UGKWP method will go to the EE approach. On the contrary, at a large cell $Kn_c$, such as the collisionless regime, the evolution of the solid phase is fully determined by tracking the particle transport, and the GKS-UGKWP becomes the EL method. At an intermediate $Kn_c$, both wave and particle formulations in UGKWP contribute the solid phase's evolution, and the number of the sampled particles in UGKWP depends on the local $Kn_c$, which ensures a smooth transition between different regimes.
In addition, in the continuum regime, the particle phase UGKWP itself will automatically converge to the kinetic theory-based Navier-Stokes flow solver, the so-called gas-kinetic scheme (GKS), which is validated in the flow, acoustic wave, and turbulence simulation, etc \cite{GKS-2001, GKS-acoustic-zhao2019, GKS-turbulence-implicitHGKS-Cao2019, CompactGKS-ji2020-unstructured, GKS-HLLC-compare-yang2022comparison}.
In the gas-solid particle fluidization, the GKS is also employed for the gas phase with assumption of continuum flow.
Therefore, the limiting EE model from GKS-UGKWP will become GKS-GKS for the coupled two fluid phases.
In conclusion, due to the coexistence of dilute/transition/dense solid particle regimes, the multiscale GKS-UGKWP can recover EE and EL formulations seamlessly in a single gas-particle two phase flow simulation.

In the gas-solid particle two phase flow, the accurate evaluation of inter-phase interaction is also essential for the accurate simulation of   fluidization. The momentum and energy exchange between gas and solid phases due to the phase interaction is modeled through the drag force, buoyancy force, etc \cite{Gasparticle-book-gidaspow1994multiphase, Gasparticle-momentmethod-Fox2013computational}. Among them, drag force plays the dominant role \cite{Gasparticle-fluidized-turbulent-gao2012experimental, Gasparticle-drag-compare-zhongwenqi-yuaibing-xie2018mp}.
The hybrid model proposed by Gidaspow can be used for both dilute and dense solid particle regimes, and is widely accepted and employed in fluidization simulation \cite{Gasparticle-book-gidaspow1994multiphase}.
Due to the heterogeneous property of gas-solid fluidization, such as the existence of clustering, the drag model modified by a scaling factor was proposed for the further improvement of accuracy \cite{Gasparticle-drag-scaling-zhang2002effects, Gasparticle-drag-scaling-mckeen2003simulation, Gasparticle-fluidized-turbulent-gao2012experimental}.
In addition, the energy-minimization multiscale (EMMS) theory was successfully developed to model the heterogeneous structures
in the gas-solid fluidization problem \cite{Gasparticle-book-EMMS-li1994particle}.
As an extension of the original EMMS method, the EMMS drag model, including the effect of local heterogeneous flow structures through EMMS theory, was proposed and successfully employed in the gas-solid fluidization simulation from the schemes, such as TFM, MP-PIC, etc \cite{Gasparticle-subgridmodel-EMMS-drag-yang2004simulation, Gasparticle-subgridmodel-EMMS-drag-wang2007simulation, Gasparticle-subgridmodel-EMMS-drag-lu2011eulerian, Gasparticle-subgridmodel-EMMS-MPPIC-li2012mp, Gasparticle-subgridmodel-EMMS-DPM-lu2014emms}.

The numerical results will be compared with the experimental measurements.
This paper is organized as follows. Section 2 introduces the governing equations for the particle phase and UGKWP method. Then, Section 3 introduces the governing equations for the gas phase and GKS method.
Section 4 introduces the numerical experiments, where
two lab-scale fluidized bed problems, such as the turbulent fluidized bed from Gao et al \cite{Gasparticle-fluidized-turbulent-gao2012experimental} and circulating fluidized bed from Horio et al \cite{Gasparticle-fluidized-circulating-horio1988solid}, will be studied by GKS-UGKWP in three-dimensional space.
The last section is the conclusion.

\section{UGKWP for solid-particle phase}
\subsection{Governing equation for particle phase}
The evolution of particle phase is governed by the following kinetic equation,
\begin{gather}\label{particle phase kinetic equ}
\frac{\partial f_{s}}{\partial t}
+ \nabla_x \cdot \left(\textbf{u}f_{s}\right)
+ \nabla_u \cdot \left(\textbf{a}f_{s}\right)
= \frac{g_{s}-f_{s}}{\tau_{s}},
\end{gather}
where $f_{s}$ is the distribution function of particle phase, $\textbf{u}$ is the particle velocity, $\textbf{a}$ is the particle acceleration caused by the external force, $\nabla_x$ is the divergence operator with respect to space, $\nabla_u$ is the divergence operator with respect to velocity, $\tau_s$ is the relaxation time for the particle phase. The equilibrium state $g_{s}$ is,
\begin{gather*}
g_{s}=\epsilon_s\rho_s\left(\frac{\lambda_s}{\pi}\right)^{\frac{3}{2}}e^{-\lambda_s \left[(\textbf{u}-\textbf{U}_s)^2\right]},
\end{gather*}
where $\epsilon_s$ is the volume fraction of particle phase, $\rho_s$ is the material density of particle phase, $\lambda_s$ is the value relevant to the granular temperature $\theta$ with $\lambda_s = \frac{1}{2\theta}$, and $\textbf{U}_s$ is the macroscopic velocity of particle phase. The sum of kinetic and thermal energy for colliding particle may not be conserved due to the inelastic collision between particles. Therefore the collision term in Eq.\eqref{particle phase kinetic equ} should satisfy the following compatibility condition \cite{UGKS-gas-particle-liu2019unified},
\begin{equation}\label{particle phase compatibility condition}
\frac{1}{\tau_s} \int g_s \bm{\psi} \text{d}\textbf{u}=
\frac{1}{\tau_s} \int f_s \bm{\psi}' \text{d}\textbf{u},
\end{equation}
where $\bm{\psi}=\left(1,\textbf{u},\displaystyle \frac{1}{2}\textbf{u}^2\right)^T$ and $\bm{\psi}'=\left(1,\textbf{u},\displaystyle \frac{1}{2}\textbf{u}^2+\frac{e^2-1}{2}\left(\textbf{u}-\textbf{U}_s\right)^2\right)^T$. The lost energy due to inelastic collision in 3D can be written as,
\begin{gather*}
Q_{loss} = \frac{\left(1-e^2\right)3p_s}{2},
\end{gather*}
where $e\in\left[0,1\right]$ is the restitution coefficient for determining the percentage of lost energy in inelastic collision. While $e=1$ means no energy loss (elastic collision), $e=0$ refers to total loss of all internal energy of particle phase $\epsilon_s\rho_se_s =\frac{3}{2}p_s$ with $p_s=\frac{\epsilon_s\rho_s}{2\lambda_s}$.

The particle acceleration $\textbf{a}$ is determined by the external force, including the force reflecting the inter-phase interaction.
In this paper, the drag force $\textbf{D}$, the buoyancy force $\textbf{F}_b$, and gravity $\textbf{G}$ are considered. Here $\textbf{D}$ and $\textbf{F}_b$ are inter-phase force, standing for the force applied on the solid particles by gas flow. The general form of drag force can be written as,
\begin{gather}\label{drag force model}
\textbf{D} = \frac{m_s}{\tau_{st}}\left(\textbf{U}_g-\textbf{u}\right),
\end{gather}
where $m_s=\rho_s \frac{4}{3}\pi\left(\frac{d_s}{2}\right)^3$ is the mass of one particle, $d_s$ is the diameter of solid particle, $\textbf{U}_g$ is the macroscopic velocity of gas phase, and $\tau_{st}$ is the particle internal response time. The more commonly-used parameter in the drag model is $\beta$, called the inter-phase momentum transfer coefficient, with the relation to $\tau_{st}$ by $\beta=\frac{\epsilon_{s}\rho_{s}}{\tau_{st}}$. The accurate evaluation of drag plays key roles for the prediction of gas-solid fluidization by numerical method. Many studies about the modeling of drag force have been conducted, such as the widely-accepted model by Gidaspow \cite{Gasparticle-book-gidaspow1994multiphase}, the modified drag model through a scaling factor \cite{Gasparticle-drag-scaling-zhang2002effects, Gasparticle-drag-scaling-mckeen2003simulation, Gasparticle-fluidized-turbulent-gao2012experimental}, the EMMS-based drag model \cite{Gasparticle-subgridmodel-EMMS-drag-yang2004simulation, Gasparticle-subgridmodel-EMMS-drag-wang2007simulation, Gasparticle-subgridmodel-EMMS-drag-lu2011eulerian}, etc. Different drag models can be employed in GKS-UGKWP, and
the drag model particularly used in this paper will be introduced in detail later.

Another interactive force considered is the buoyancy force, which can be modeled as,
\begin{gather}\label{buoyancy force model}
\textbf{F}_b = -\frac{m_s}{\rho_{s}} \nabla_x p_g,
\end{gather}
where $p_g$ is the pressure of gas phase. Then, the particle's acceleration can be obtained as,
\begin{gather*}\label{particle phase acceleration term}
\textbf{a}=\frac{\textbf{D} + \textbf{F}_b}{m_s} + \textbf{G}.
\end{gather*}

\subsection{UGKWP method}
In this subsection, the UGKWP for the evolution of solid particle phase is introduced. Generally, the kinetic equation of particle phase Eq.\eqref{particle phase kinetic equ} is split as,
\begin{align}
\label{particle phase kinetic equ without acce}
\mathcal{L}_{s1} &:~~ \frac{\partial f_{s}}{\partial t}
+ \nabla_x \cdot \left(\textbf{u}f_{s}\right)
= \frac{g_{s}-f_{s}}{\tau_{s}}, \\
\label{particle phase kenetic equ only acce}
\mathcal{L}_{s2} &:~~ \frac{\partial f_{s}}{\partial t}
+ \nabla_u \cdot \left(\textbf{a}f_{s}\right)
= 0,
\end{align}
and splitting operator is used to solve Eq.\eqref{particle phase kinetic equ}. Firstly, we focus on $\mathcal{L}_{s1}$ part, the particle phase kinetic equation without external force,
\begin{gather*}
\frac{\partial f_{s}}{\partial t}
+ \nabla_x \cdot \left(\textbf{u}f_{s}\right)
= \frac{g_{s}-f_{s}}{\tau_{s}}.
\end{gather*}
For brevity, the subscript $s$ standing for the solid particle phase will be neglected in this subsection. The integration solution of the kinetic equation can be written as,
\begin{equation}\label{particle phase integration solution}
f(\textbf{x},t,\textbf{u})=\frac{1}{\tau}\int_0^t g(\textbf{x}',t',\textbf{u} )e^{-(t-t')/\tau}\text{d}t'\\
+e^{-t/\tau}f_0(\textbf{x}-\textbf{u}t, \textbf{u}),
\end{equation}
where $\textbf{x}'=\textbf{x}+\textbf{u}(t'-t)$ is the trajectory of particles, $f_0$ is the initial gas distribution function at time $t=0$, and $g$ is the corresponding equilibrium state.

In UGKWP, both macroscopic conservative variables and microscopic gas distribution function need to be updated. Generally, in the finite volume framework, the cell-averaged macroscopic variables $\textbf{W}_i$ of cell $i$ can be updated by the conservation law,
\begin{gather}
\textbf{W}_i^{n+1} = \textbf{W}_i^n - \frac{1}{\Omega_i} \sum_{S_{ij}\in \partial \Omega_i}\textbf{F}_{ij}S_{ij} + \Delta t \textbf{S}_{i},
\end{gather}
where $\textbf{W}_i=\left(\rho_i, \rho_i \textbf{U}_i, \rho_i E_i\right)$ is the cell-averaged macroscopic variables,
\begin{gather*}
\textbf{W}_i = \frac{1}{\Omega_{i}}\int_{\Omega_{i}} \textbf{W}\left(\textbf{x}\right) \text{d}\Omega,
\end{gather*}
$\Omega_i$ is the volume of cell $i$, $\partial\Omega_i$ denotes the set of cell interfaces of cell $i$, $S_{ij}$ is the area of the $j$-th interface of cell $i$, $\textbf{F}_{ij}$ denotes the macroscopic fluxes across the interface $S_{ij}$, which can be written as
\begin{align}\label{particle phase Flux equation}
\textbf{F}_{ij}=\int_{0}^{\Delta t} \int \textbf{u}\cdot\textbf{n}_{ij} f_{ij}(\textbf{x},t,\textbf{u}) \bm{\psi} \text{d}\textbf{u}\text{d}t,
\end{align}
where $\textbf{n}_{ij}$ is the normal unit vector of interface $S_{ij}$, $f_{ij}\left(t\right)$ is the time-dependent distribution function on the interface $S_{ij}$, and $\bm{\psi}=(1,\textbf{u},\displaystyle \frac{1}{2}\textbf{u}^2)^T$. $\textbf{S}_{i}$ is the source term due to inelastic collision inside each control volume, where the solid-particle's internal energy has not been taken into account in the above equation.

Substituting the time-dependent distribution function Eq.\eqref{particle phase integration solution} into Eq.\eqref{particle phase Flux equation}, the fluxes can be obtained,
\begin{align*}
\textbf{F}_{ij}
&=\int_{0}^{\Delta t} \int \textbf{u}\cdot\textbf{n}_{ij} f_{ij}(\textbf{x},t,\textbf{u}) \bm{\psi} \text{d}\textbf{u}\text{d}t\\
&=\int_{0}^{\Delta t} \int\textbf{u}\cdot\textbf{n}_{ij} \left[ \frac{1}{\tau}\int_0^t g(\textbf{x}',t',\textbf{u})e^{-(t-t')/\tau}\text{d}t' \right] \bm{\psi} \text{d}\textbf{u}\text{d}t\\
&+\int_{0}^{\Delta t} \int\textbf{u}\cdot\textbf{n}_{ij} \left[ e^{-t/\tau}f_0(\textbf{x}-\textbf{u}t,\textbf{u})\right] \bm{\psi} \text{d}\textbf{u}\text{d}t\\
&\overset{def}{=}\textbf{F}^{eq}_{ij} + \textbf{F}^{fr}_{ij}.
\end{align*}

The procedure of obtaining the local equilibrium state $g_0$ at the cell interface as well as the construction of $g\left(t\right)$ is the same as that in GKS \cite{GKS-2001}.
For a second-order accuracy, the equilibrium state $g$ around the cell interface is written as,
\begin{gather*}
g\left(\textbf{x}',t',\textbf{u}\right)=g_0\left(\textbf{x},\textbf{u}\right)
\left(1 + \overline{\textbf{a}} \cdot \textbf{u}\left(t'-t\right) + \bar{A}t'\right),
\end{gather*}
where $\overline{\textbf{a}}=\left[\overline{a_1}, \overline{a_2}, \overline{a_3}\right]^T$, $\overline{a_i}=\frac{\partial g}{\partial x_i}/g$, $i=1,2,3$,  $\overline{A}=\frac{\partial g}{\partial t}/g$, and $g_0$ is the local equilibrium on the interface.
Specifically, the coefficients of spatial derivatives $\overline{a_i}$ can be obtained from the corresponding derivatives of the macroscopic variables,
\begin{equation*}
\left\langle \overline{a_i}\right\rangle=\partial \textbf{W}_0/\partial x_i,
\end{equation*}
where $i=1,2,3$, and $\left\langle...\right\rangle$ means the moments of the Maxwellian distribution functions,
\begin{align*}
\left\langle...\right\rangle=\int \bm{\psi}\left(...\right)g\text{d}\textbf{u}.
\end{align*}
The coefficients of temporal derivative $\overline{A}$ can be determined by the compatibility condition,
\begin{equation*}
\left\langle \overline{\textbf{a}} \cdot \textbf{u}+\overline{A} \right\rangle =
\left[\begin{array}{c}
0\\
\textbf{0}\\
-\frac{Q_{loss}}{\tau_s}
\end{array}\right].
\end{equation*}
where $Q_{loss}=\frac{\left(1-e^2\right)3p_s}{2}$ is the energy lose due to particle-particle inelastic collision. Now, all the coefficients in the equilibrium state $g\left(\textbf{x}',t',\textbf{u}\right)$ have been determined, and its integration becomes,
\begin{gather}
f^{eq}(\textbf{x},t,\textbf{u}) \overset{def}{=} \frac{1}{\tau}\int_0^t g(\textbf{x}',t',\textbf{u})e^{-(t-t')/\tau}\text{d}t' \nonumber\\
= c_1 g_0\left(\textbf{x},\textbf{u}\right)
+ c_2 \overline{\textbf{a}} \cdot \textbf{u} g_0\left(\textbf{x},\textbf{u}\right)
+ c_3 A g_0\left(\textbf{x},\textbf{u}\right),
\end{gather}
with coefficients,
\begin{align*}
c_1 &= 1-e^{-t/\tau}, \\
c_2 &= \left(t+\tau\right)e^{-t/\tau}-\tau, \\
c_3 &= t-\tau+\tau e^{-t/\tau},
\end{align*}
and thereby the integrated flux over a time step for equilibrium state can be obtained,
\begin{gather*}
\textbf{F}^{eq}_{ij}
=\int_{0}^{\Delta t} \int \textbf{u}\cdot\textbf{n}_{ij} f_{ij}^{eq}(\textbf{x},t,\textbf{u})\bm{\psi}\text{d}\textbf{u}\text{d}t.
\end{gather*}

Besides, the flux contribution from the particle free transport $f_0$ in Eq.\eqref{particle phase integration solution} is calculated by tracking the particles sampled from $f_0$. Therefore, the updating of the cell-averaged macroscopic variables can be written as,
\begin{gather}\label{particle phase equ_updateW_ugkp}
\textbf{W}_i^{n+1} = \textbf{W}_i^n - \frac{1}{\Omega_i} \sum_{S_{ij}\in \partial \Omega_i}\textbf{F}^{eq}_{ij}S_{ij}
+ \frac{\textbf{w}_{i}^{fr}}{\Omega_{i}}
+ \Delta t \textbf{S}_{i},
\end{gather}
where $\textbf{w}^{fr}_i$ is the net free streaming flow of cell $i$, standing for the flux contribution of the free streaming of particles, and the term $\textbf{S}_{i} = \left[0,\textbf{0},-\frac{Q_{loss}}{\tau_s}\right]^T$ is the source term due to the inelastic collision for solid particle phase.

The net free streaming flow $\textbf{w}^{fr}_i$ is determined in the following. The evolution of particle should also satisfy the integral solution of the kinetic equation, which can be written as,
\begin{equation}
f(\textbf{x},t,\textbf{u})
=\left(1-e^{-t/\tau}\right)g^{+}(\textbf{x},t,\textbf{u})
+e^{-t/\tau}f_0(\textbf{x}-\textbf{u}t,\textbf{u}),
\end{equation}
where $g^{+}$ is named as the hydrodynamic distribution function with analytical formulation. The initial distribution function $f_0$ has a probability of $e^{-t/\tau}$ to free transport and $(1-e^{-t/\tau})$ to colliding with other particles. The post-collision particles satisfies the distribution $g^+\left(\textbf{x},\textbf{u},t\right)$. The free transport time before the first collision with other particles is denoted as $t_c$. The cumulative distribution function of $t_c$ is,
\begin{gather}\label{particle phase wp cumulative distribution}
F\left(t_c < t\right) = 1 - e^{-t/ \tau},
\end{gather}
and therefore $t_c$ can be sampled as $t_c=-\tau \text{ln}\left(\eta\right)$, where $\eta$ is a random number generated from a uniform distribution $U\left(0,1\right)$. Then, the free streaming time $t_f$ for each particle is determined separately by,
\begin{gather}
t_f = min\left[-\tau\text{ln}\left(\eta\right), \Delta t\right],
\end{gather}
where $\Delta t$ is the time step. Therefore, within one time step, all particles can be divided into two groups: the collisionless particle and the collisional particle, and they are determined by the relation between of time step $\Delta t$ and free streaming time $t_f$. Specifically, if $t_f=\Delta t$ for one particle, it is collisionless one, and the trajectory of this particle is fully tracked in the whole time step. On the contrary, if $t_f<\Delta t$ for one particle, it is collisional particle, and its trajectory will be tracked until $t_f$. The collisional particle is eliminated at $t_f$ in the simulation and the associated mass, momentum and energy carried by this particle are merged into the updated macroscopic quantities of all annihilated particles in the relevant cell. More specifically, the particle trajectory in the free streaming process within time $t_f$ is tacked by,
\begin{gather}
\textbf{x} = \textbf{x}^n + \textbf{u}^n t_f .
\end{gather}
The term $\textbf{w}_{i}^{fr}$ can be calculated by counting the particles passing through the interfaces of cell $i$,
\begin{gather}
\textbf{w}_{i}^{fr} = \sum_{k\in P\left(\partial \Omega_{i}^{+}\right)} \bm{\phi}_k - \sum_{k\in P\left(\partial \Omega_{i}^{-}\right)} \bm{\phi}_k,
\end{gather}
where $P\left(\partial \Omega_{i}^{+}\right)$ is the particle set moving into the cell $i$ during one time step, $P\left(\partial \Omega_{i}^{-}\right)$ is the particle set moving out of the cell $i$ during one time step, $k$ is the particle index in one specific set, and $\bm{\phi}_k=\left[m_{k}, m_{k}\textbf{u}_k, \frac{1}{2}m_{k}(\textbf{u}^2_k)\right]^T$ is the mass, momentum and energy carried by particle $k$. Therefore, $\textbf{w}_{i}^{fr}/\Omega_{i}$ is the net conservative quantities caused by the free stream of the tracked particles. Now, all the terms in Eq.\eqref{particle phase equ_updateW_ugkp} have been determined and the macroscopic variables $\textbf{W}_i$ can be updated.

The trajectories of all particles have been tracked during the time interval $\left(0, t_f\right)$. For the collisionless particles with $t_f=\Delta t$, they still survive at the end of one time step; while the collisional particles with $t_f<\Delta t$ are deleted after their first collision and they are supposed to go to the equilibrium state in that cell. Therefore, the macroscopic variables of the collisional particles in cell $i$ at the end of each time step can be directly obtained based on the conservation law,
\begin{gather}
\textbf{W}^h_i = \textbf{W}^{n+1}_i - \textbf{W}^p_i,
\end{gather}
where $\textbf{W}^{n+1}_i$ is the updated conservative variables in Eq.\eqref{particle phase equ_updateW_ugkp} and  $\textbf{W}^p_i$ are the  mass, momentum, and energy of remaining collisionless particles in the cell at the end of the time step.
Besides, the macroscopic variables $\textbf{W}^h_i$ account for all eliminated collisional particles to the equilibrium state,
and these particles can be re-sampling from $\textbf{W}^h_i$ based on the overall Maxwellian distribution at the beginning of the next time step.
Now the updates of both macroscopic variables and the microscopic particles have been presented. The above method is the so-called unified gas-kinetic particle (UGKP) method.

The above UGKP can be further developed to UGKWP method.
In UGKP method, all particles are divided into collisionless and collisional particles in each time step. The collisional particles are deleted after the first collision and re-sampled from $\textbf{W}^h_i$ at the beginning of the next time step.
However, only the collisionless part of the re-samples particles can survive in the next time step, and all collisional ones will be deleted again.
Actually, the transport fluxes from these collisional particles can be evaluated analytically without using particles.
According to the cumulative distribution Eq.\eqref{particle phase wp cumulative distribution}, the proportion of the collisionless particles is $e^{-\Delta t/\tau}$, and therefore in UGKWP only the collisionless particles from the hydrodynamic variables $\textbf{W}^{h}_i$ in cell $i$ will  be re-sampled with the total mass, momentum, and energy,
\begin{gather}
\textbf{W}^{hp}_i = e^{-\Delta t/\tau} \textbf{W}^{h}_i.
\end{gather}
Then, the free transport time of all the re-sampled particles will be $t_f=\Delta t$ in UGKWP.
The fluxes $\textbf{F}^{fr,wave}$ from these un-sampled collisional particle of $ (1- e^{-\Delta t/\tau} )\textbf{W}^{h}_i$ can be evaluated
analytically \cite{WP-first-liu2020unified, WP-second-zhu-unstructured-mesh-zhu2019unified}.
Now, same as UGKP, the net flux $\textbf{w}_{i}^{fr,p}$ by the free streaming of the particles, which include remaining particles from previous time step and re-sampled collisionless ones, in UGKWP can be calculated by
\begin{gather}
\textbf{w}_{i}^{fr,p} = \sum_{k\in P\left(\partial \Omega_{i}^{+}\right)} \bm{\phi}_k - \sum_{k\in P\left(\partial \Omega_{i}^{-}\right)} \bm{\phi}_k.
\end{gather}
Then, the macroscopic flow variables in UGKWP are updated by
\begin{gather}\label{particle phase wp final update W}
\textbf{W}_i^{n+1} = \textbf{W}_i^n
- \frac{1}{\Omega_i} \sum_{S_{ij}\in \partial \Omega_i}\textbf{F}^{eq}_{ij}S_{ij}
- \frac{1}{\Omega_i} \sum_{S_{ij}\in \partial \Omega_i}\textbf{F}^{fr,wave}_{ij}S_{ij}
+ \frac{\textbf{w}_{i}^{fr,p}}{\Omega_{i}}
+ \Delta t \textbf{S}_{i},
\end{gather}
where $\textbf{F}^{fr,wave}_{ij}$ is the flux function from the un-sampled collisional particles \cite{WP-first-liu2020unified, WP-second-zhu-unstructured-mesh-zhu2019unified, UGKS-book-framework-xu2021cambridge}, which can be written as,
\begin{align*}
\textbf{F}^{fr,wave}_{ij}
&=\textbf{F}^{fr,UGKS}_{ij}(\textbf{W}^h_i) - \textbf{F}^{fr,DVM}_{ij}(\textbf{W}^{hp}_i) \\
&=\int_{0}^{\Delta t} \int \textbf{u} \cdot \textbf{n}_{ij} \left[ e^{-t/\tau}f_0(\textbf{x}-\textbf{u}t,\textbf{u})\right] \bm{\psi} \text{d}\textbf{u}\text{d}t\\
&-e^{-\Delta t/\tau}\int_{0}^{\Delta t} \int \textbf{u} \cdot \textbf{n}_{ij} \left[g_0^h\left(\textbf{x},\textbf{u} \right) - t\textbf{u} \cdot g_\textbf{x}^h\left(\textbf{x},\textbf{u} \right) \right] \bm{\psi}\text{d}\textbf{u}\text{d}t\\
&=\int \textbf{u} \cdot \textbf{n}_{ij} \left[ \left(q_4  - \Delta t e^{-\Delta t/\tau}\right) g_0^h \left(\textbf{x},\textbf{u} \right)
+ \left(q_5 + \frac{\Delta t^2}{2}e^{-\Delta t/\tau}\right) \textbf{u} \cdot g_\textbf{x}^h\left(\textbf{x},\textbf{u} \right) \right]\bm{\psi}\text{d}\textbf{u},
\end{align*}
with the coefficients,
\begin{align*}
q_4&=\tau\left(1-e^{-\Delta t/\tau}\right), \\
q_5&=\tau\Delta te^{-\Delta t/\tau} - \tau^2\left(1-e^{-\Delta t/\tau}\right).
\end{align*}

The second part $\mathcal{L}_{s2}$ in Eq.\eqref{particle phase kenetic equ only acce} accounts for the external acceleration,
\begin{gather*}
\frac{\partial f_{s}}{\partial t}
+ \nabla_u \cdot \left(\textbf{a}f_{s}\right)
= 0,
\end{gather*}
where the velocity-dependent acceleration term caused by inter-phase forces and solid particle's gravity has the following form,
\begin{gather*}
\textbf{a} = \frac{\textbf{U}_g - \textbf{u}}{\tau_{st}} - \frac{1}{\rho_{s}} \nabla_x p_g + \textbf{G}.
\end{gather*}
Taking moment $\bm{\psi}$ to Eq.\eqref{particle phase kenetic equ only acce},
\begin{gather*}
\int \bm{\psi}
\left( \frac{\partial f_{s}}{\partial t}
+ \textbf{a} \cdot \nabla_u f_{s}
+ f_{s}\nabla_u \cdot \textbf{a}
\right) \text{d}\textbf{u} = 0,
\end{gather*}
and in the Euler regime with $f_s = g_s + \mathcal{O}\left(\tau_{s}\right)$, we can obtain,
\begin{gather*}
\frac{\partial \textbf{W}_s}{\partial t} + \textbf{Q}_s= 0,
\end{gather*}
where
\begin{gather*}
\textbf{W}_s=\left[\begin{array}{c}
\epsilon_s\rho_s\\
\epsilon_s\rho_s \textbf{U}_s\\
\epsilon_s\rho_s E_s
\end{array}
\right], ~~
\textbf{Q}_s=\left[\begin{array}{c}
0 \\
\frac{\epsilon_s\rho_s\left(\textbf{U}_s-\textbf{U}_g\right)}{\tau_{st}}
+\epsilon_s \nabla_x p_g
- \epsilon_{s}\rho_{s} \textbf{G} \\
\frac{\epsilon_s\rho_{s}\textbf{U}_s \cdot \left(\textbf{U}_s-\textbf{U}_g\right)}{\tau_{st}} +3\frac{p_s}{\tau_{st}}
+ \epsilon_s\textbf{U}_s \cdot \nabla_x p_g
- \epsilon_{s}\rho_{s} \textbf{U}_s \cdot \textbf{G}
\end{array}\right].
\end{gather*}
When the first-order forward Euler method is employed for time marching, the cell-averaged macroscopic variable can be updated by,
\begin{gather}\label{update macroscopic variable of acceleration wave}
\textbf{W}^{n+1}_s = \textbf{W}_s - \textbf{Q}_s \Delta t,
\end{gather}
and the modifications on velocity and location of the remaining free transport particles can be written as,
\begin{align}
\textbf{u}^{n+1} &= \textbf{u} + \textbf{a}t_f,\\
\textbf{x}^{n+1} &= \textbf{x} + \frac{\textbf{a}}{2} t_f^2.\label{displacement by acceleartion term}
\end{align}
Now the update of the solid particle phase in one time step has been finished. In the following, specific variables determination for the solid-particle
phase will be presented.

\subsection{Particle phase Knudsen number}
The particle phase Knudsen number $Kn$ is defined by the ratio of collision time $\tau_{s}$ to the characteristic time of macroscopic flow $t_{ref}$,
\begin{gather}\label{particle phase Kn_s}
Kn = \frac{\tau_s}{t_{ref}},
\end{gather}
where $t_{ref}$ is the characteristic time, defined as the ratio of flow characteristic length to the flow characteristic velocity, $t_{ref}=L_{ref}/U_{ref}$, and $\tau_s$ is the time interval between collisions of solid particles. In this paper, $\tau_s$ is taken as \cite{Gasparticle-MOM-Fox-passalacqua2010fully, Gasparticle-momentmethod-Fox2013computational},
\begin{gather}\label{particle phase tau_s}
\tau_s = \frac{\sqrt{\pi}d_s}{12\epsilon_sg_r \sqrt{\theta}}  ,
\end{gather}
where $d_s$ is the diameter of solid particle, $\epsilon_s$ is the volume fraction of solid phase, $\theta$ is the granular temperature. $g_r$ is the radial distribution function with the following form,
\begin{gather}\label{radial distribution g0}
g_r = \frac{2-c}{2\left(1-c\right)^3},
\end{gather}
where $c=\epsilon_s/\epsilon_{s,max}$ is the ratio of the volume fraction $\epsilon_{s}$ to the allowed maximum value $\epsilon_{s,max}$. A typical feature of the gas-solid flow in fluidized bed is that the instantaneously coexistence of the dilute and dense zones. Generally, in the dilute zone, the collision frequency between solid particles is low, leading to a large $Kn$, and in UGKWP particles will be sampled and tracked to model the transport behavior of solid particles; on the contrary, for the dense flow, the high-frequency inter-particle collisions usually make the solid phase in equilibrium state, so in UGKWP the evolution can be fully determined the wave formula in Eq.\eqref{particle phase wp final update W}, and there is no need for particle sampling. The solid particles' behaviors and flow states can be directly modeled in the UGKWP based on $Kn$, ensuring the consistence of numerical scheme with flow physics.

\subsection{Hydrodynamic equations in continuum flow regime}
When the collision between solid particles are elastic with $e=1$, in the continuum flow regime with $f_s = g_s + \mathcal{O}\left(\tau_s\right)$, the hydrodynamic equations becomes the Euler equations coupled with the momentum and energy exchange terms, which can be obtained based on the Chapman-Enskog asymptotic analysis for the kinetic equation Eq.\eqref{particle phase kinetic equ} \cite{CE-expansion},
\begin{align}\label{particle phase Euler equ}
&\frac{\partial \left(\epsilon_s\rho_s\right)}{\partial t}
+ \nabla_x \cdot \left(\epsilon_s\rho_s \textbf{U}_s\right) = 0,\nonumber \\
&\frac{\partial \left(\epsilon_s\rho_s \textbf{U}_s\right)}{\partial t}
+ \nabla_x \cdot \left(\epsilon_s\rho_s \textbf{U}_s \textbf{U}_s + p_s \mathbb{I} \right)
= \frac{\epsilon_{s}\rho_{s}\left(\textbf{U}_g - \textbf{U}_s\right)}{\tau_{st}}
- \epsilon_{s} \nabla_x p_g
+ \epsilon_{s}\rho_{s} \textbf{G} , \\
&\frac{\partial \left(\epsilon_s\rho_s E_s\right)}{\partial t}
+ \nabla_x \cdot \left(\left(\epsilon_s\rho_s E_s  + p_s\right) \textbf{U}_s \right)
= \frac{\epsilon_{s}\rho_{s}\textbf{U}_s \cdot \left(\textbf{U}_g - \textbf{U}_s\right)}{\tau_{st}}
- \frac{3p_s}{\tau_{st}}
- \epsilon_{s} \textbf{U}_s \cdot \nabla_x p_g
+ \epsilon_{s}\rho_{s} \textbf{U}_s \cdot \textbf{G}.\nonumber
\end{align}

With the increasing of solid volume fraction, the inter-particle interaction becomes more complex, and a precise evaluation of solid phase's pressure becomes difficult. The pressure term $p_s$ in Eq.\eqref{particle phase Euler equ} is the so-called kinetic pressure, which plays the dominant role in the dilute and moderately dense regime. Besides the kinetic pressure part $p_s$, the collisional pressure $p_c$ closed by KTGF and the frictional pressure $p_f$ reflecting the effect of enduring inter-particle contact and frictions are widely accepted and employed in TFM, which shows excellent performance in the gas-solid fluidization problems \cite{Gasparticle-KTGF-lun1984kinetic, Gasparticle-review-WangJunwu2020continuum}. Many studies to improve the accuracy of pressure/stress terms are conducted \cite{Gasparticle-pressure-friction-schneiderbauer2012comprehensive, Gasparticle-pressue-all-chialvo2013modified, Gasparticle-pressue-all-luhuilin-zhao2020comprehensive}. To the authors' knowledge, however, no such a model giving accurate kinetic/collisional/frictional pressure in a multi-scale solver for dilute/moderately dense/dense flow has been proposed.
So in this paper as the first attempt, the models of $p_c$ and $p_f$ widely used in TFM are directly added to the macroscopic variables in the UGKWP method. The collisional pressure $p_c$, proposed by Lun et al. \cite{Gasparticle-KTGF-lun1984kinetic}, is widely employed in the gas-solid flow in fluidized beds, which is used in this paper and can be written as,
\begin{equation*}
p_c = 2\left(1+e\right)\epsilon_{s}^2 \rho_s \theta g_r,
\end{equation*}
where $e$ is the restitution coefficient, taken as 0.8 in this paper unless special notification, and $g_r$ is the radial distribution function given by Eq.\eqref{radial distribution g0}.
The $p_{f}$ accounts for the enduring inter-particle contacts and frictions, which plays important roles when the solid phase is in the near-packing state. Some models of $p_{f}$ have been proposed \cite{Gasparticle-KTGF-pressure-friction-johnson1987frictional, Gasparticle-pressure-friction-srivastava2003analysis, Gasparticle-pressure-friction-schneiderbauer2012comprehensive}. In this paper, the correlation proposed by Johnson and Jackson is employed \cite{Gasparticle-KTGF-pressure-friction-johnson1987frictional, Gasparticle-TFM-compressible-houim2016multiphase},
\begin{align}
p_{f} = \left\{\begin{aligned}
&~~~~~~~~ 0 &  ,   & \epsilon_{s} \le \epsilon_{s,crit}, \\
&0.1 \epsilon_{s} \frac{\left(\epsilon_{s} - \epsilon_{s,crit}\right)^2}{\left(\epsilon_{s,max} -  \epsilon_{s}\right)^5}&  ,   & \epsilon_{s} > \epsilon_{s,crit},
\end{aligned} \right.
\end{align}
where $p_{f}$ is with unit of $Pa$. $\epsilon_{s,crit}$ is the critical volume fraction of particle flow, and it takes a value $0.5$ in this paper unless special notification. In this paper, both $p_c$ and $p_f$ are considered to recover a more realistic physics. Finally, the momentum equation under continuum limiting regime can be written as,
\begin{gather}\label{particle phase momentum equ equ with p_fr}
\frac{\partial \left(\epsilon_s\rho_s \textbf{U}_s\right)}{\partial t}
+ \nabla_x \cdot \left(\epsilon_s\rho_s \textbf{U}_s \textbf{U}_s + p_s \mathbb{I} + p_{c} \mathbb{I} + p_{f} \mathbb{I}\right)
= \frac{\epsilon_{s}\rho_{s}\left(\textbf{U}_g - \textbf{U}_s\right)}{\tau_{st}}
- \epsilon_{s} \nabla_x p_g
+ \epsilon_{s}\rho_{s} \textbf{G}.
\end{gather}
\begin{gather}\label{particle phase energy equ equ with p_fr}
\frac{\partial \left(\epsilon_s\rho_s E_s\right)}{\partial t}
+ \nabla_x \cdot \left(\left(\epsilon_s\rho_s E_s  + p_s + p_{c} + p_{f}\right) \textbf{U}_s \right)
= \frac{\epsilon_{s}\rho_{s}\textbf{U}_s \cdot \left(\textbf{U}_g - \textbf{U}_s\right)}{\tau_{st}}
- \frac{3p_s}{\tau_{st}}
- \epsilon_{s} \textbf{U}_s \cdot \nabla_x p_g \nonumber\\
~~~~~~~~~~~~~~~~~~~~~~~~~~
+ \epsilon_{s}\rho_{s} \textbf{U}_s \cdot \textbf{G}.
\end{gather}
The terms relevant to collisional pressure, $\nabla_x \cdot \left(p_{c} \mathbb{I}\right)$, $\nabla_x \cdot \left(p_{c} \textbf{U}_s\right)$, and frictional pressure, $\nabla_x \cdot \left(p_{f} \mathbb{I}\right)$, $\nabla_x \cdot \left(p_{f} \textbf{U}_s\right)$, are solved as source terms in this paper.
To avoid the solid volume fraction $\epsilon_{s}$ exceeding its maximum value $\epsilon_{s,max}$, the flux limiting model near the packing condition, proposed in our previous work, is employed in UGKWP method for solid phase and isn't reiterated here \cite{WP-gasparticle-dense-yang2022unified}.

\section{GKS for gas phase}
\subsection{Governing equation for gas phase}
The gas phase is regarded as continuum flow and the governing equations are the Navier-Stokes equations with source terms reflecting the inter-phase interaction \cite{Gasparticle-book-gidaspow1994multiphase, Gasparticle-book-ishii2010thermo},
\begin{align}\label{gas phase macroscopic equ}
&\frac{\partial \left(\widetilde{\rho_g}\right)}{\partial t}
+ \nabla_x \cdot \left(\widetilde{\rho_g} \textbf{U}_g\right)= 0,\nonumber \\
&\frac{\partial \left(\widetilde{\rho_g} \textbf{U}_g\right)}{\partial t}
+ \nabla_x \cdot \left(\widetilde{\rho_g} \textbf{U}_g \textbf{U}_g + \widetilde{p_g}\mathbb{I}\right)
- \epsilon_{g} \nabla_x \cdot \left(\mu_g \bm{\sigma}\right)
=
p_g \nabla_x \epsilon_{g}
-\frac{\epsilon_{s}\rho_{s}\left(\textbf{U}_g - \textbf{U}_s\right)}{\tau_{st}}
+ \rho_g \textbf{G}, \\
&\frac{\partial \left(\widetilde{\rho_g} E_g\right)}{\partial t}
+ \nabla_x \cdot \left(\left(\widetilde{\rho_g} E_g  + \widetilde{p_g}\right) \textbf{U}_g \right)
- \epsilon_{g} \nabla_x \cdot \left(\mu_g \bm{\sigma}\cdot\textbf{U}_g - \kappa \nabla_x T_g \right)
=
- p_{g} \frac{\partial \epsilon_{g}}{\partial t} \nonumber \\
& ~~~~~~~~~~~~~~~~~~~~~~~~~~~~~~~~~~~~~~~~~~~~~~~~~~
-\frac{\epsilon_{s}\rho_{s}\textbf{U}_s \cdot \left(\textbf{U}_g - \textbf{U}_s\right)}{\tau_{st}}
+ \frac{3p_s}{\tau_{st}}
+ \rho_g \textbf{U}_g \cdot \textbf{G}, \nonumber
\end{align}
where $\widetilde{\rho_g}=\epsilon_{g}\rho_g$ is the apparent density of gas phase, $p_g=\rho_gRT_g$ is the pressure of gas phase and $\widetilde{p_g}=\widetilde{\rho_g}RT_g$, the strain rate tensor $\bm{\sigma}$ is
\begin{gather*}
\bm{\sigma} = \nabla_x\textbf{U}_g + \left(\nabla_x\textbf{U}_g\right)^T
- \frac{2}{3} \nabla_x \cdot \textbf{U}_g \mathbb{I},
\end{gather*}
and
\begin{gather*}
\mu_g = \tau_{g} p_g, ~~~~ \kappa = \frac{5}{2} R \tau_{g} p_g.
\end{gather*}
In particular, at the right hand side in Eq.\eqref{gas phase macroscopic equ}, the term $p_{g} \nabla_x \epsilon_{g}$ is called ``nozzle" term, and the associated work term $- p_{g} \frac{\partial \epsilon_{g}}{\partial t}$ is called $pDV$ work term, since it is similar to the $pDV$ term in the quasi-one-dimensional gas nozzle flow equations \cite{Gasparticle-TFM-compressible-houim2016multiphase}. Unphysical pressure fluctuations might occurs if the ``nozzle" term and $pDV$ term are not solved correctly. According to \cite{Toro2013book}, Eq.\eqref{gas phase macroscopic equ} can be written as the following form,
\begin{align}\label{gas phase macroscopic equ final}
&\frac{\partial \left(\rho_g\right)}{\partial t}
+ \nabla_x \cdot \left(\rho_g \textbf{U}_g\right)= C_{\epsilon_g}\rho_g,\nonumber \\
&\frac{\partial \left(\rho_g \textbf{U}_g\right)}{\partial t}
+ \nabla_x \cdot \left(\rho_g \textbf{U}_g \textbf{U}_g + p_g\mathbb{I} - \mu_g \bm{\sigma}\right)
=
C_{\epsilon_g} \rho_g \textbf{U}_g
-\frac{\epsilon_{s}\rho_{s}\left(\textbf{U}_g - \textbf{U}_s\right)}{\epsilon_g \tau_{st}}
+ \frac{\rho_g\textbf{G}}{\epsilon_{g}}, \\
&\frac{\partial \left(\rho_g E_g\right)}{\partial t}
+ \nabla_x \cdot \left(\left(\rho_g E_g  + p_g\right) \textbf{U}_g
- \mu_g \bm{\sigma}\cdot\textbf{U}_g + \kappa \nabla_x T_g \right)
=
C_{\epsilon_g} \left(\rho_g E_g + p_g\right) \nonumber \\
& ~~~~~~~~~~~~~~~~~~~~~~~~~~~~~~~~~~~~~~~~~~
-\frac{\epsilon_{s}\rho_{s}\textbf{U}_s \cdot \left(\textbf{U}_g - \textbf{U}_s\right)}{\epsilon_g \tau_{st}}
+ \frac{3p_s}{\epsilon_g \tau_{st}}
+ \frac{\rho_g \textbf{U}_g \cdot \textbf{G}}{\epsilon_{g}}, \nonumber
\end{align}
where, $C_{\epsilon_g} = -\frac{1}{\epsilon_{g}}\frac{\text{d}\epsilon_{g}}{\text{d}t}$ with $\frac{\text{d}\epsilon_{g}}{\text{d}t}=\frac{\partial \epsilon_{g}}{\partial t}+\textbf{U}_g \cdot \nabla\epsilon_{g}$, and how to solve $C_{\epsilon_{g}}$ in this paper will be introduced later.

\subsection{GKS for gas evolution}
This subsection introduces the evolution of gas phase in gas-particle two-phase system. The gas flow is governed by the Navier-Stokes equations with the inter-phase interaction, and the corresponding  GKS is a limiting scheme of UGKWP in the continuum flow regime. In general, the evolution of gas phase Eq.\eqref{gas phase macroscopic equ final} can be split into two parts,

\begin{align}
\mathcal{L}_{g1}&:~~
\left\{
\begin{array}{lr}
\frac{\partial \left(\rho_g\right)}{\partial t}
+ \nabla_x \cdot \left(\rho_g \textbf{U}_g\right)= 0, & \vspace{1ex}\\
\frac{\partial \left(\rho_g \textbf{U}_g\right)}{\partial t}
+ \nabla_x \cdot \left(\rho_g \textbf{U}_g \textbf{U}_g + p_g\mathbb{I} - \mu_g \bm{\sigma}\right)
= 0, & \vspace{1ex}\\
\frac{\partial \left(\rho_g E_g\right)}{\partial t}
+ \nabla_x \cdot \left(\left(\rho_g E_g  + p_g\right) \textbf{U}_g
- \mu_g \bm{\sigma}\cdot\textbf{U}_g + \kappa \nabla_x T_g \right) = 0, &
\end{array}
\right. \\
\nonumber\\
\mathcal{L}_{g2}&:~~
\left\{
\begin{array}{lr}
\frac{\partial \left(\rho_g\right)}{\partial t} = C_{\epsilon_g}\rho_g, & \vspace{1ex}\\
\frac{\partial \left(\rho_g \textbf{U}_g\right)}{\partial t} =
C_{\epsilon_g} \rho_g \textbf{U}_g
-\frac{\epsilon_{s}\rho_{s}\left(\textbf{U}_g - \textbf{U}_s\right)}{\epsilon_g \tau_{st}}
+ \frac{\rho_g\textbf{G}}{\epsilon_{g}}, & \vspace{1ex}\\
\frac{\partial \left(\rho_g E_g\right)}{\partial t} =
C_{\epsilon_g} \left(\rho_g E_g + p_g\right)
-\frac{\epsilon_{s}\rho_{s}\textbf{U}_s \cdot \left(\textbf{U}_g - \textbf{U}_s\right)}{\epsilon_g \tau_{st}}
+ \frac{3p_s}{\epsilon_g \tau_{st}}
+ \frac{\rho_g \textbf{U}_g \cdot \textbf{G}}{\epsilon_{g}}. &
\end{array}
\right.
\end{align}
The GKS is constructed to solve $\mathcal{L}_{g1}$ and $\mathcal{L}_{g2}$ separately.
Firstly, the kinetic equation without acceleration term for gas phase $\mathcal{L}_{g1}$ is,
\begin{equation}\label{gas phase kinetic equ without acce}
\frac{\partial f_{g}}{\partial t}
+ \nabla_x \cdot \left(\textbf{u}f_{g}\right)
= \frac{g_{g}-f_{g}}{\tau_{g}},
\end{equation}
where $\textbf{u}$ is the velocity, $\tau_g$ is the relaxation time for gas phase, $f_{g}$ is the distribution function of gas phase, and $g_{g}$ is the corresponding equilibrium state (Maxwellian distribution). The local equilibrium state $g_{g}$ can be written as,
\begin{gather*}
g_{g}=\rho_g\left(\frac{\lambda_g}{\pi}\right)^{\frac{K+3}{2}}e^{-\lambda_g\left[(\textbf{u}-\textbf{U}_g)^2+\bm{\xi}^2\right]},
\end{gather*}
where $\rho_g$ is the density of gas phase, $\lambda_g$ is determined by gas temperature through $\lambda_g = \frac{m_g}{2k_BT_g}$, $m_g$ is the molecular mass, and $\textbf{U}_g$ is the macroscopic velocity of gas phase. Here $K$ is the internal degree of freedom with $K=(5-3\gamma)/(\gamma-1)$ for three-dimensional diatomic gas, where $\gamma=1.4$ is the specific heat ratio.
The collision term satisfies the compatibility condition
\begin{equation}\label{gas phase compatibility condition}
\int \frac{g_g-f_g}{\tau_g} \bm{\psi} \text{d}\Xi=0,
\end{equation}
where $\bm{\psi}=\left(1,\textbf{u},\displaystyle \frac{1}{2}(\textbf{u}^2+\bm{\xi}^2)\right)^T$, the internal variables $\bm{\xi}^2=\xi_1^2+...+\xi_K^2$, and $\text{d}\Xi=\text{d}\textbf{u}\text{d}\bm{\xi}$.

For Eq.\eqref{gas phase kinetic equ without acce}, the integral solution of $f$ at the cell interface can be written as,
\begin{equation}\label{gas phase equ_integral1}
f(\textbf{x},t,\textbf{u},\bm{\xi})=\frac{1}{\tau}\int_0^t g(\textbf{x}',t',\textbf{u},\bm{\xi})e^{-(t-t')/\tau}\text{d}t'\\
+e^{-t/\tau}f_0(\textbf{x}-\textbf{u}t,\textbf{u},\bm{\xi}),
\end{equation}
where $\textbf{x}'=\textbf{x}+\textbf{u}(t'-t)$ is the trajectory of particles, $f_0$ is the initial gas distribution function at time $t=0$, and $g$ is the corresponding equilibrium state. The initial NS gas distribution function $f_0$ in Eq.\eqref{gas phase equ_integral1} can be constructed as
\begin{equation}\label{gas phase equ_f0}
f_0=f_0^l(\textbf{x},\textbf{u})H(x)+f_0^r(\textbf{x},\textbf{u})(1-H(x)),
\end{equation}
where $H(x)$ is the Heaviside function, $f_0^l$ and $f_0^r$ are the
initial gas distribution functions on the left and right side of one cell interface.
More specifically, the initial gas distribution function $f_0^k$, $k=l,r$, is constructed as
\begin{equation*}
f_0^k=g^k\left(1+\textbf{a}^k \cdot \textbf{x}-\tau(\textbf{a}^k \cdot \textbf{u}+A^k)\right),
\end{equation*}
where $g^l$ and $g^r$ are the Maxwellian distribution functions on the left and right hand sides of a cell interface, and they can be determined by the corresponding conservative variables $\textbf{W}^l$ and $\textbf{W}^r$. The coefficients $\textbf{a}^l=\left[a^l_1, a^l_2, a^l_3\right]^T$, $\textbf{a}^r=\left[a^r_1, a^r_2, a^r_3\right]^T$, are related to the spatial derivatives in normal and tangential directions, which can be obtained from the corresponding derivatives of the initial macroscopic variables,
\begin{equation*}
\left\langle a^l_i\right\rangle=\partial \textbf{W}^l/\partial x_i,
\left\langle a^r_i\right\rangle=\partial \textbf{W}^r/\partial x_i,
\end{equation*}
where $i=1,2,3$, and $\left\langle...\right\rangle$ means the moments of the Maxwellian distribution functions,
\begin{align*}
\left\langle...\right\rangle=\int \bm{\psi}\left(...\right)g\text{d}\Xi.
\end{align*}
Based on the Chapman-Enskog expansion, the non-equilibrium part of the distribution function satisfies,
\begin{equation*}
\left\langle \textbf{a}^l \cdot\textbf{u}+A^l\right\rangle = 0,~
\left\langle \textbf{a}^r \cdot\textbf{u}+A^r\right\rangle = 0,
\end{equation*}
and therefore the coefficients $A^l$ and $A^r$ can be fully determined. The equilibrium state $g$ around the cell interface is modeled as,
\begin{equation}\label{gas phase equ_g}
g=g_0\left(1+\overline{\textbf{a}}\cdot\textbf{x}+\bar{A}t\right),
\end{equation}
where $\overline{\textbf{a}}=\left[\overline{a}_1, \overline{a}_2, \overline{a}_3\right]^T$, $g_0$ is the local equilibrium of the cell interface. More specifically, $g$ can be determined by the compatibility condition,
\begin{align*}
\int\bm{\psi} g_{0}\text{d}\Xi=\textbf{W}_0
&=\int_{u>0}\bm{\psi} g^{l}\text{d}\Xi+\int_{u<0}\bm{\psi} g^{r}\text{d}\Xi, \nonumber \\
\int\bm{\psi} \overline{a_i} g_{0}\text{d}\Xi=\partial \textbf{W}_0/\partial x_i
&=\int_{u>0}\bm{\psi} a^l_i g^{l}\text{d}\Xi+\int_{u<0}\bm{\psi} a^r_i g^{r}\text{d}\Xi,
\end{align*}
$i=1,2,3$, and
\begin{equation*}
\left\langle \overline{\textbf{a}} \cdot \textbf{u}+\bar{A}\right\rangle = 0.
\end{equation*}
After determining all parameters in the initial gas distribution function $f_0$ and the equilibrium state $g$, substituting Eq.\eqref{gas phase equ_f0} and Eq.\eqref{gas phase equ_g} into Eq.\eqref{gas phase equ_integral1}, the time-dependent distribution function $f(\textbf{x}, t, \textbf{u},\bm{\xi})$ at a cell interface can be expressed as,
\begin{align}\label{gas phase equ_finalf}
f(\textbf{x}, t, \textbf{u},\bm{\xi})
&=c_1 g_0+ c_2 \overline{\textbf{a}}\cdot\textbf{u}g_0 +c_3 {\bar{A}} g_0\nonumber\\
&+\left[c_4 g^r +c_5 \textbf{a}^r\cdot\textbf{u} g^r + c_6 A^r g^r\right] (1-H(u)) \\
&+\left[c_4 g^l +c_5 \textbf{a}^l\cdot\textbf{u} g^l + c_6 A^l g^l\right] H(u) \nonumber.
\end{align}
with coefficients,
\begin{align*}
c_1 &= 1-e^{-t/\tau}, \\
c_2 &= \left(t+\tau\right)e^{-t/\tau}-\tau, \\
c_3 &= t-\tau+\tau e^{-t/\tau}, \\
c_4 &= e^{-t/\tau}, \\
c_5 &= -\left(t+\tau\right)e^{-t/\tau}, \\
c_6 &= -\tau e^{-t/\tau}.
\end{align*}
Then, the integrated flux over a time step can be obtained,
\begin{align}
\textbf{F}_{ij} =\int_{0}^{\Delta t} \int\textbf{u}\cdot\textbf{n}_{ij} f_{ij}(\textbf{x},t,\textbf{u},\bm{\xi})\bm{\psi}\text{d}\Xi\text{d}t,
\end{align}
where $\textbf{n}_{ij}$ is the normal vector of the cell interface.
Then, the cell-averaged conservative variables of cell $i$ can be updated as follows,
\begin{gather}
\textbf{W}_i^{n+1} = \textbf{W}_i^n
- \frac{1}{\Omega_i} \sum_{S_{ij}\in \partial \Omega_i}\textbf{F}_{ij}S_{ij},
\end{gather}
where $\Omega_i$ is the volume of cell $i$, $\partial\Omega_i$ denotes the set of interface of cell $i$, $S_{ij}$ is the area of $j$-th interface of cell $i$, $\textbf{F}_{ij}$ denotes the projected macroscopic fluxes in the normal direction, and $\textbf{W}_{g}=\left[\rho_g,\rho_g \textbf{U}_g, \rho_g E_g\right]^T$ are the cell-averaged conservative flow variables for gas phase.

The second part, $\mathcal{L}_{g2}$, is from the inter-phase interaction. The increased macroscopic variables for gas phase in 3D can be calculated as
\begin{gather}
\textbf{W}^{n+1}_g = \textbf{W}_g + \textbf{Q}\Delta t,
\end{gather}
where
\begin{gather*}
\textbf{W}_g=\left[\begin{array}{c}
\rho_g\\
\rho_g \textbf{U}_g\\
\rho_g E_g
\end{array}
\right], ~~
\textbf{Q}=\left[\begin{array}{c}
C_{\epsilon_g}\rho_g \\
C_{\epsilon_g} \rho_g \textbf{U}_g
-\frac{\epsilon_{s}\rho_{s}\left(\textbf{U}_g - \textbf{U}_s\right)}{\epsilon_g \tau_{st}}
+ \frac{\rho_g\textbf{G}}{\epsilon_{g}}\\
C_{\epsilon_g} \left(\rho_g E_g + p_g\right)
-\frac{\epsilon_{s}\rho_{s}\textbf{U}_s \cdot \left(\textbf{U}_g - \textbf{U}_s\right)}{\epsilon_g \tau_{st}}
+ \frac{3p_s}{\epsilon_g \tau_{st}}
+ \frac{\rho_g \textbf{U}_g \cdot \textbf{G}}{\epsilon_{g}}
\end{array}\right],
\end{gather*}
with $C_{\epsilon_g} = -\frac{1}{\epsilon_{g}}\frac{\text{d}\epsilon_{g}}{\text{d}t}$ and $\frac{\text{d}\epsilon_{g}}{\text{d}t}=\frac{\partial \epsilon_{g}}{\partial t}+\textbf{U}_g \cdot \nabla\epsilon_{g}$. In this paper, $\frac{\partial \epsilon_{g}}{\partial t}$ is evaluated,
\begin{equation}
\frac{\partial \epsilon_{g}}{\partial t} = \frac{\epsilon_{g}^{n+1} - \epsilon_{g}^n}{\Delta t}.
\end{equation}
Here $\nabla\epsilon_{g}$ is the cell-averaged volume fraction gradient of gas phase in the cell. For example, $\frac{\partial \epsilon_{g}}{\partial x}$ is calculated by,
\begin{equation}
\frac{\partial \epsilon_{g,i}}{\partial x} = \frac{\epsilon_{g,i+\frac{1}{2}} - \epsilon_{g,i-\frac{1}{2}}}{\Delta x},
\end{equation}
where $\epsilon_{g,i-\frac{1}{2}}$ and $\epsilon_{g,i+\frac{1}{2}}$ are volume fractions of gas phase at the left and right interface of cell $i$, which can be obtained from the reconstructed $\epsilon_{s}$ according to $\epsilon_{s} + \epsilon_{g} = 1$. Note that the gravity $\textbf{G}$ of gas phase is ignored in this paper. Now the update for the gas phase in one time step has been finished.

\section{Numerical experiments}
In the following cases, the time step of gas phase $\Delta t_g$ is determined by,
\begin{equation*}
\Delta t_g = \text{CFL} \times \left[\frac{\Delta_c}{\textbf{U}_g + \sqrt{\gamma R T_g}}\right]_{min},
\end{equation*}
where $\Delta_c$ is the cell size. Similarly, the time step of solid phase $\Delta t_s$ is determined by
\begin{equation*}
\Delta t_s = \text{CFL} \times \left[\frac{\Delta_c}{\textbf{U}_s + a\sqrt{\theta}}\right]_{min},
\end{equation*}
where $a$ is taken as 3, and $\text{CFL}$ is taken as 0.5 in this paper. For most fluidized bed problems, $\Delta t_s$ is larger than $\Delta t_g$, more than one order. Therefore, in this paper, two time steps $\Delta t_s$ and $\Delta t_g$ are used in the evolution of solid and gas phase, respectively; since $\Delta t_s > \Delta t_g$, the solid phase will be frozen when the gas phase is evolved by $\Delta t_g$.

\subsection{Turbulent fluidized bed problem}
\begin{figure}[htbp]
	\centering
	\subfigure[]{		
		\includegraphics[height=11.0cm]{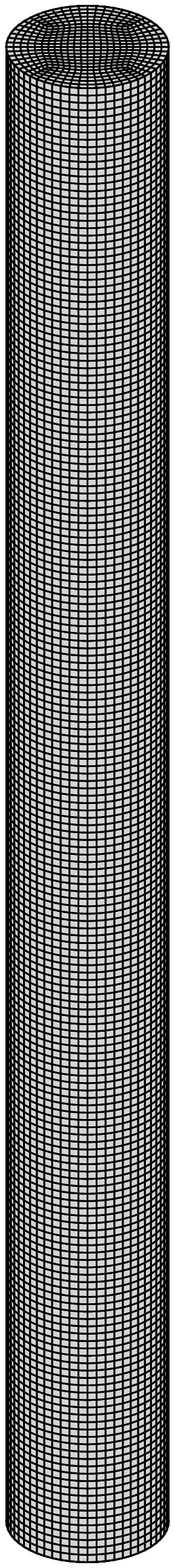}			
	}
	\quad
	\subfigure[]{
		\raisebox{0.19\textwidth}{			
			\includegraphics[height=4.5cm]{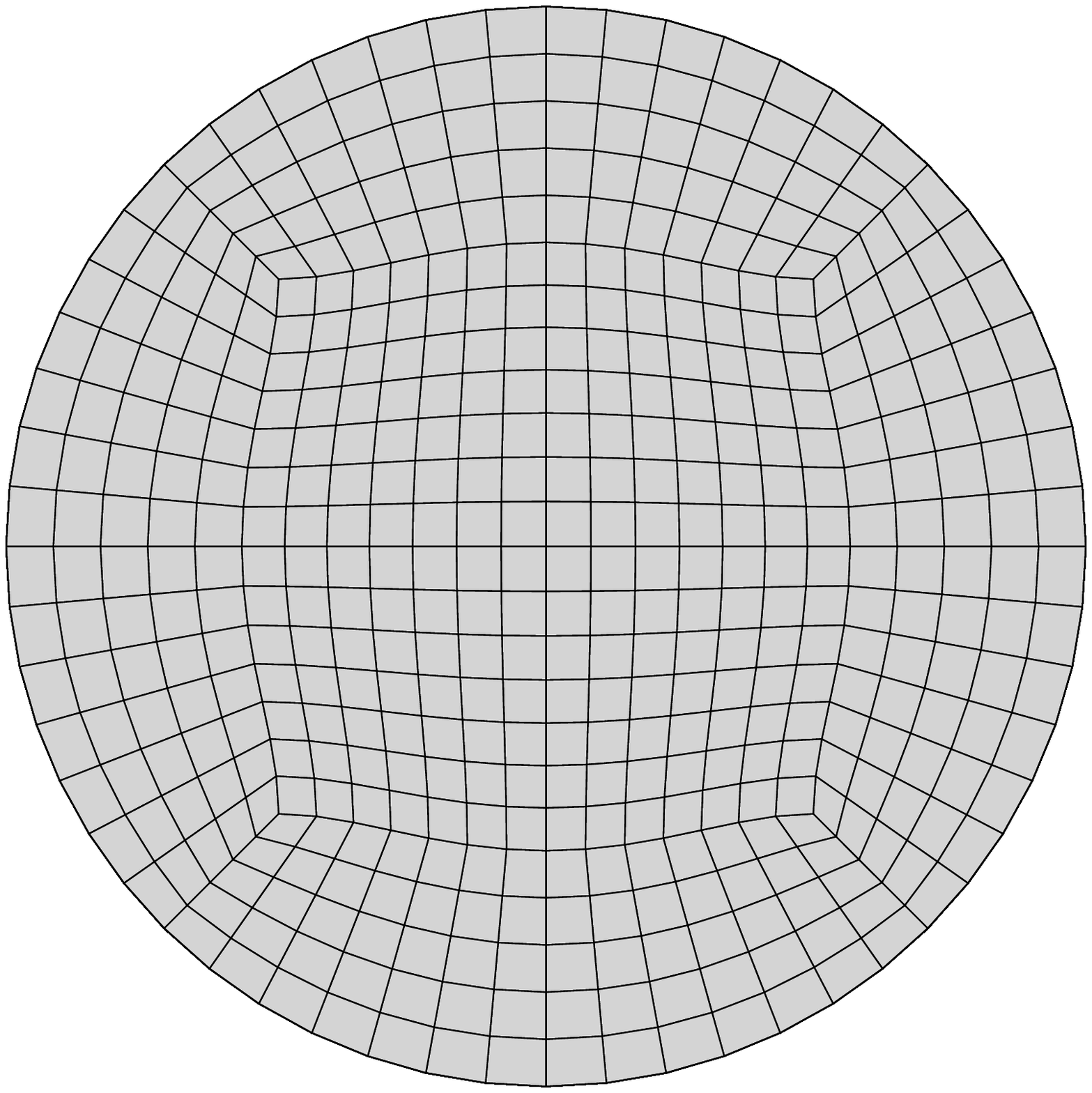}
		}	
	}
	\quad
	\subfigure[]{
		\includegraphics[height=11.0cm]{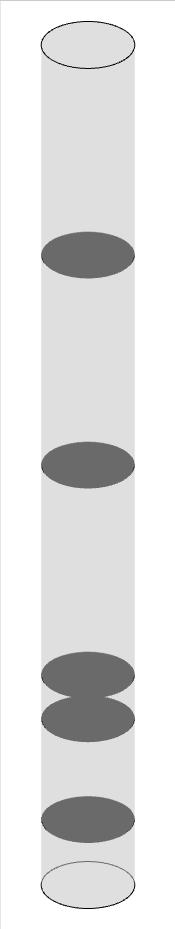}	
	}	
	\quad
	\subfigure[]{
		\includegraphics[height=11.0cm]{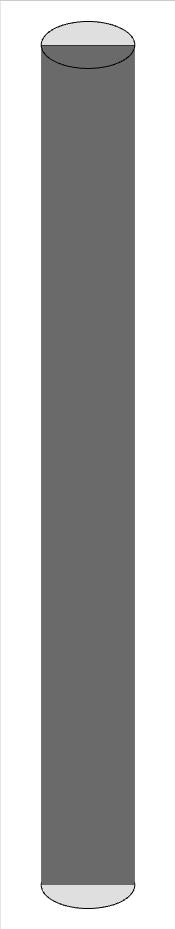}	
	}	
	\caption{Sketch of mesh and cross section used in post-processing: (a) the mesh employed in 3D view, (b) the mesh with 476 cells in one 2D cross section, (c) the horizontal cross sections at heights $0.078m$, $0.198m$, $0.25m$, $0.50m$ and $0.75m$, where the flow distributions will be shown at these cross sections later, and (d) the vertical cross section.}	
	\label{FBGao mesh and cross section sketch}		
\end{figure}

The first case is a turbulent fluidized bed problem studied experimentally by Gao et al.\cite{Gasparticle-fluidized-turbulent-gao2012experimental}. This experiment was conducted on a fluidized system, including a fluidizing column, an expanded column, and a recycling system.
In this paper, only the fluidizing column is simulated, as the previous study by CFD model \cite{Gasparticle-fluidized-turbulent-gao2012experimental}. The computational domain is a three-dimensional cylinder with diameter $D=0.095m$ and height $H=1m$. Figure \ref{FBGao mesh and cross section sketch}(a) and Figure \ref{FBGao mesh and cross section sketch}(b) present the sketches of the employed mesh in three-dimensional view and two-dimensional cross section, respectively. The mesh cells are hexahedrons with a total  number of 95200 control volumes, which are composed of a horizontal 476 cells with 200 layers in the vertical direction.
The cells are nearly uniformly distributed in the whole domain and  the approximate cell size is $3.86\times10^{-3}m$ horizontally and $5.00\times10^{-3}m$ vertically. The material density and diameter of solid particles are $\rho_{s}=2400kg/m^3$ and $d=0.139mm$, and the maximum solid volume fraction $\epsilon_{s,max}$ is 0.63. In this paper, the case with initial bed height $H_0=0.096m$ and inlet gas velocity $U_g=1.25m/s$ is calculated by GKS-UGKWP. Initially the equivalent solid mass is uniformly distributed in the computational domain; in the simulation, the solid particles are free to leave at the top boundary, and the escaped solid mass is recirculated to the computational domain through the bottom boundary to maintain a constant solid inventory in the riser. The gas blows into the fluidized bed with a uniform vertical velocity $U_g$ and a pressure  $\Delta p=\epsilon_{s,max}\left(\rho_s-\rho_g\right)GH_0$. The non-slip wall and slip wall boundary condition are employed for gas phase and solid phase respectively for the riser wall. For the turbulent fluidized bed, the particle distributions composed of dense bottom, transition middle, and dilute top zone are commonly observed in the previous numerical and experimental studies. Therefore, the drag model proposed by Gao et al \cite{Gasparticle-fluidized-turbulent-gao2012experimental} in all aforementioned zones is employed in this turbulent fluidized bed study, and the inter-phase momentum transfer coefficient $\beta$ in this drag model can be written as follows,
\begin{equation}\label{beta by Gao}
\beta=
\left\{\begin{aligned}
&0.001 \left(\frac{17.3}{Re_s}+0.336\right)\frac{\rho_g|\textbf{U}_g-\textbf{u}|}{d_s} \epsilon_{s} \epsilon_{g}^{-1.8},  & \epsilon_{g} \le 0.94, \\
&\frac{3}{4}C_d\frac{\rho_g\epsilon_{s}\epsilon_{g}|\textbf{U}_g-\textbf{u}|}{d_s} \epsilon_{g}^{-2.65}, & 0.94 < \epsilon_{g} \le 0.99,
\\
&\frac{3}{4}C_d\frac{\rho_g\epsilon_{g}|\textbf{U}_g-\textbf{u}|}{d_s}, & 0.99 < \epsilon_{g} \le 1.0,
\end{aligned}\right.
\end{equation}
where $Re_s = \epsilon_{g}|\textbf{U}_g-\textbf{u}| d_s/\nu_g$ is the particle Reynolds number, $\nu_g$ is the kinematic viscosity of gas phase, and $C_d$ is the $Re_s$ dependent drag coefficient,
\begin{equation}\label{Cd}
C_d = \left\{\begin{aligned}
&\frac{24}{Re_s}\left(1+0.15 Re_s^{0.687}\right), &  & Re_s \le 1000, \\
&0.44, &  & Re_s > 1000.
\end{aligned} \right.
\end{equation}

\begin{figure}[htbp]
	\centering
	\subfigure[]{
		\includegraphics[height=6.5cm]{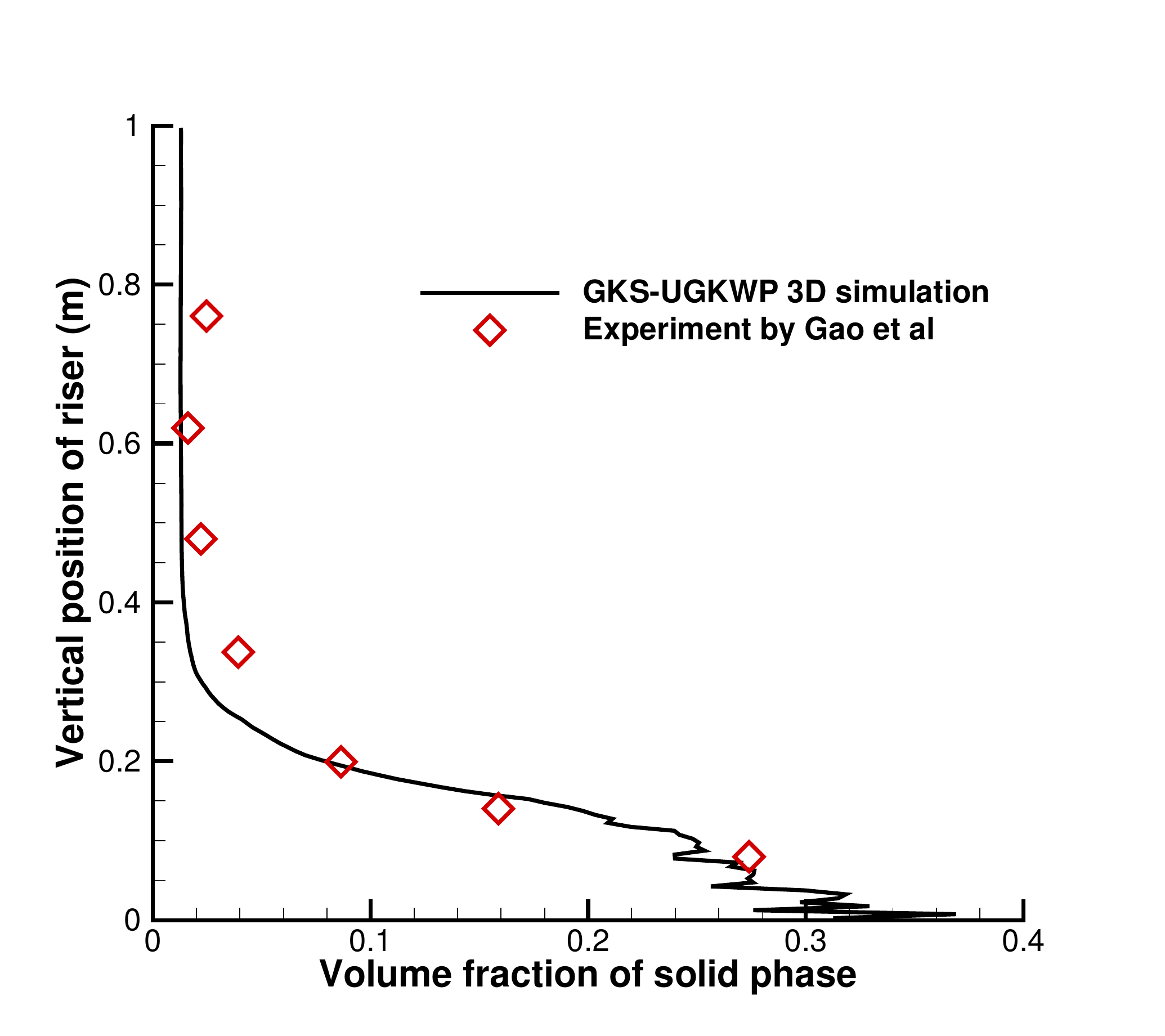}	
	}
	\quad
	\subfigure[]{
		\includegraphics[height=6.5cm]{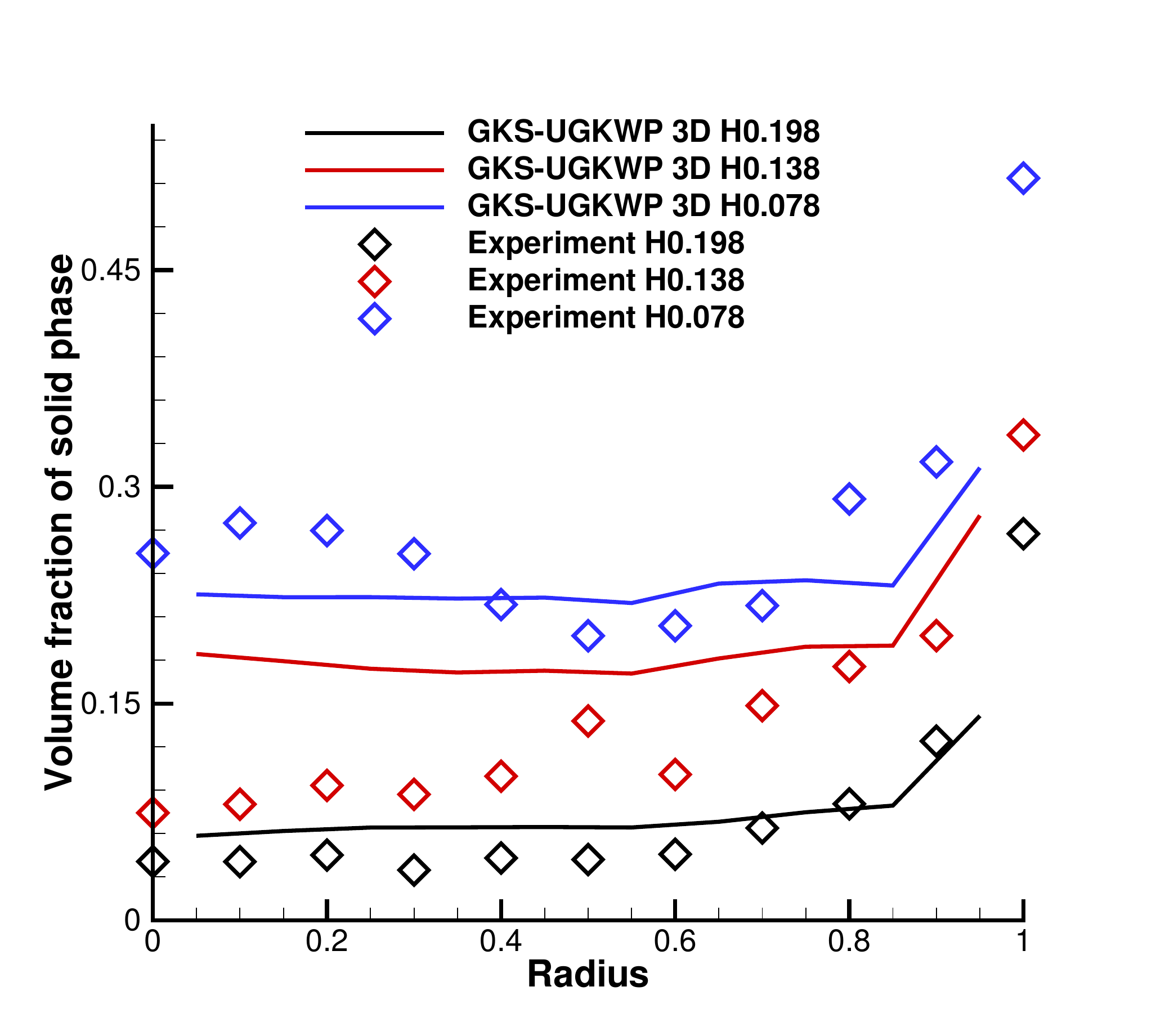}	
	}
	\caption{Time-averaged solid volume fraction $\epsilon_{s}$ and the comparison with experimental measurement. (a): $\epsilon_{s}$ distribution versus vertical riser height. (b): $\epsilon_{s}$ distribution versus horizontal radius at three different heights, $0.078m$, $0.138m$, and $0.198m$.}	
	\label{FBGao epsilon 3D2D profile}		
\end{figure}

\begin{figure}[htbp]
	\centering
	\subfigure{
		\centering
		\includegraphics[height=1.4cm]{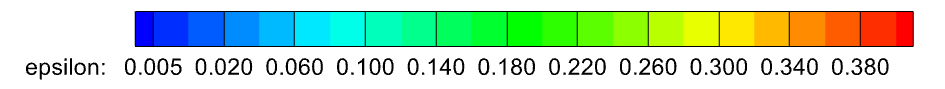}
	}
	
	\subfigure{		
		\centering		
		\begin{minipage}[c]{.12\textwidth}
			\centering
			\includegraphics[height=11.0cm]{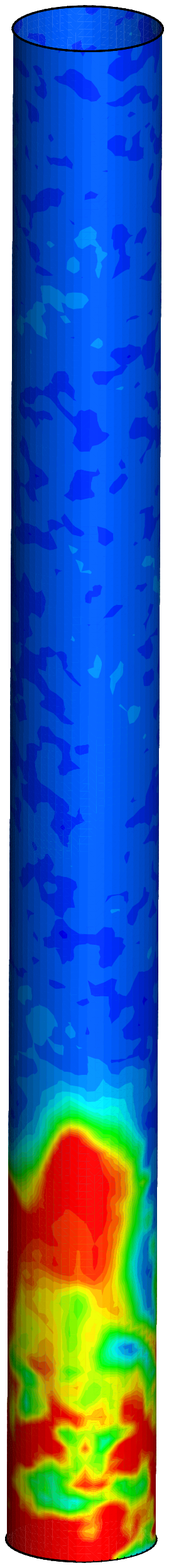}		
		\end{minipage}%
		\begin{minipage}[c]{.15\textwidth}
			\centering
			\includegraphics[height=2.0cm]{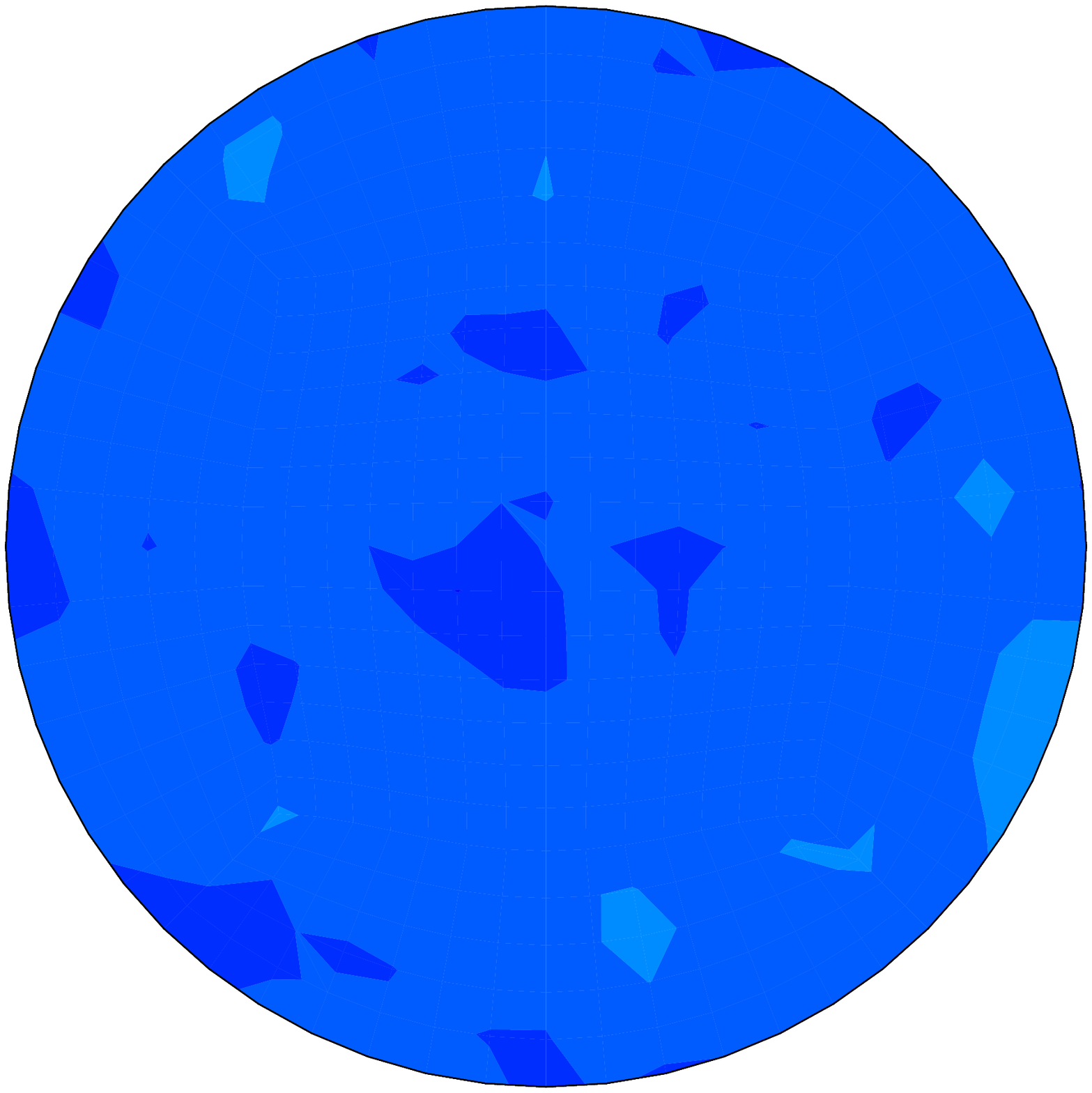}
			\includegraphics[height=2.0cm]{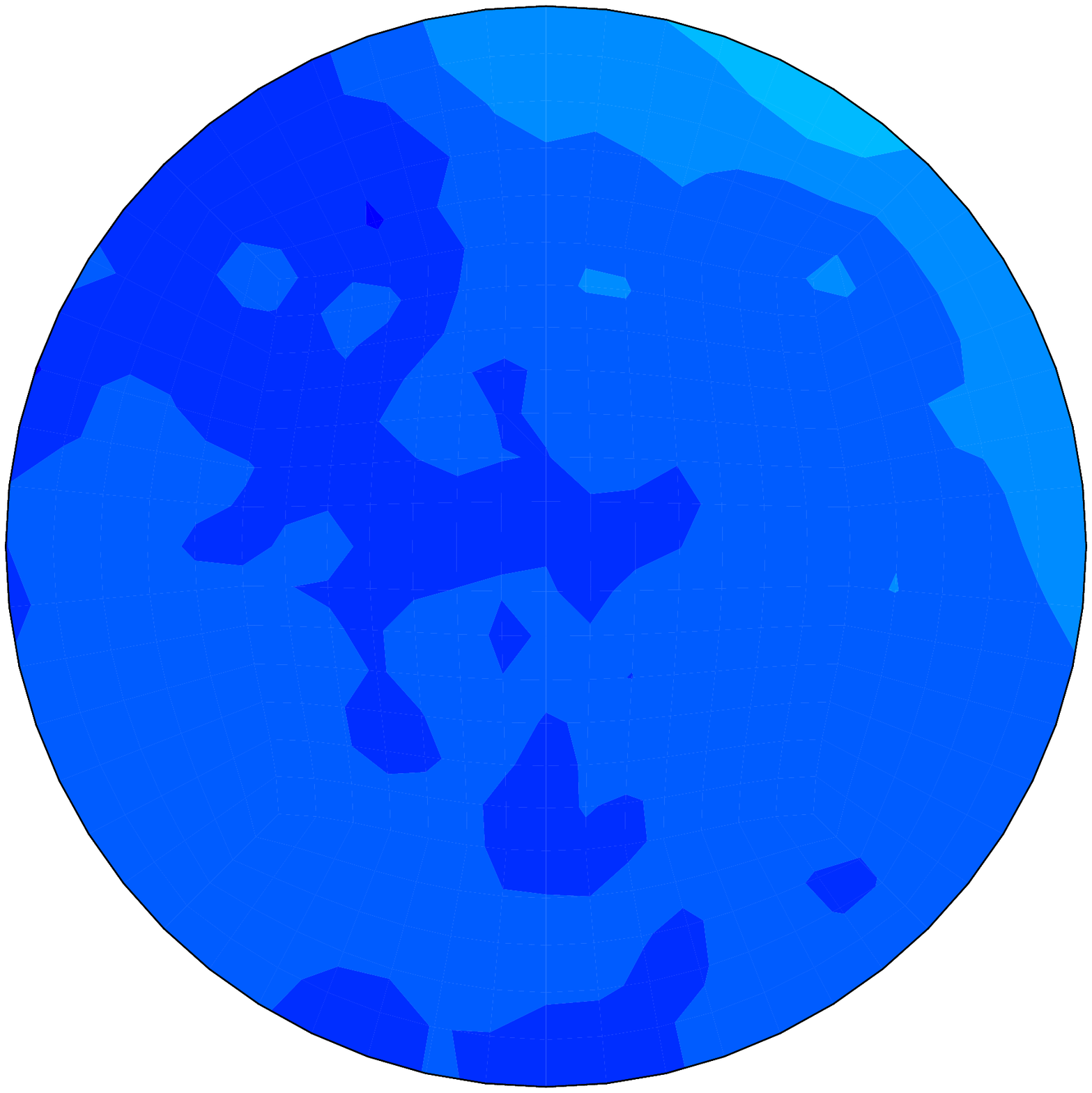}
			\includegraphics[height=2.0cm]{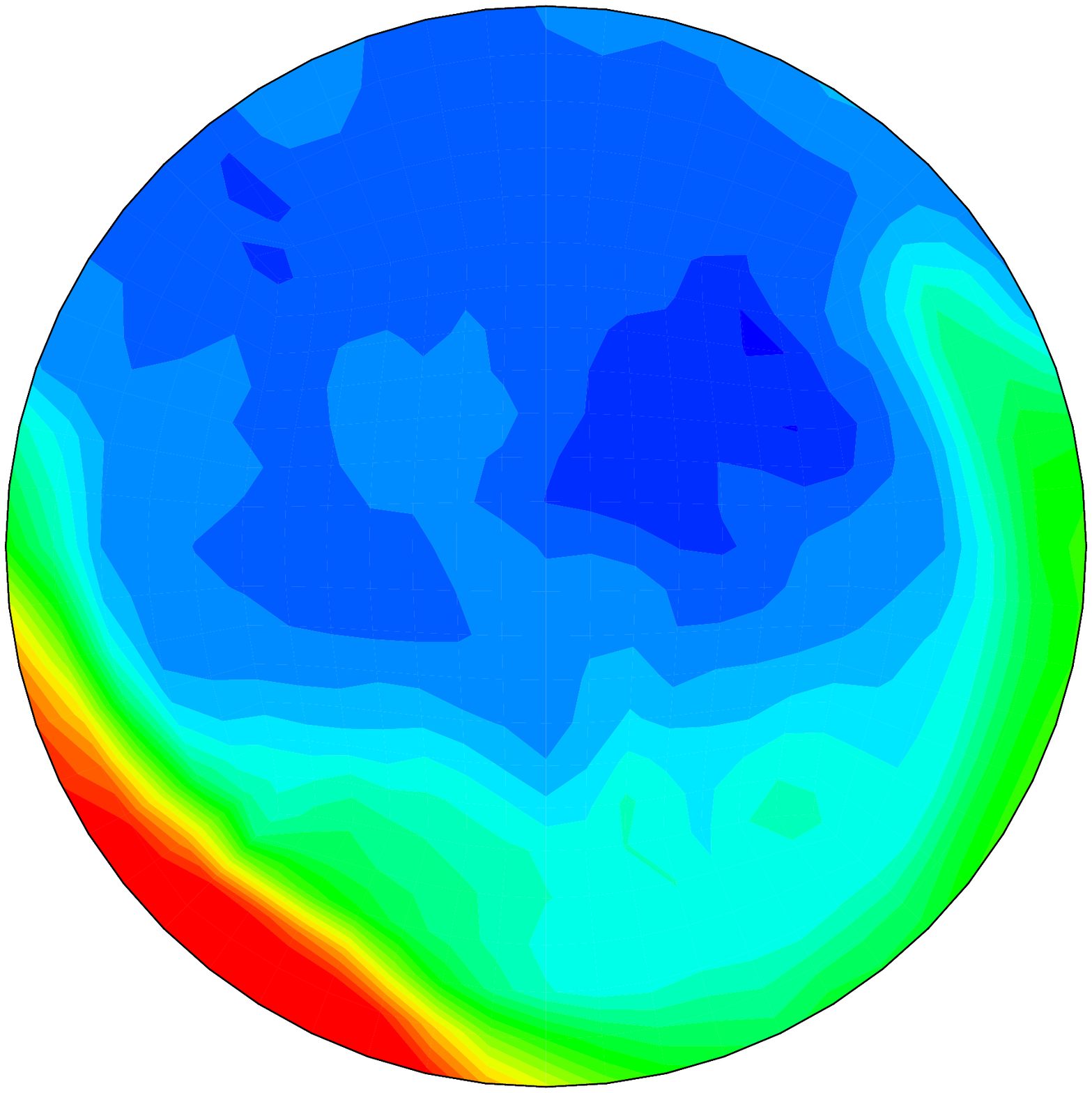}
			\includegraphics[height=2.0cm]{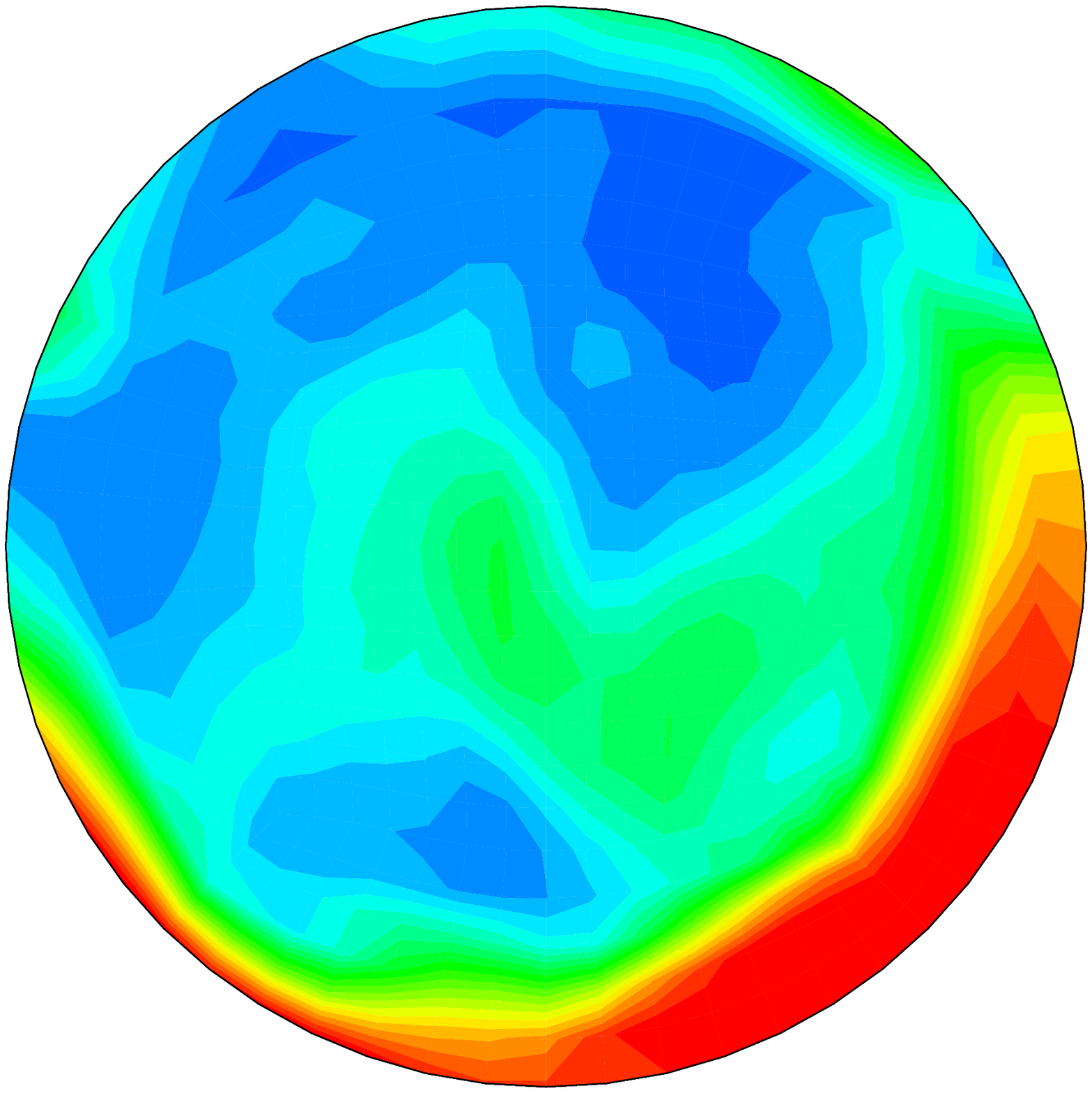}	
			\includegraphics[height=2.0cm]{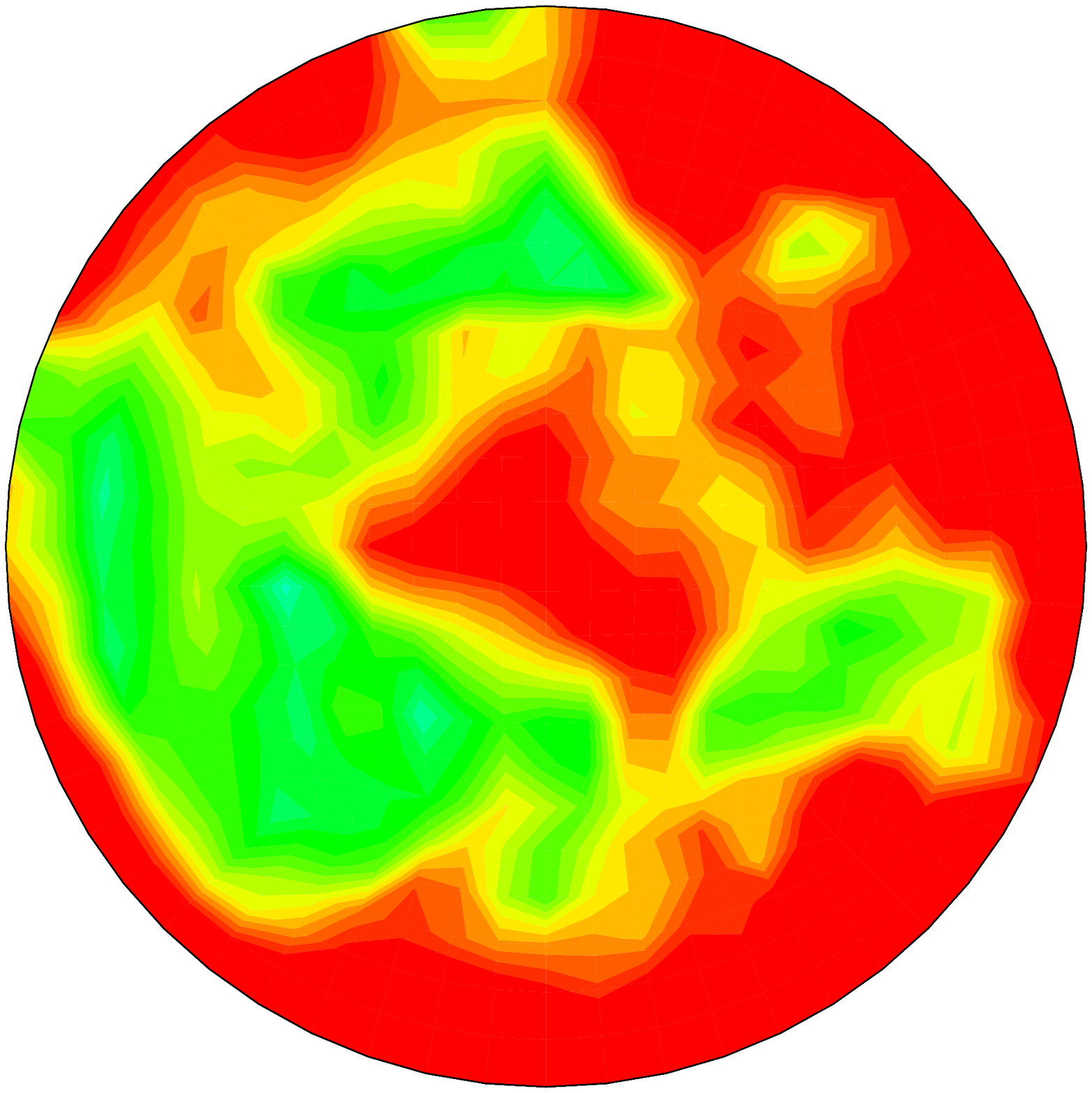}
		\end{minipage}	
	}
	\quad
	\subfigure{
		\centering	
		\begin{minipage}[c]{.12\textwidth}
			\centering
			\includegraphics[height=11.0cm]{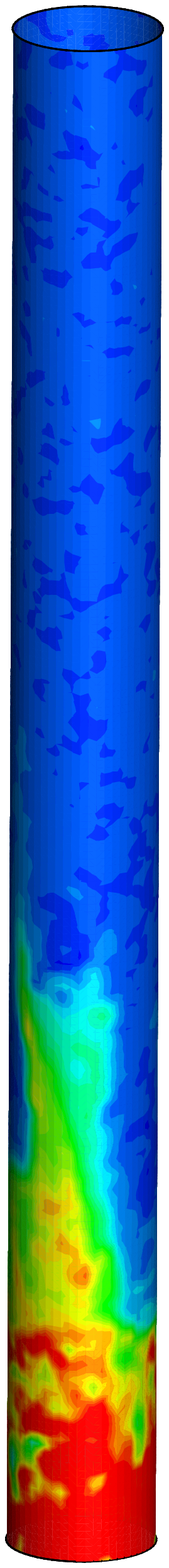}		
		\end{minipage}%
		\begin{minipage}[c]{.15\textwidth}
			\centering
			\includegraphics[height=2.0cm]{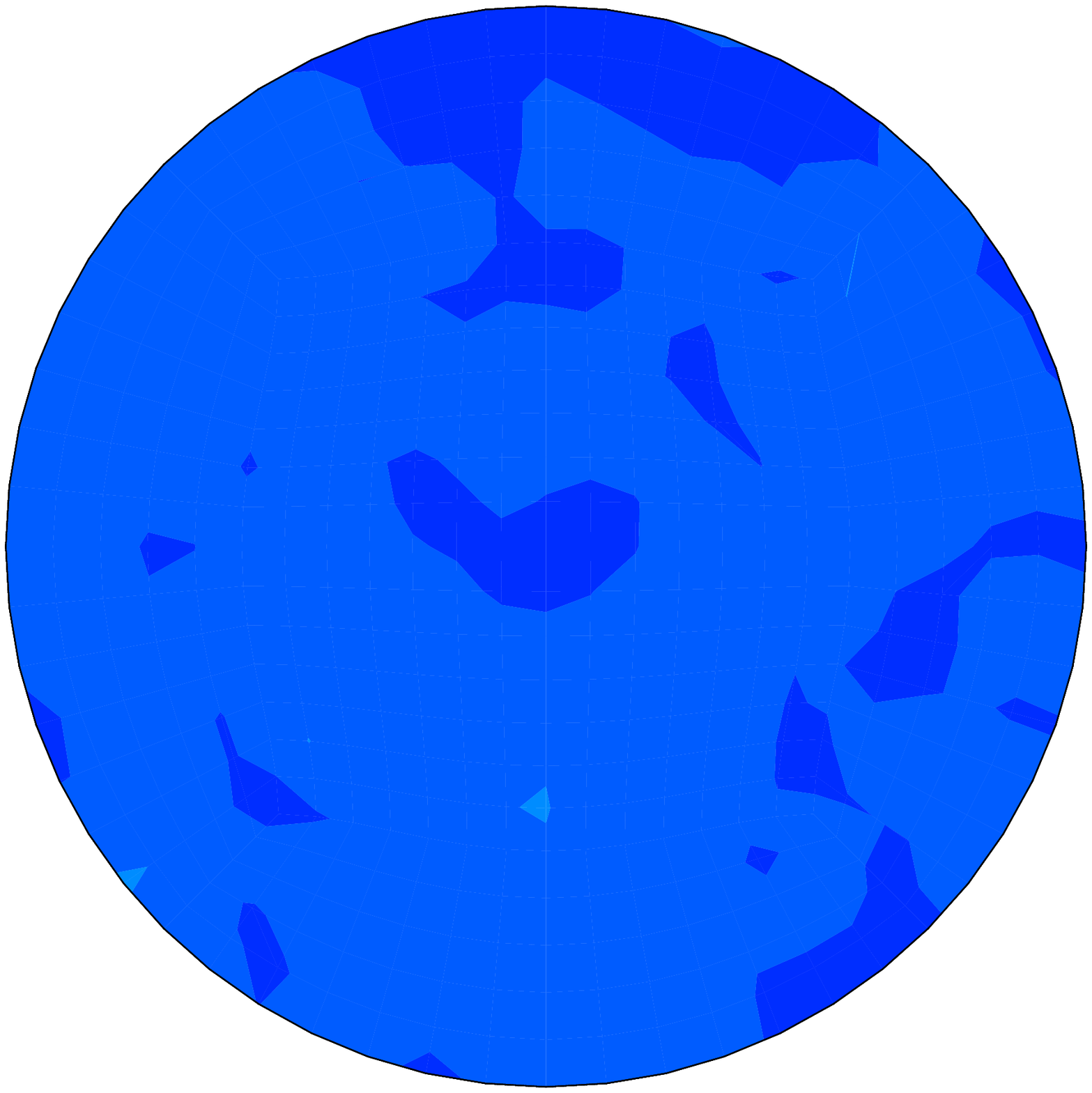}
			\includegraphics[height=2.0cm]{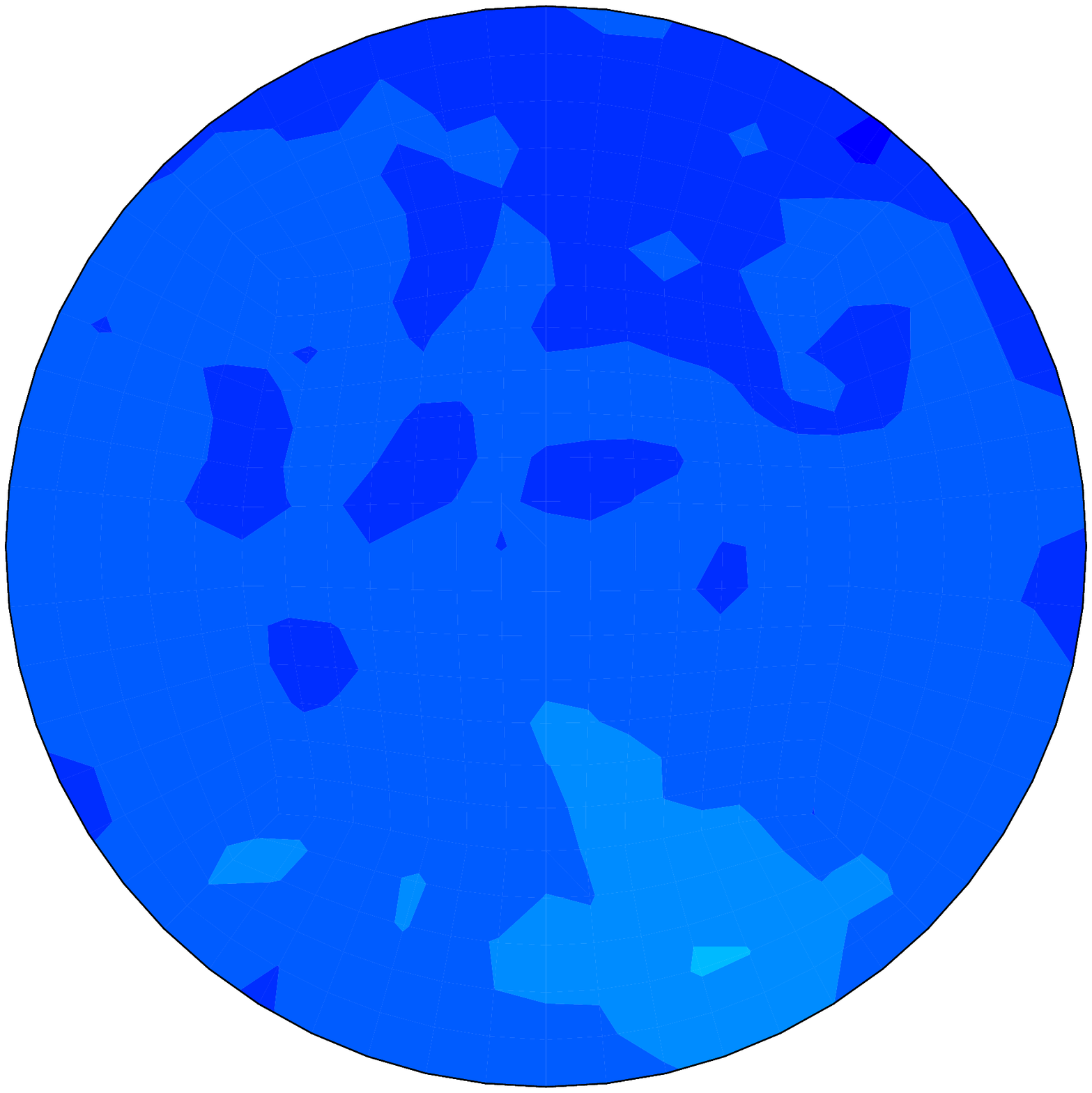}
			\includegraphics[height=2.0cm]{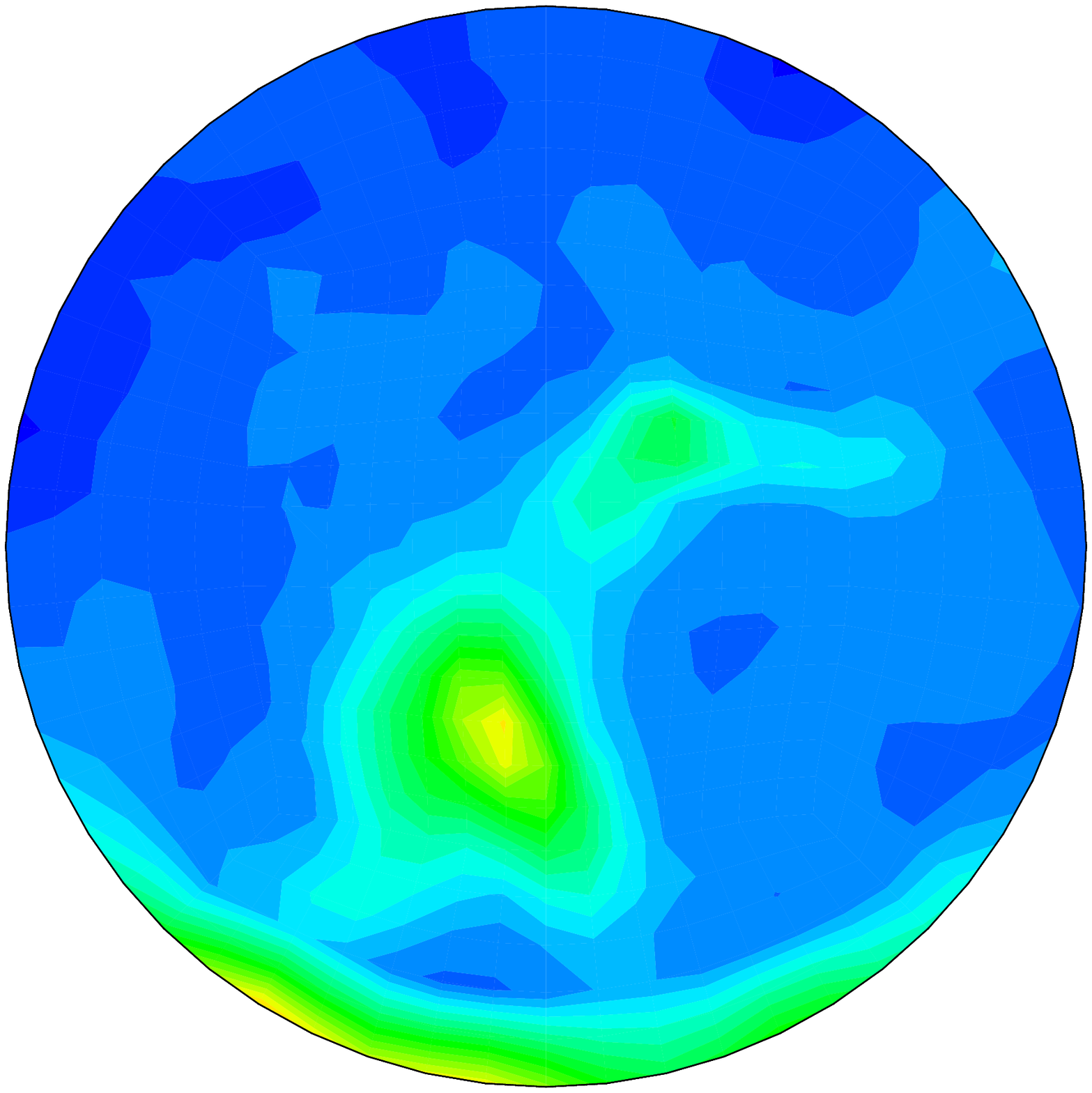}
			\includegraphics[height=2.0cm]{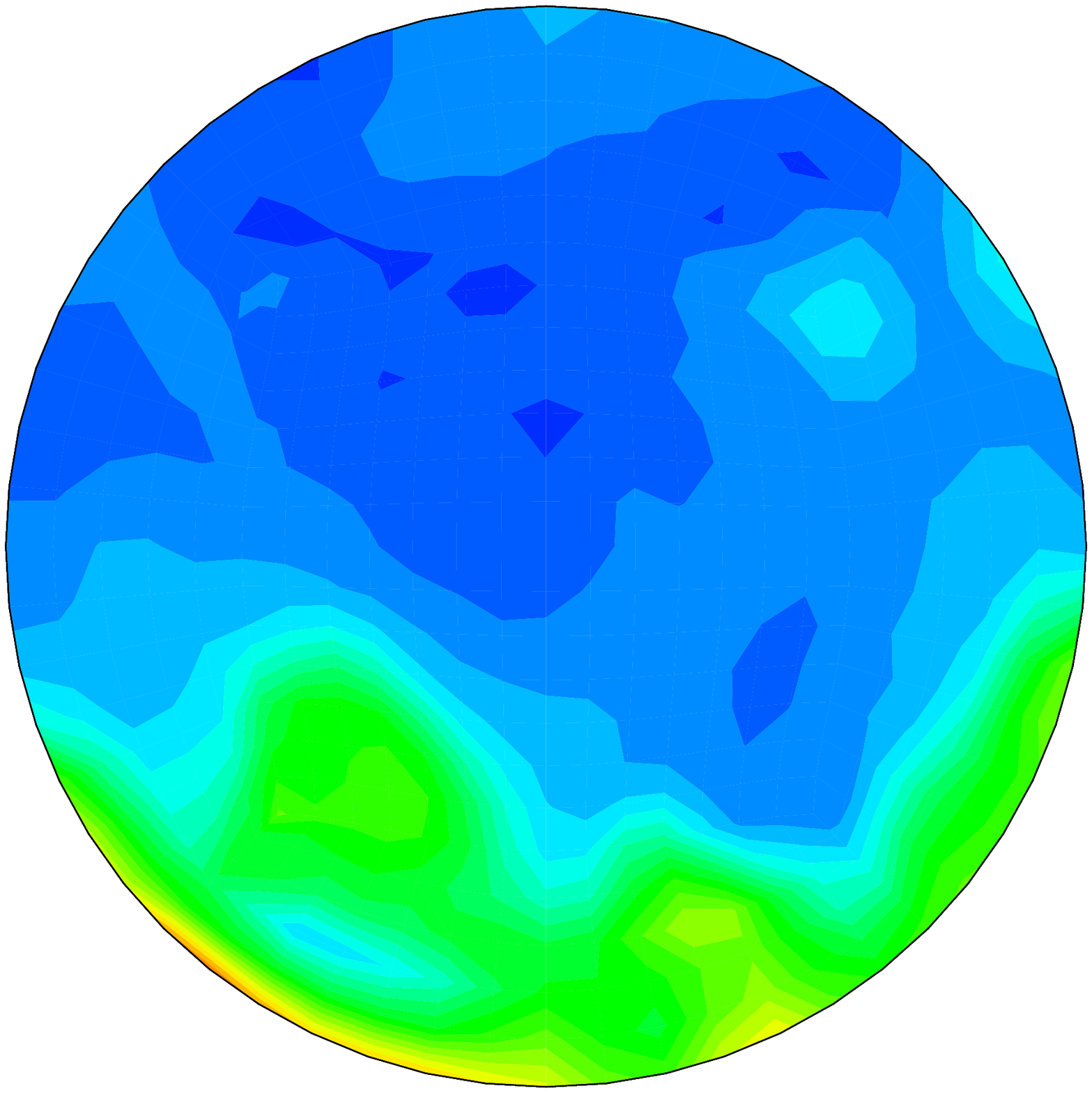}	
			\includegraphics[height=2.0cm]{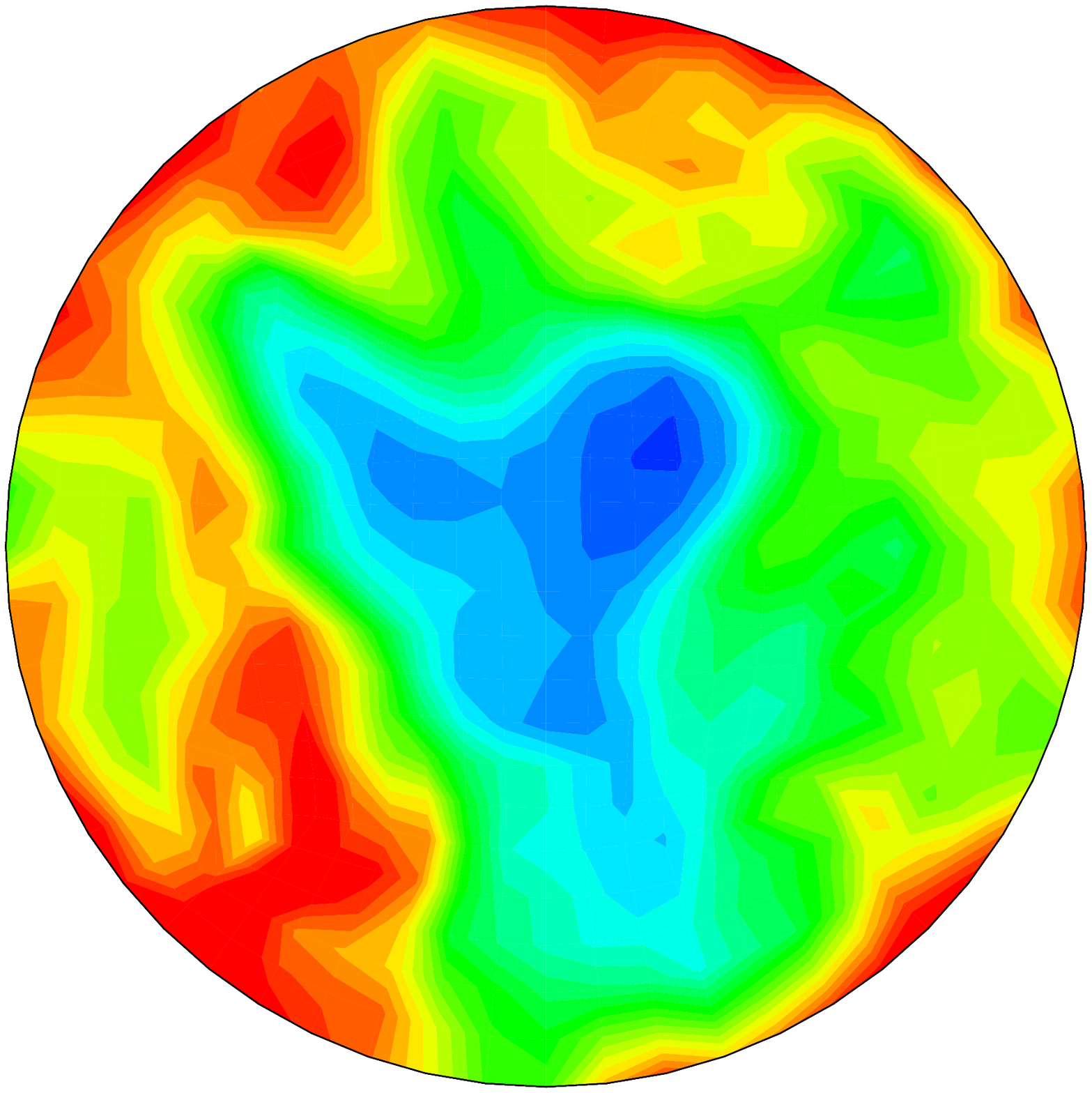}
		\end{minipage}		
	}
	\quad
	\subfigure{
		\centering	
		\begin{minipage}[c]{.12\textwidth}
			\centering
			\includegraphics[height=11.0cm]{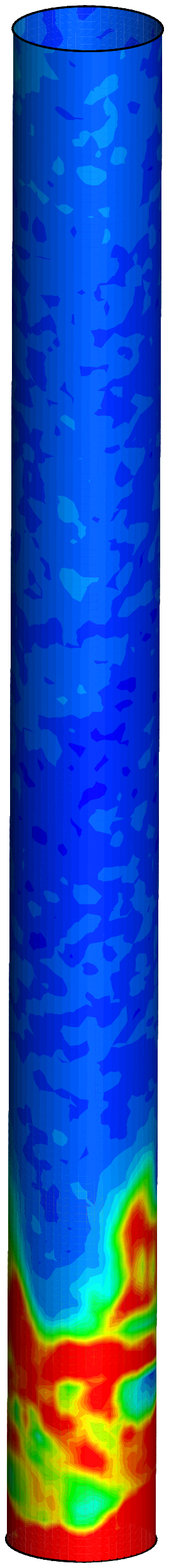}		
		\end{minipage}%
		\begin{minipage}[c]{.15\textwidth}
			\centering
			\includegraphics[height=2.0cm]{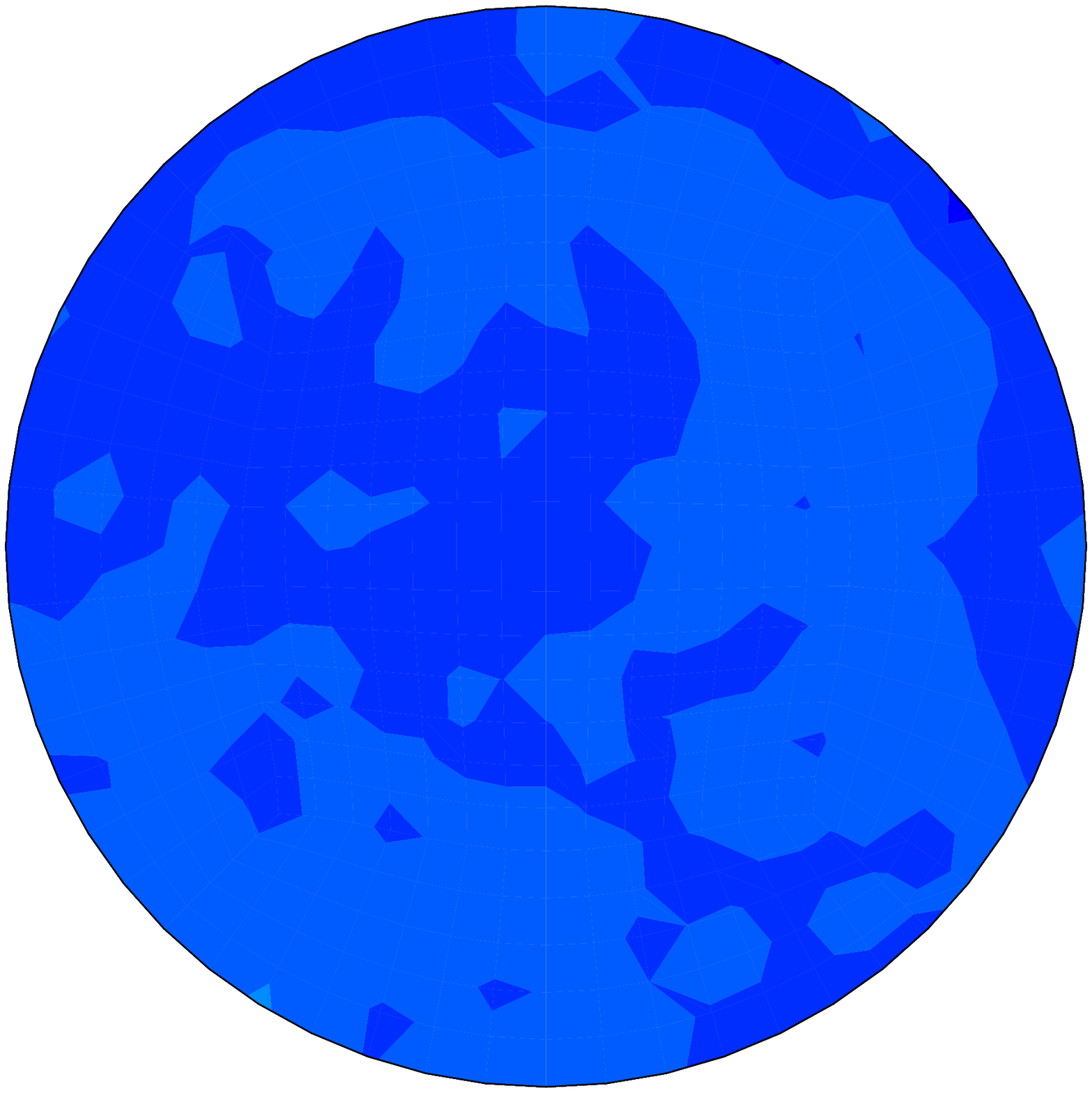}
			\includegraphics[height=2.0cm]{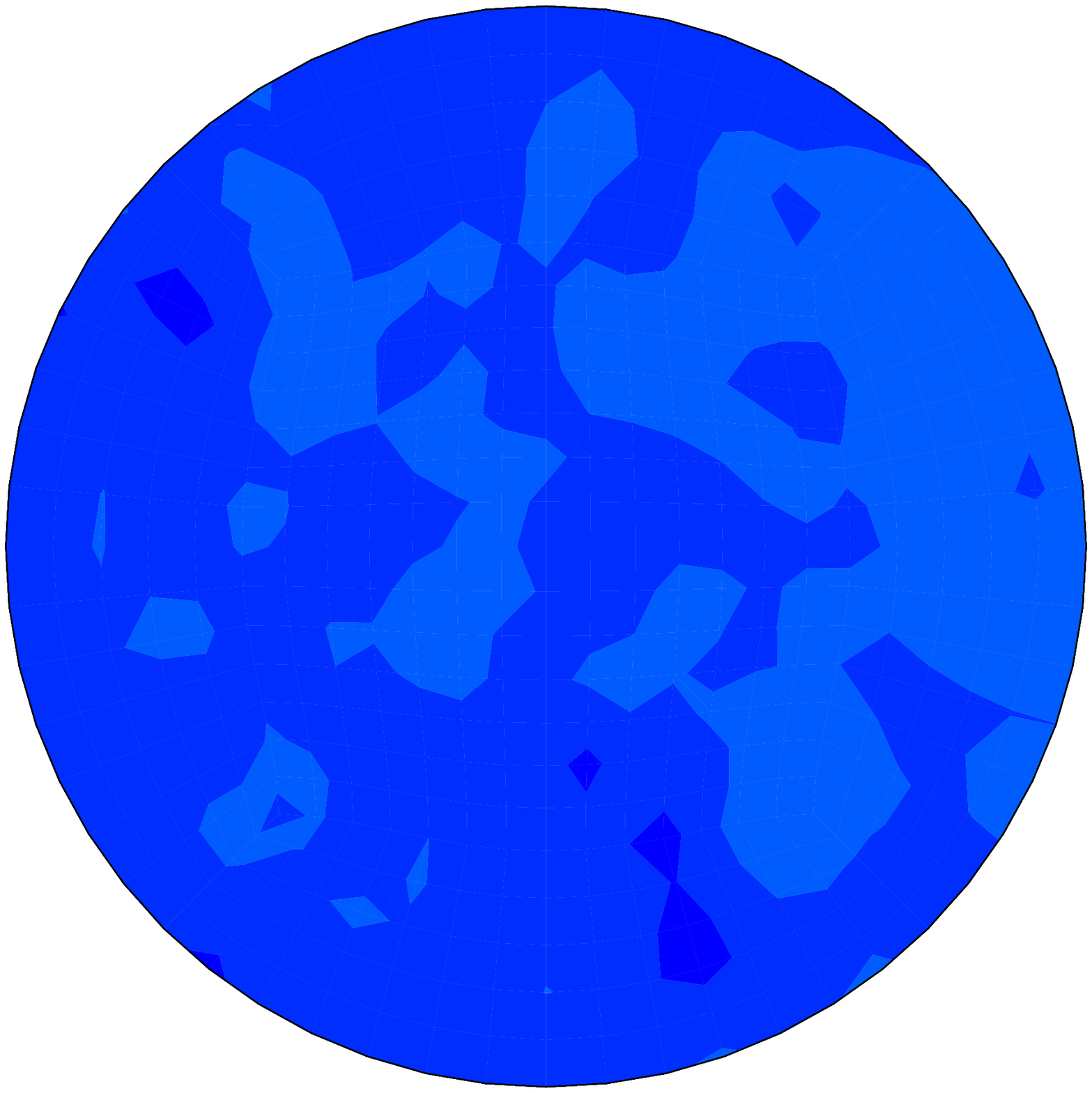}
			\includegraphics[height=2.0cm]{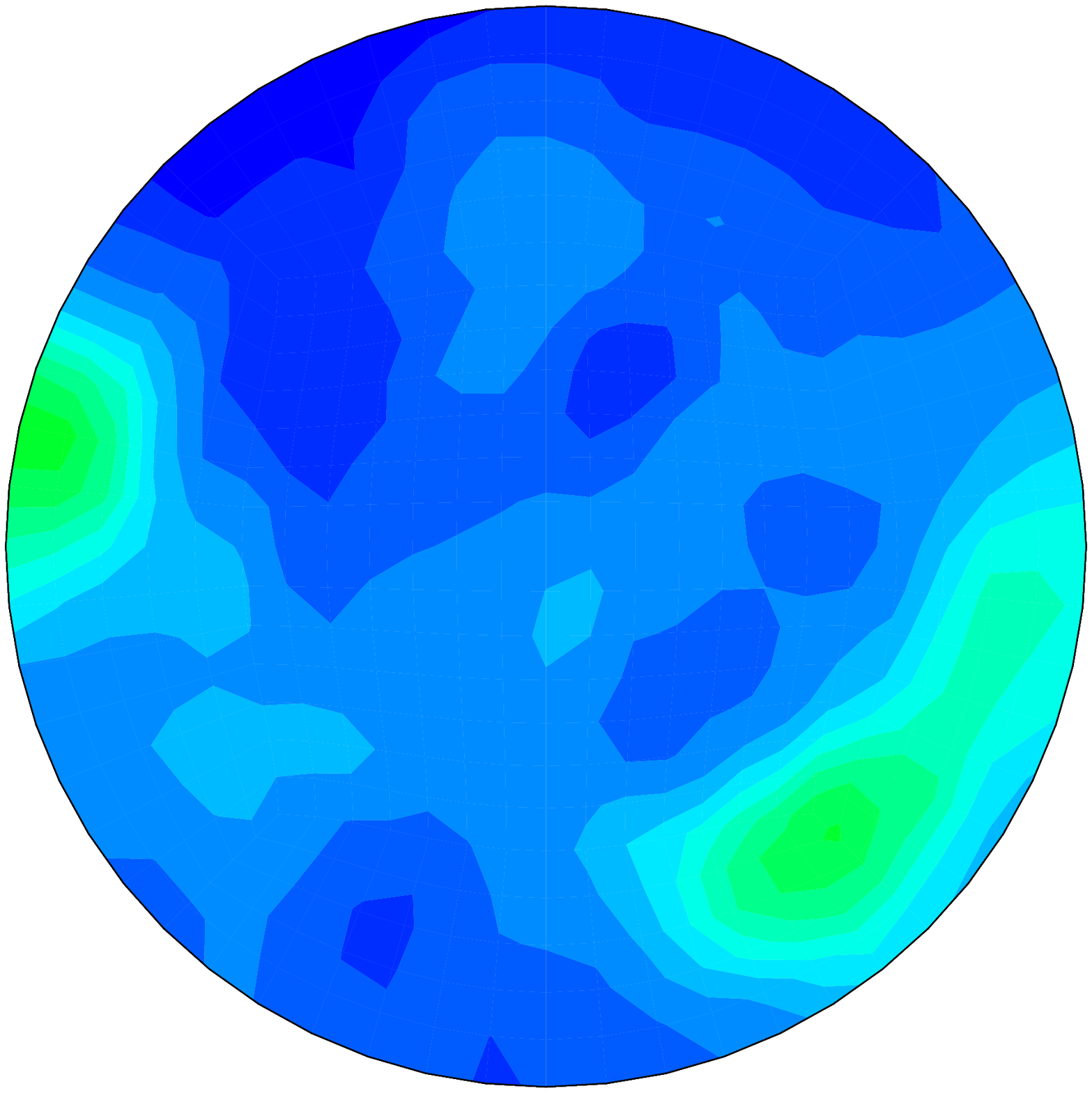}
			\includegraphics[height=2.0cm]{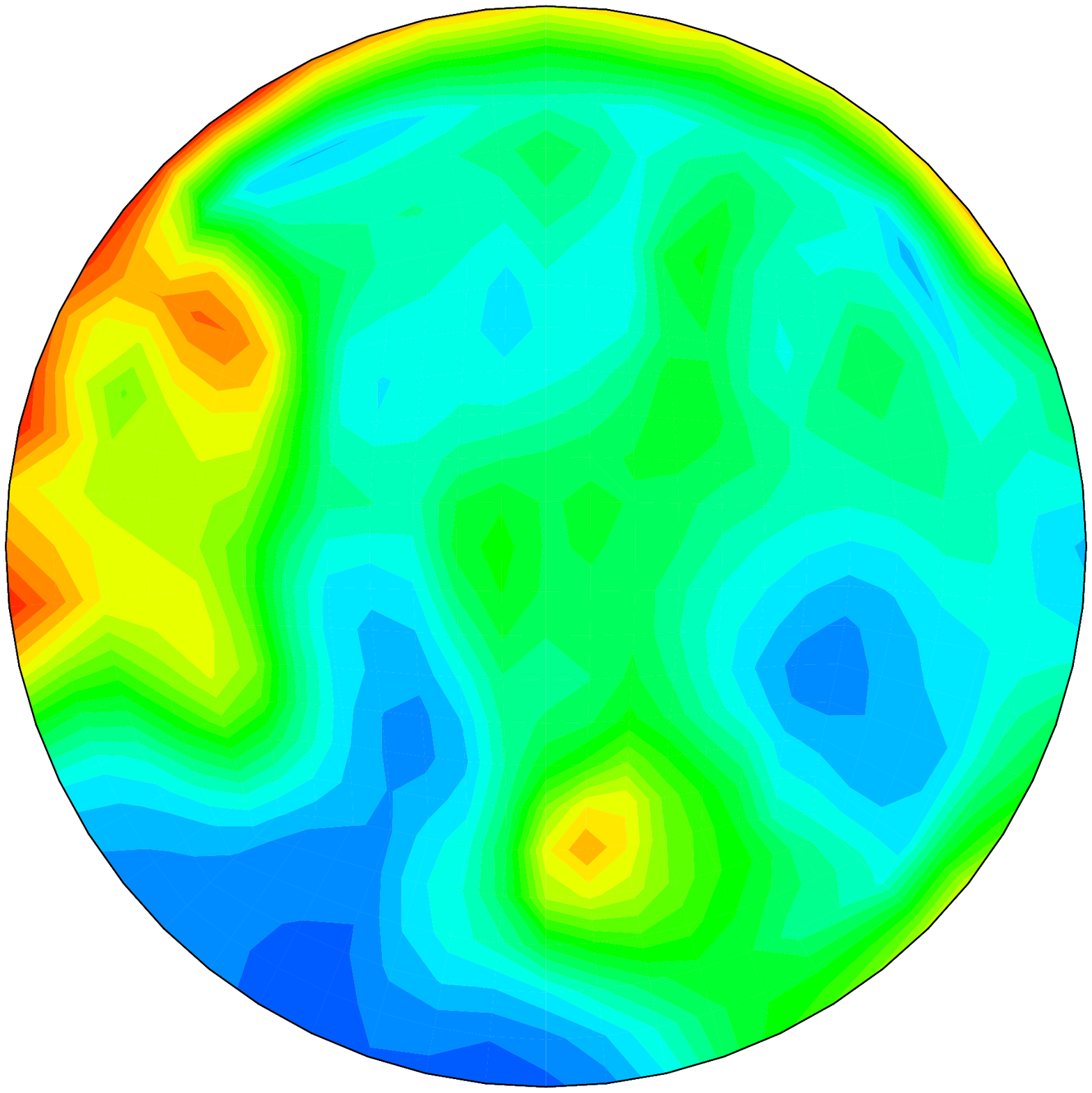}	
			\includegraphics[height=2.0cm]{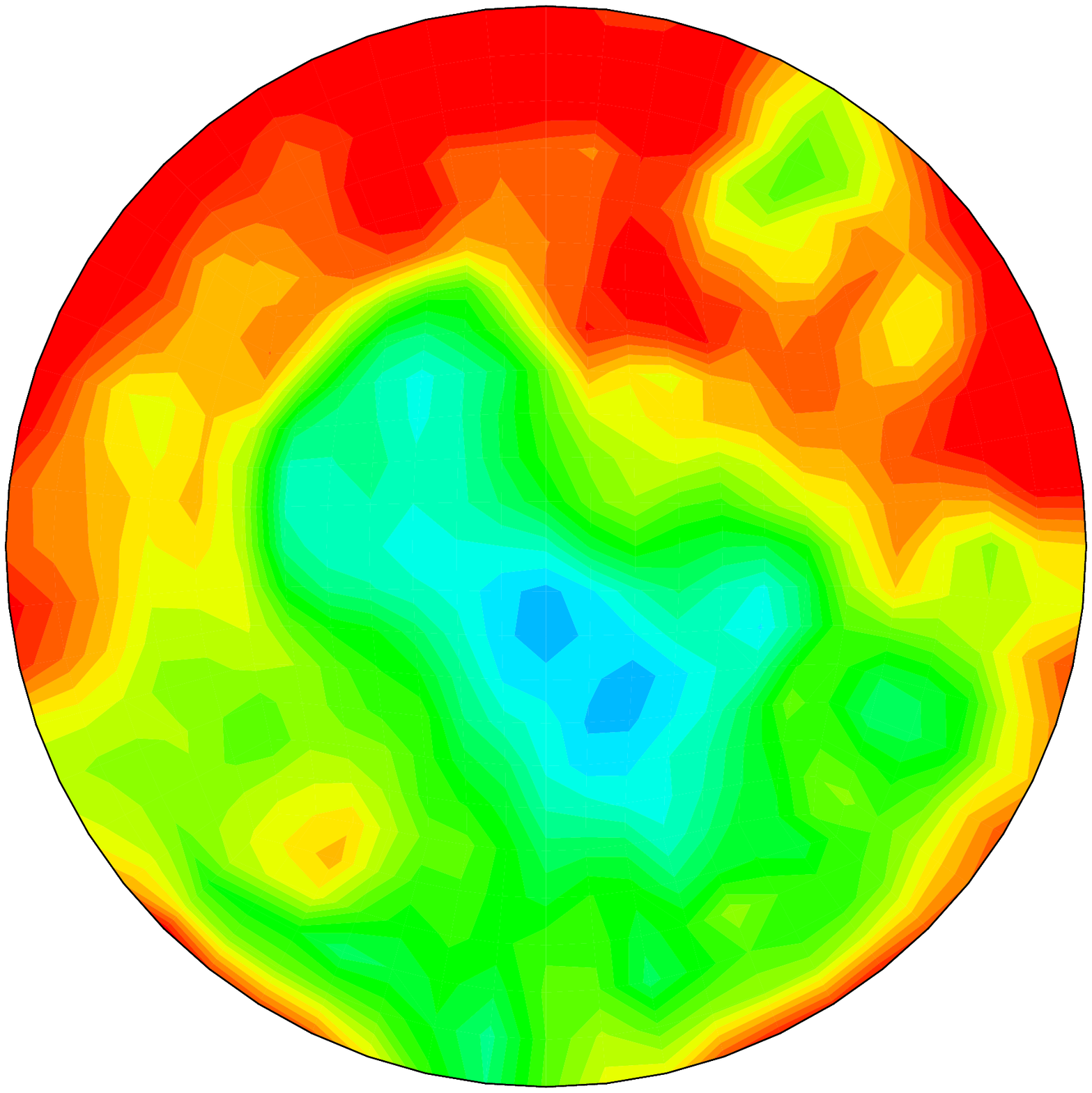}
		\end{minipage}		
	}
	\begin{center}
		\footnotesize ~~(a) $t=2.5s$ \qquad\qquad\qquad\qquad\quad ~(b) $t=3.5s$ \qquad\qquad\qquad\qquad\qquad (c) $t=4.5s$
	\end{center}
	\caption{Instantaneous snapshots of solid volume fraction $\epsilon_{s}$ by GKS-UGKWP at different times: (a) $t=2.5s$, (b) $t=3.5s$, (c) $t=4.5s$.
	 $\epsilon_{s}$ distributions near the surface of cylinder wall and at the horizontal cross sections at heights $0.078m$, $0.198m$, $0.25m$, $0.50m$ and $0.75m$, see Figure \ref{FBGao mesh and cross section sketch}(c), are presented.}	
	\label{FBGao epsilon 3D cicle shown}		
\end{figure}

The time-averaged solid volume fraction $\epsilon_{s}$ in time $2.0\sim5.5s$ is shown in Figure \ref{FBGao epsilon 3D2D profile}, and the experimental measurement results are also presented for comparison. The $\epsilon_{s}$ profiles along with the riser height and the radius agree well with the experiment data, although some derivations exist in the radial profile at height $0.138m$.
The possible reason for the deviation is most likely coming from the inaccurate drag force model, and it is
difficult to develop an accurate drag model which is suitable in all flow regimes.
Besides, in Figure \ref{FBGao epsilon 3D2D profile}(b), the presented $\epsilon_{s}$ near the wall by GKS-UGKWP is lower than that by experiment measurement at three heights. Close to the surface of the cylinder, the first experimental probe is located at $9.5\times10^{-4}m$ away from the wall; while in the computation the numerical cell size around the wall is about $3.9\times10^{-3}m$, which cannot resolve the point value observed at the probe. Therefore, with the consideration of large gradient of $\epsilon_{s}$ in the near-wall region, the low value of $\epsilon_{s}$ in numerical solution is somehow reasonable. Figure \ref{FBGao epsilon 3D cicle shown} presents the instantaneous snapshots of solid volume fraction $\epsilon_{s}$ at times $2.5s$, $3.5s$, and $4.5s$. The distributions of $\epsilon_s$ at different horizontal cross-sections, at the locations in Figure \ref{FBGao mesh and cross section sketch}(c), are shown. The results show the solid particle dense region (bottom), transition region (middle), and dilute region (top). Besides, the radial heterogeneous structure of solid particles can be found in the horizontal cross-sections. In general, the solid particles concentrate in the near-wall region.
These typical flow features are also observed in the previous experimental and numerical simulation \cite{Gasparticle-fluidized-turbulent-gao2012experimental}. Overall, GKS-UGKWP can give a reasonable prediction for this fluidized bed problem.

\begin{figure}[htbp]
	\centering
	\subfigure{
		\centering
		\includegraphics[height=1.4cm]{legendhor}
	}
	
	\subfigure{
		\centering
		\includegraphics[height=1.25cm]{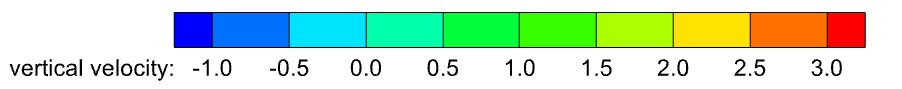}
	}

	\subfigure{				
		\centering		
		\begin{minipage}[c]{.08\textwidth}
			\centering
			\includegraphics[height=10.0cm]{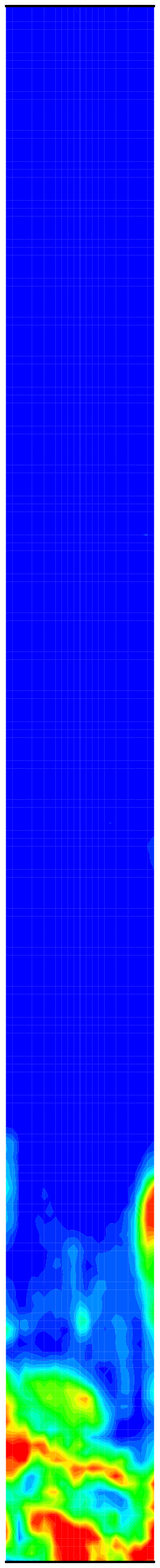}
		\end{minipage}
		\begin{minipage}[c]{.08\textwidth}
			\centering
			\includegraphics[height=10.0cm]{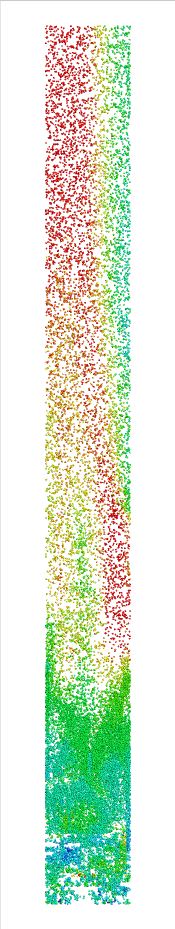}		
		\end{minipage}
		\begin{minipage}[c]{.08\textwidth}
			\centering
			\includegraphics[height=10.0cm]{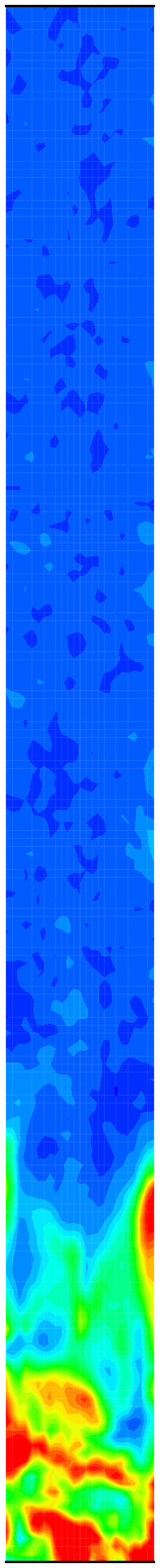}		
		\end{minipage}
	}	
	\qquad
	\subfigure{		
		\centering		
		\begin{minipage}[c]{.08\textwidth}
			\centering
			\includegraphics[height=10.0cm]{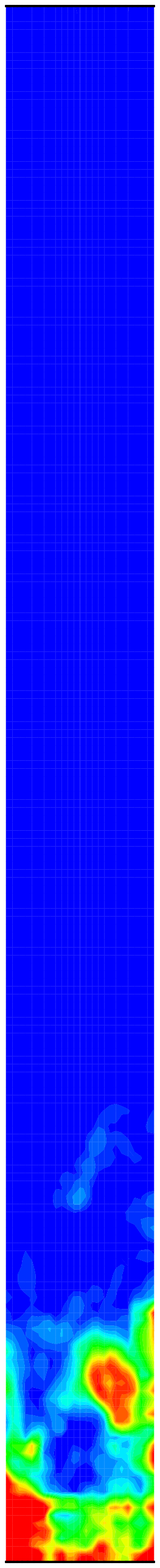}
		\end{minipage}
		\begin{minipage}[c]{.08\textwidth}
			\centering
			\includegraphics[height=10.0cm]{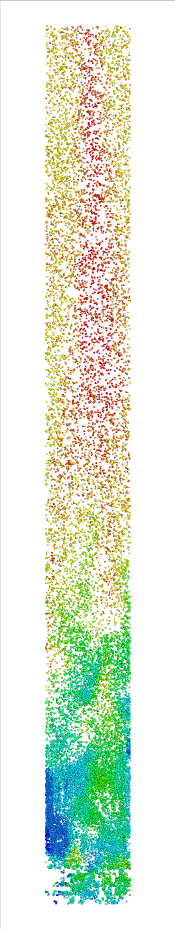}		
		\end{minipage}
		\begin{minipage}[c]{.08\textwidth}
			\centering
			\includegraphics[height=10.0cm]{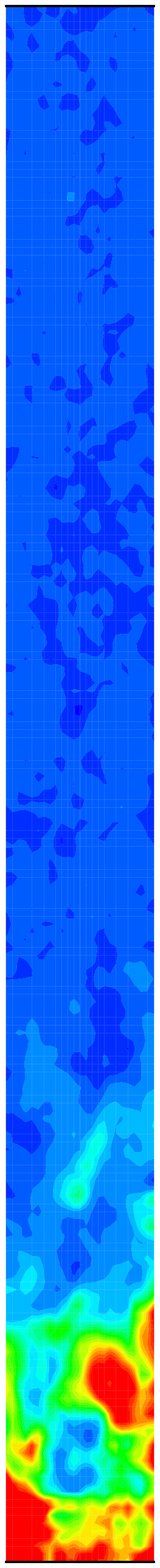}		
		\end{minipage}
	}	
	\qquad
	\subfigure{		
		\centering		
		\begin{minipage}[c]{.08\textwidth}
			\centering
			\includegraphics[height=10.0cm]{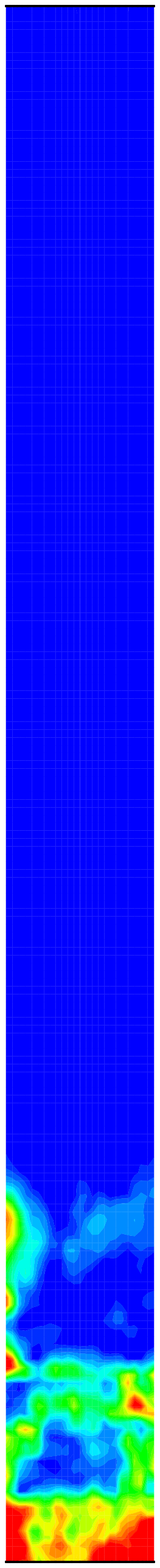}
		\end{minipage}
		\begin{minipage}[c]{.08\textwidth}
			\centering
			\includegraphics[height=10.0cm]{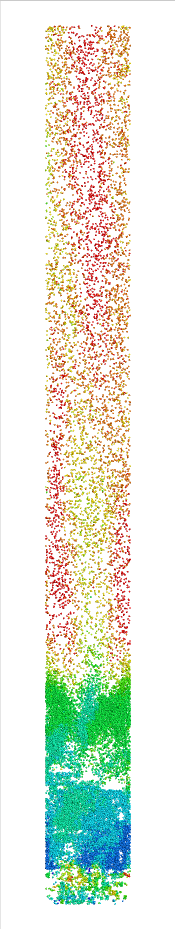}		
		\end{minipage}
		\begin{minipage}[c]{.08\textwidth}
			\centering
			\includegraphics[height=10.0cm]{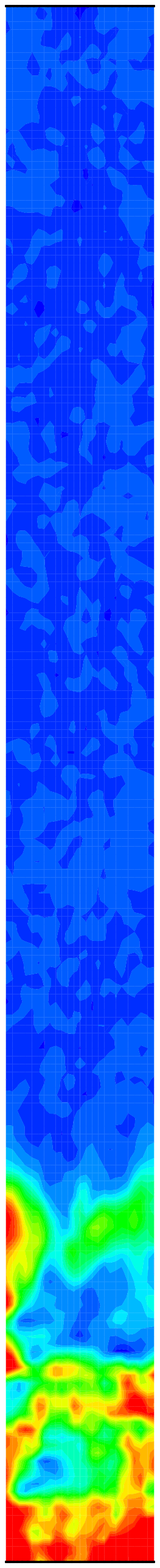}		
		\end{minipage}
	}
	\begin{center}
		\vspace{-10pt}		
		\footnotesize \quad ~~wave~~~~particle~~~~total \qquad\qquad\quad ~wave~~~~particle~~~~total \qquad\qquad\quad ~wave~~~~particle~~~~total
	\end{center}	
	\begin{center}		
		\footnotesize ~(a) $t=2.5s$ \qquad\qquad\qquad\qquad\qquad ~(b) $t=3.5s$ \qquad\qquad\qquad\qquad\qquad ~~~(c) $t=4.5s$
	\end{center}		
	\caption{Solid volume fraction $\epsilon_s$ represented by wave and discrete particle (colored by vertical velocity) at different times: (a) $t=2.5s$, (b) $t=3.5s$, (c) $t=4.5s$, on the vertical symmetric cross-section shown in Figure \ref{FBGao mesh and cross section sketch}(d).
The sub-figures give the individual \textbf{wave} and \textbf{particle} decomposition, and the \textbf{total} $\epsilon$.
The \textbf{wave} composition ($\epsilon_s^{wave}$) is colored by the epsilon-legend.
The discrete \textbf{particle} composition is colored by the vertical velocity-legend. The \textbf{total} $\epsilon_s$ is also colored by the
epsilon-legend. Note that $\epsilon_{s}$ is exactly the sum of wave and discrete particle decomposition, such as \textbf{wave} $+$ \textbf{particle} $=$ \textbf{total}.}	
	\label{FBGao epsilon 3D plane shown wave and particle1}		
\end{figure}

In order to show the coupled evolution of wave and particle in UGKWP for solid particle flow, the constitution of wave and sampled particle at different times are shown in Figure \ref{FBGao epsilon 3D plane shown wave and particle1}.
The solid particle distributions from the decompositions of wave and discrete particle are clearly presented.
For this turbulent fluidized bed, the solid phase is generally dilute in the riser's top region (above 0.3m in height), where
the particle free transport is dominant. The particle is in a non-equilibrium state driven by the gas flow.
On the other hand, in the bottom region the particles are highly concentrated, especially in the zones near riser wall.
In this region, the intensive particle-particle collision pushes the particle distribution to an equilibrium state,
and the particle phase evolution is mainly controlled by the wave component through the hydrodynamic flow variables.
Even with abundant solid particles, few particles will appear in UGKWP in this region.
Besides the above limiting cases, in the transition regions with an intermediate $\epsilon_s$, such as the layers with height $0.1m \sim 0.2m$,
both wave and discrete solid particle influence the evolution and the solution update depends
on the fluxes, as shown in Eq.\eqref{particle phase wp final update W}, from both hydrodynamic wave (EE) and discrete particles transport (EL).
The modeling in UGKWP captures the multi-scale nature of solid particle transport and presents
a smooth transition to cover the dense, transition, and dilute particle regions in a fluidized bed.

\subsection{Circulating fluidized bed case}
\begin{figure}[htbp]
	\centering
	\subfigure[]{		
		\includegraphics[height=10.0cm]{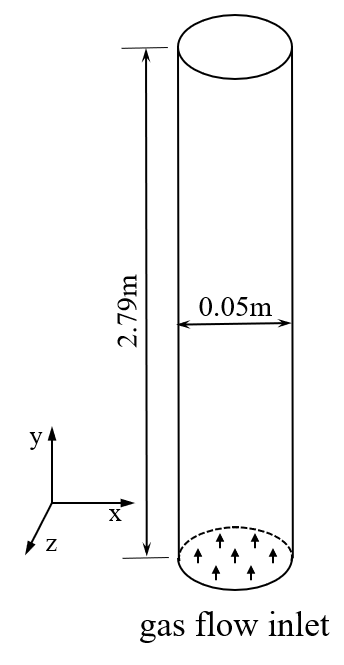}			
	}
	\quad	
	\subfigure[]{
		\raisebox{0.02\textwidth}{			
			\includegraphics[height=10.0cm]{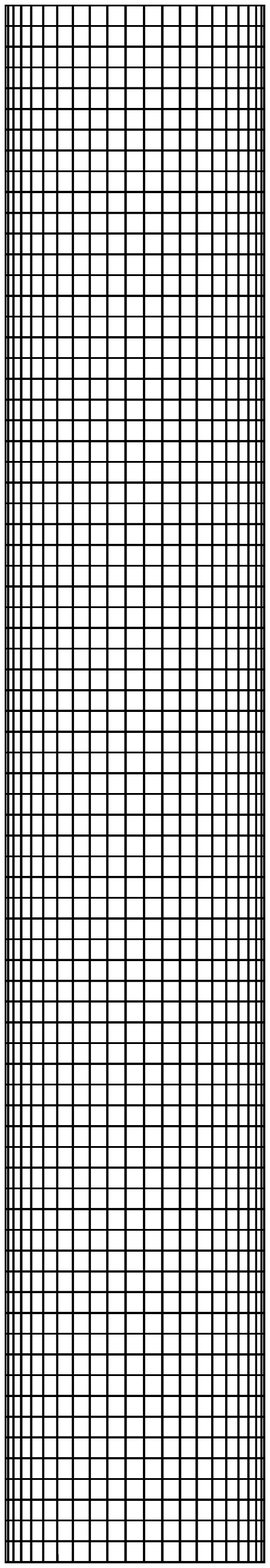}
		}	
	}
	\quad
	\subfigure[]{
		\raisebox{0.17\textwidth}{			
			\includegraphics[height=4.5cm]{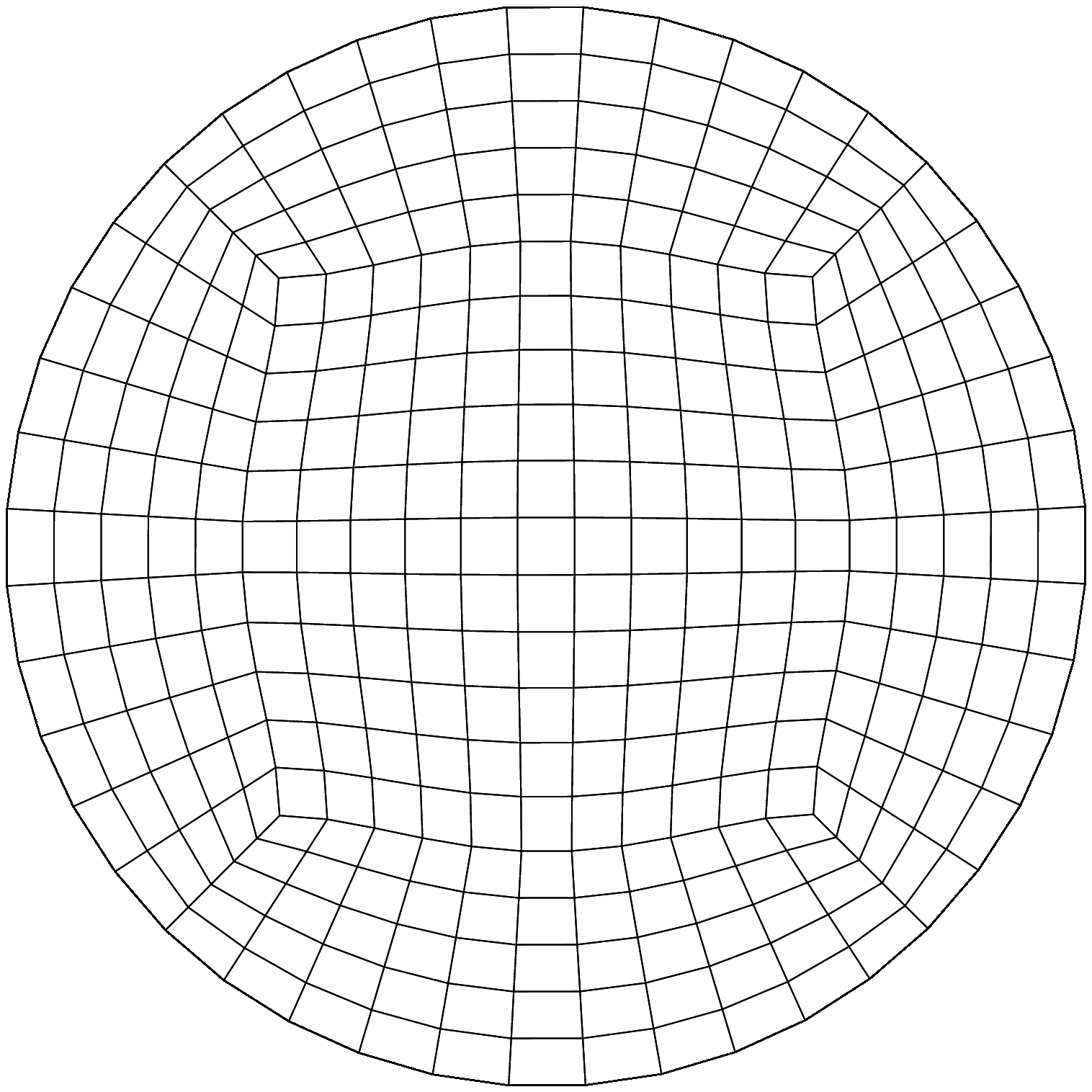}
		}	
	}	
	\caption{Sketch of geometry and mesh of Horio riser: (a) geometry of the riser employed in this paper, (b) the mesh from the front view (with a section of $0.3m$ in height), (c) the mesh on each cross-section.}	
	\label{FBHorio gemotery and mesh sketch}		
\end{figure}

In the following, the 3D circulating fluidized bed of Horio et al. is simulated by GKS-UGKWP \cite{Gasparticle-fluidized-circulating-horio1988solid}. This case has been taken as a typical example to validate the numerical methods for gas-particle two-phase flow \cite{Gasparticle-subgridmodel-EMMS-vs-Filtered-TFM-hongkong2016fine, Gasparticle-subgridmodel-EMMS-MPPIC-li2012mp}. The sketch of the riser is shown in Figure \ref{FBHorio gemotery and mesh sketch}(a), and the computational domain is a cylinder with a diameter $R=0.05m$ and a height $H=2.80m$, which is $0.01m$ higher than the actual size. The numerical cells are hexahedron. Figure \ref{FBHorio gemotery and mesh sketch}(b) and Figure \ref{FBHorio gemotery and mesh sketch}(c) present the mesh from the front (a section of $0.3m$ in height) and top view, respectively. The total number of cells is 238700 with 341 in horizontal surface and 700 in the vertical direction.
The cell size is approximately $\Delta=2.26\times10^{-3}m$ horizontally and $\Delta=4.00\times10^{-3}m$ vertically.
The solid particles employed in the experiment have material density $\rho_{s}=1000kg/m^3$ and diameter $d_s=60um$.
In the numerical simulation, initially the solid particles are uniformly distributed in the whole riser with a solid volume fraction $\epsilon_{s,0}=0.086$. For the gas phase, the top boundary is set as the outlet pressure, and the air blows from the bottom into the riser with the uniform velocity $U_g=1.17m/s$ and pressure $\epsilon_{s,0}\left(\rho_s-\rho_g\right)GH$. Same as the previous case, solid particles can escape from the riser at the top boundary, and come back into the riser at the bottom boundary, ensuring a constant solid material in the riser.
At the cylinder surface, the non-slip wall boundary condition is used for the gas phase; while for the solid phase, the mixed boundary condition proposed by Johnson et al. is employed \cite{Gasparticle-KTGF-pressure-friction-johnson1987frictional}.

\begin{figure}[htbp]
	\centering
	\subfigure[]{
		\includegraphics[height=6.5cm]{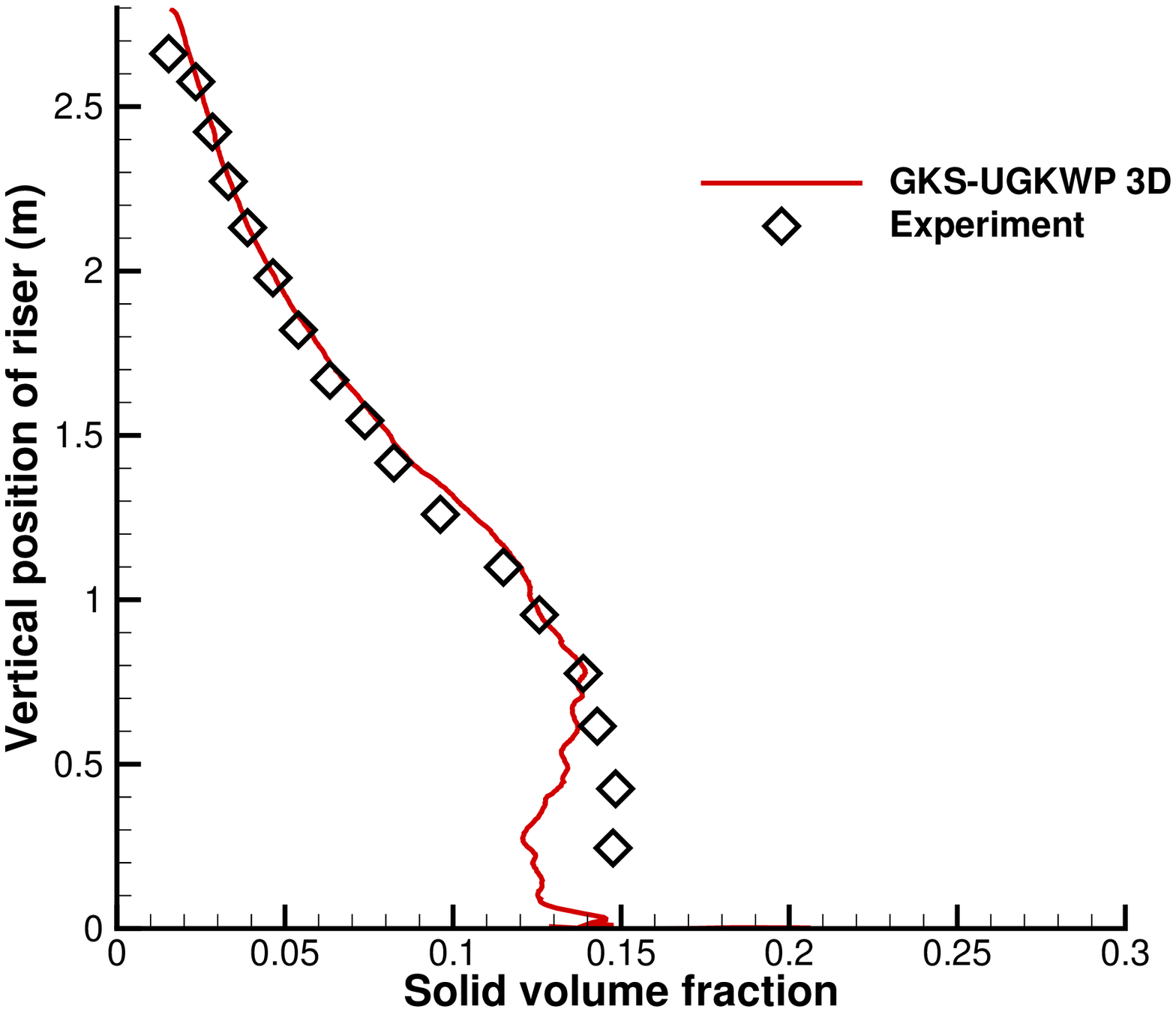}	
	}	
	
	\subfigure[]{
		\includegraphics[height=6.5cm]{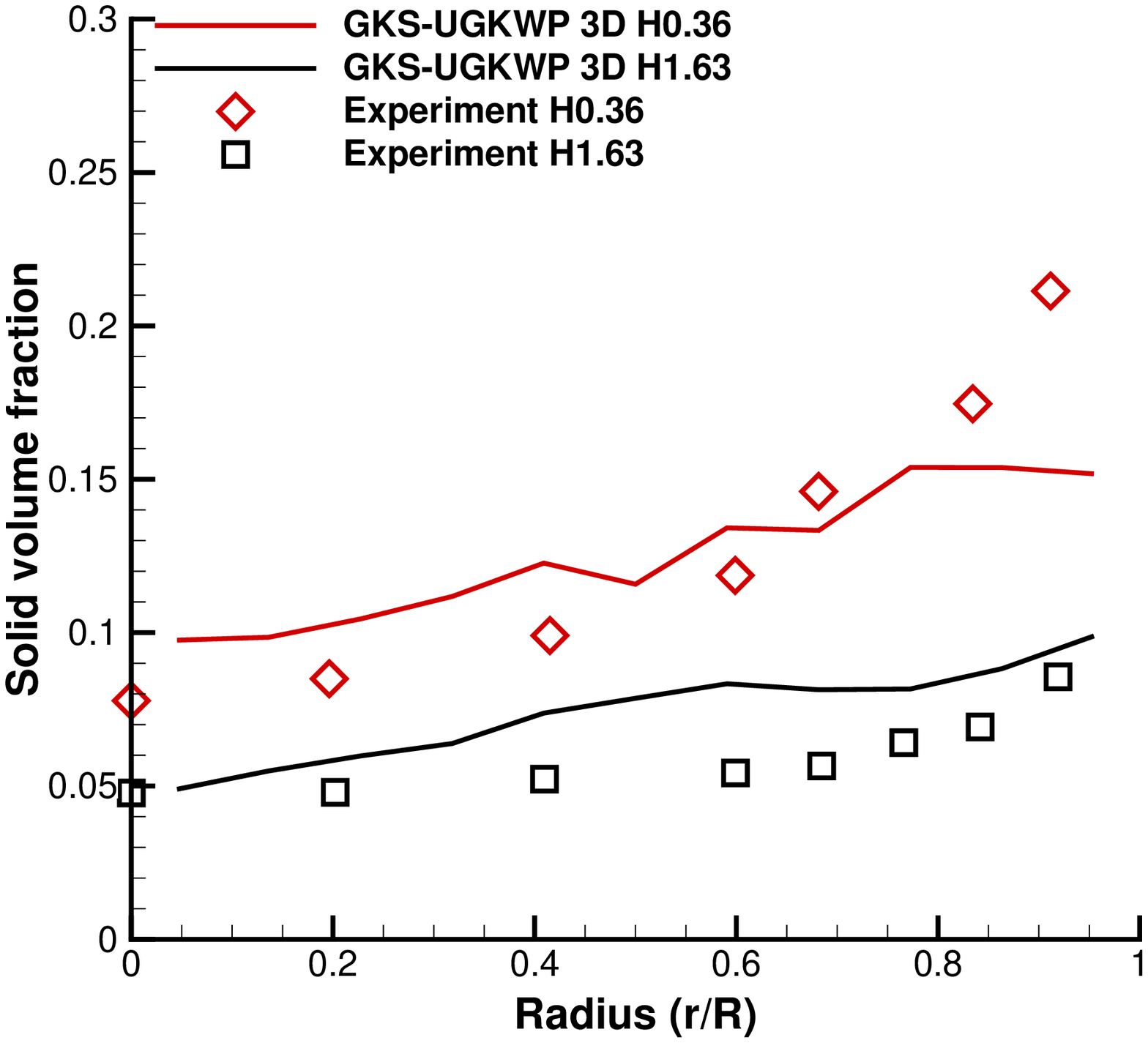}	
	}
	\subfigure[]{
		\includegraphics[height=6.5cm]{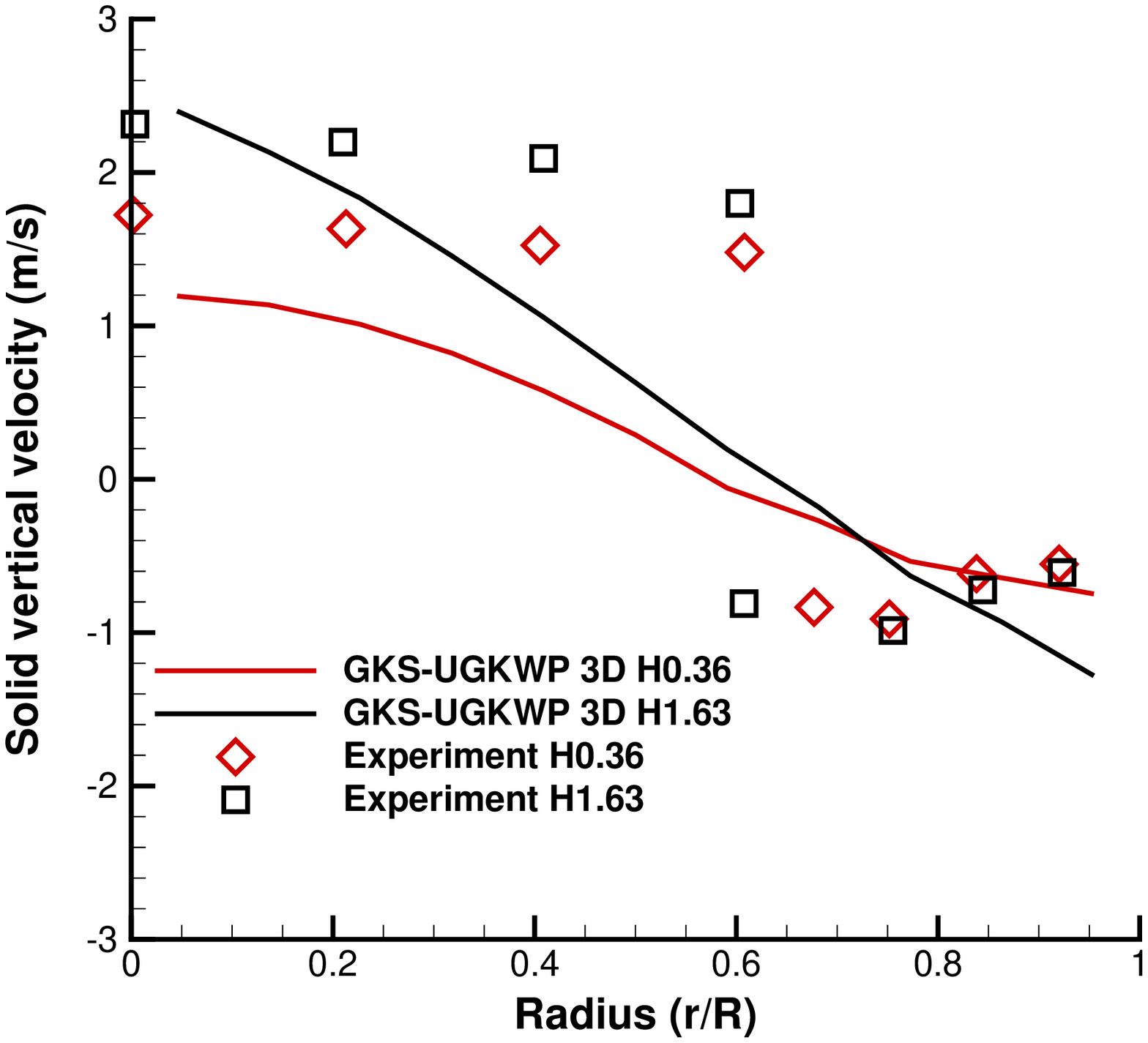}	
	}		
	\caption{Time-averaged volume fraction $\epsilon_{s}$ and vertical velocity $V_s$ of solid phase from GKS-UGKWP and experiment measurement: (a) $\epsilon_{s}$ distribution along the riser height, (b) $\epsilon_{s}$ distribution along the riser radius at $0.36m$ and $1.63m$ respectively, (c) $V_s$ distribution along the riser radius at $0.36m$ and $1.63m$ respectively.}	
	\label{FBHorio 3D epsilon and velocity profile}		
\end{figure}

The MP-PIC method coupled with EMMS drag force was employed to study this case, and the results showed obvious improvement than the traditional homogeneous drag model proposed by Gidaspow \cite{Gasparticle-subgridmodel-EMMS-MPPIC-li2012mp}.
In the current study, the EMMS drag force model is used in GKS-UGKWP method for this circulating fluidized bed riser,
\begin{equation}\label{beta by EMMS}
\beta=
\frac{3}{4}C_d\frac{\rho_g\epsilon_{s}\epsilon_{g}|\textbf{U}_g-\textbf{u}|}{d_s} \epsilon_{g}^{-2.7} H_{D},
\end{equation}
where $Re_s$ is the particle Reynolds number, and $C_d$ is the drag coefficient calculated by Eq.\eqref{Cd}. The $H_{D}$ is the so-called heterogeneity index, which is defined as,
\begin{equation}
H_{D} = a\left(Re_{s}+b\right)^c,
\end{equation}
and $a$, $b$, and $c$ are the model parameters dependent on the solid volume fraction $\epsilon_{s}$ and with the consideration of
local heterogeneous flow structures.
The specific values of model parameters ($a$, $b$, $c$) are listed in Appendix A, and more detailed introduction about EMMS drag force can refer to the previous work \cite{Gasparticle-subgridmodel-EMMS-drag-lu2011eulerian, Gasparticle-subgridmodel-EMMS-MPPIC-li2012mp}.

The time-averaged distributions of solid particles and the vertical solid velocity from the time interval $4.0s\sim7.5s$ are shown in Figure \ref{FBHorio 3D epsilon and velocity profile}. Figure \ref{FBHorio 3D epsilon and velocity profile}(a) presents the profile of $\epsilon_s$ along with the riser height, which covers heterogeneous feature of the solid flow from bottom dense region, across middle transition region, and up to the top dilute region. This flow feature is similar to that in the turbulent fluidized bed. However, in the circulating fluidized bed, the transition between dense and dilute regions is moderate, as shown in Figure \ref{FBGao epsilon 3D2D profile}(a) and Figure \ref{FBHorio 3D epsilon and velocity profile}(a). Although a slight deviation exists at the bottom region (below 0.5m), the vertical $S$-shaped curve of $\epsilon_s$ is captured by GKS-UGKWP for this circulating fluidized bed problem, and it agrees with the experimental measurement very well.
It may come from accurate inter-phase interaction EMMS drag model.
The time-averaged distribution of solid volume fraction $\epsilon_{s}$ and solid vertical velocity $V_s$ along the riser radius at height $0.36m$ and $1.63m$ are shown in Figure \ref{FBHorio 3D epsilon and velocity profile}(b) and Figure \ref{FBHorio 3D epsilon and velocity profile}(c), respectively. At both bottom region at height $0.36m$ and top region at height $1.63m$, the solid particles show higher concentration in the near-wall region than that in the central region. At the same time, the solid particle vertical velocity shows the upward movement in the central region and downward motion in the near-wall region. This so-called core-annular flow structure is widely observed in the circulating fluidized bed riser. The obtained $\epsilon_s$ and $V_s$ by GKS-UGKWP agree well with the experiment data, and the deviations deserve further investigation.

\begin{figure}[htbp]
	\centering
	\subfigure[]{
		\includegraphics[height=11.0cm]{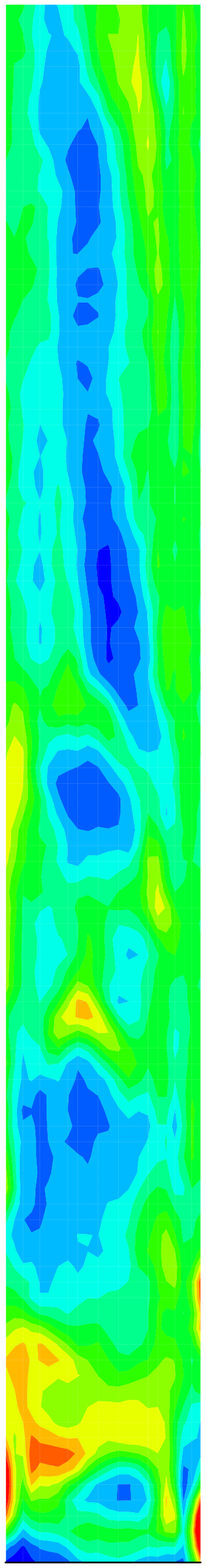}	
	}
	\hspace{1mm}
	\subfigure[]{
		\includegraphics[height=11.0cm]{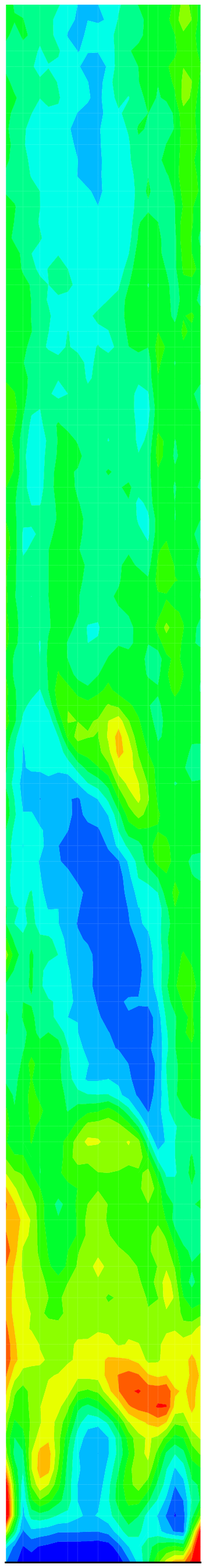}	
	}
	\hspace{1mm}
	\subfigure[]{
		\includegraphics[height=11.0cm]{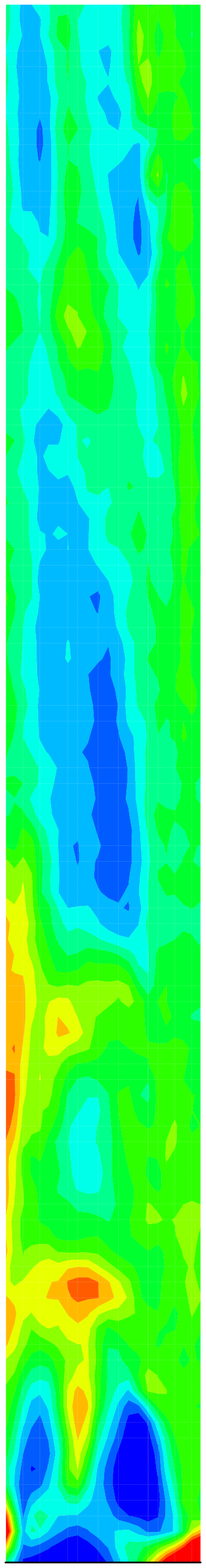}	
	}
	\hspace{1mm}
	\subfigure[]{
		\includegraphics[height=11.0cm]{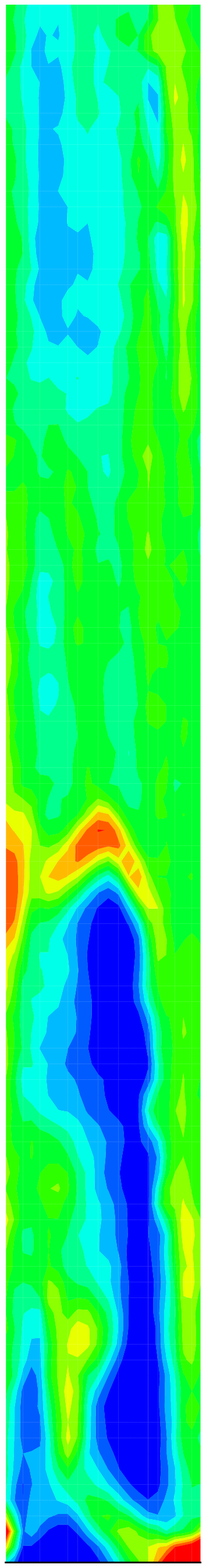}	
	}
	\hspace{1mm}
	\subfigure[]{
		\includegraphics[height=11.0cm]{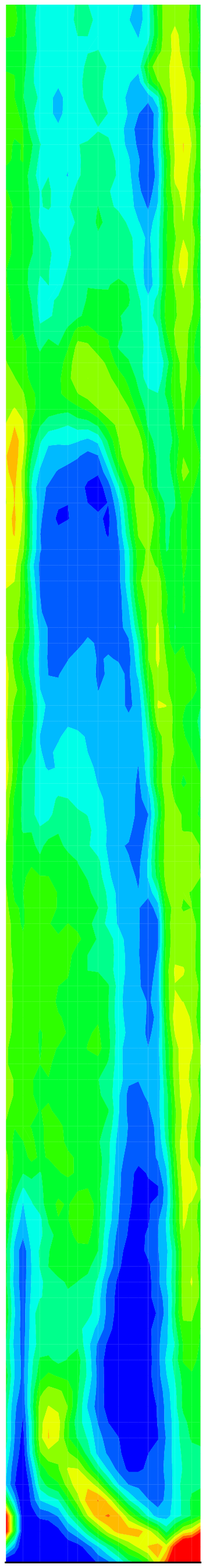}	
	}
	\hspace{1mm}	
	\subfigure{
		\includegraphics[height=8.0cm]{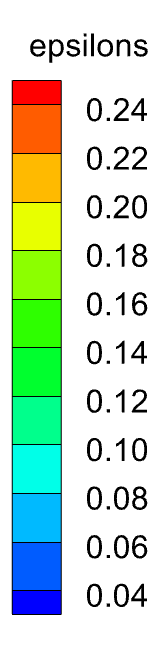}
	}
	\caption{Instantaneous snapshots of solid volume fraction $\epsilon_{s}$ on the vertical symmetric plane in a zone with heights from 0 to 0.4m at different times: (a) $t=7.0s$, (b) $t=7.1s$, (c) $t=7.2s$, (d) $t=7.3s$, (e) $t=7.4s$. }	
	\label{FBHorio 3D epsilon shown}		
\end{figure}

\begin{figure}[htbp]
	\centering
	\subfigure{
		\centering
		\includegraphics[height=1.4cm]{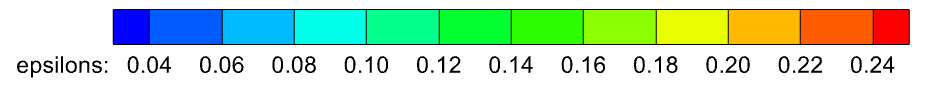}
	}
	
	\subfigure{
		\hspace{-5mm}
		\includegraphics[height=1.3cm]{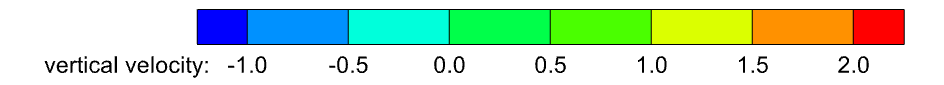}
	}	
	
	\subfigure{				
		\centering		
		\begin{minipage}[c]{0.13\textwidth}
			\centering
			\includegraphics[height=10.0cm]{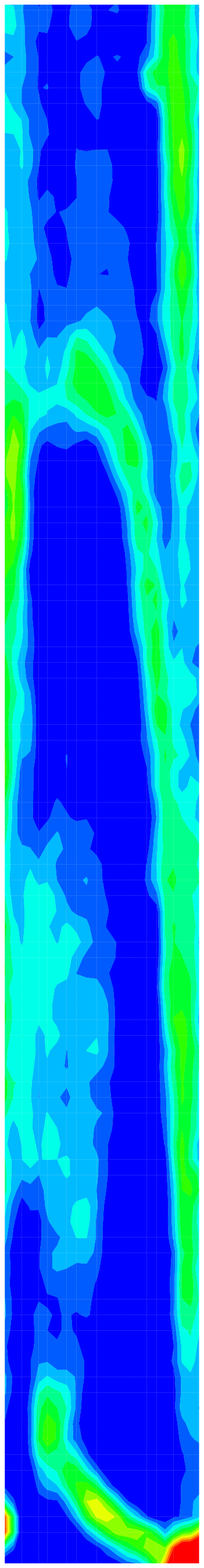}
		\end{minipage}
		\begin{minipage}[c]{0.13\textwidth}
			\centering
			\includegraphics[height=10.0cm]{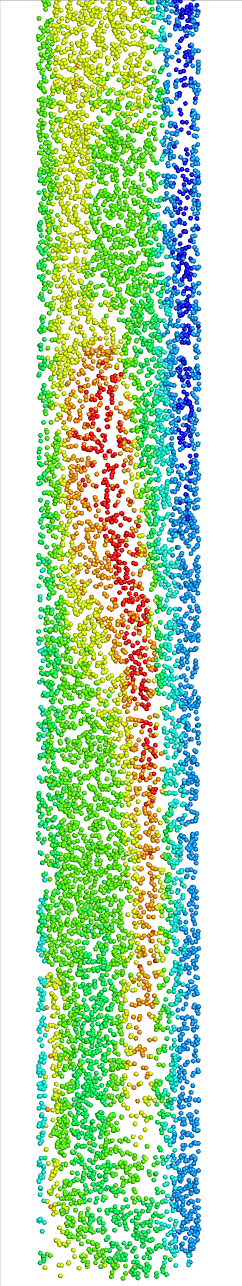}		
		\end{minipage}
		\begin{minipage}[c]{0.13\textwidth}
			\centering
			\includegraphics[height=10.0cm]{epsT0d47.eps}		
		\end{minipage}
	}	
	\hspace{10mm}
	\subfigure{				
		\centering		
		\begin{minipage}[c]{0.13\textwidth}
			\centering
			\includegraphics[height=10.0cm]{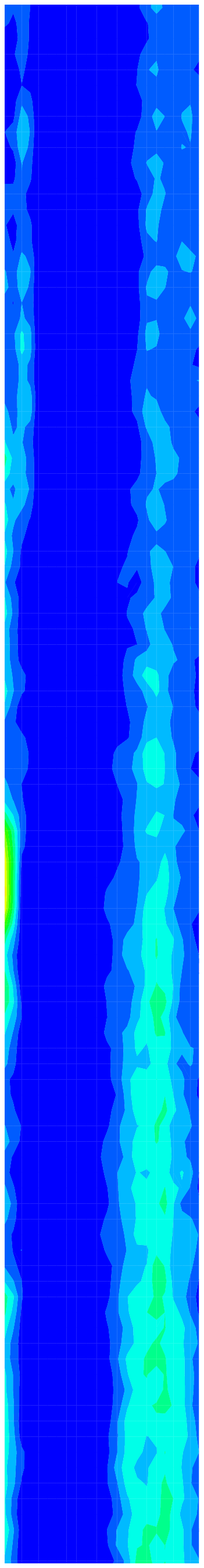}
		\end{minipage}
		\begin{minipage}[c]{0.13\textwidth}
			\centering
			\includegraphics[height=10.0cm]{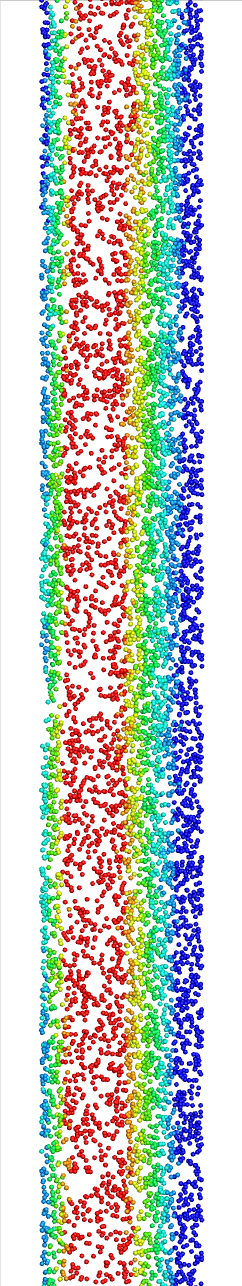}		
		\end{minipage}
		\begin{minipage}[c]{0.13\textwidth}
			\centering
			\includegraphics[height=10.0cm]{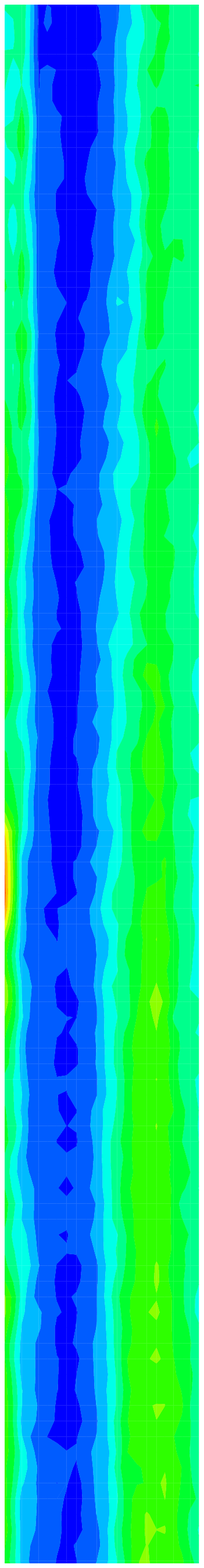}		
		\end{minipage}
	}
	\begin{center}
		\vspace{-7pt}		
		\footnotesize  wave~\qquad\quad~particle~\qquad\quad~total \qquad\qquad\qquad\quad\quad wave~\qquad\quad~particle~\qquad\quad~total~
	\end{center}	
	\begin{center}
		\footnotesize (a) $0\sim0.4m$ \qquad\qquad\qquad\qquad\qquad\qquad\qquad\qquad\qquad~ (b) $1.4\sim1.8m$
	\end{center}				
	\caption{Solid volume fraction $\epsilon_s$ of wave decomposition and discrete particles at $t=7.4s$ in different zones: (a) $0\sim0.4m$, (b) $1.4\sim1.8m$.
The sub-figures give the individual \textbf{wave} and \textbf{particle} decomposition, and the \textbf{total} $\epsilon$.
The \textbf{wave} composition ($\epsilon_s^{wave}$) is colored by the epsilon-legend.
The discrete \textbf{particle} composition is colored by the vertical velocity-legend. The \textbf{total} $\epsilon_s$ is also colored by the
epsilon-legend. Note that $\epsilon_{s}$ is exactly the sum of wave and discrete particle decomposition, such as \textbf{wave} $+$ \textbf{particle} $=$ \textbf{total}. }	
	\label{FBHorio 3D epsilon 2D vertical plane shown}		
\end{figure}

\begin{figure}[htbp]
	\subfigure{
		\centering
		\includegraphics[height=1.4cm]{legendEpsHor}
	}
	
	\subfigure{
		\hspace{-5mm}
		\includegraphics[height=1.3cm]{legendVsHor}
	}	

	\subfigure{		
		\centering		
		\begin{minipage}[c]{.15\textwidth}
			\centering
			\includegraphics[height=2.5cm]{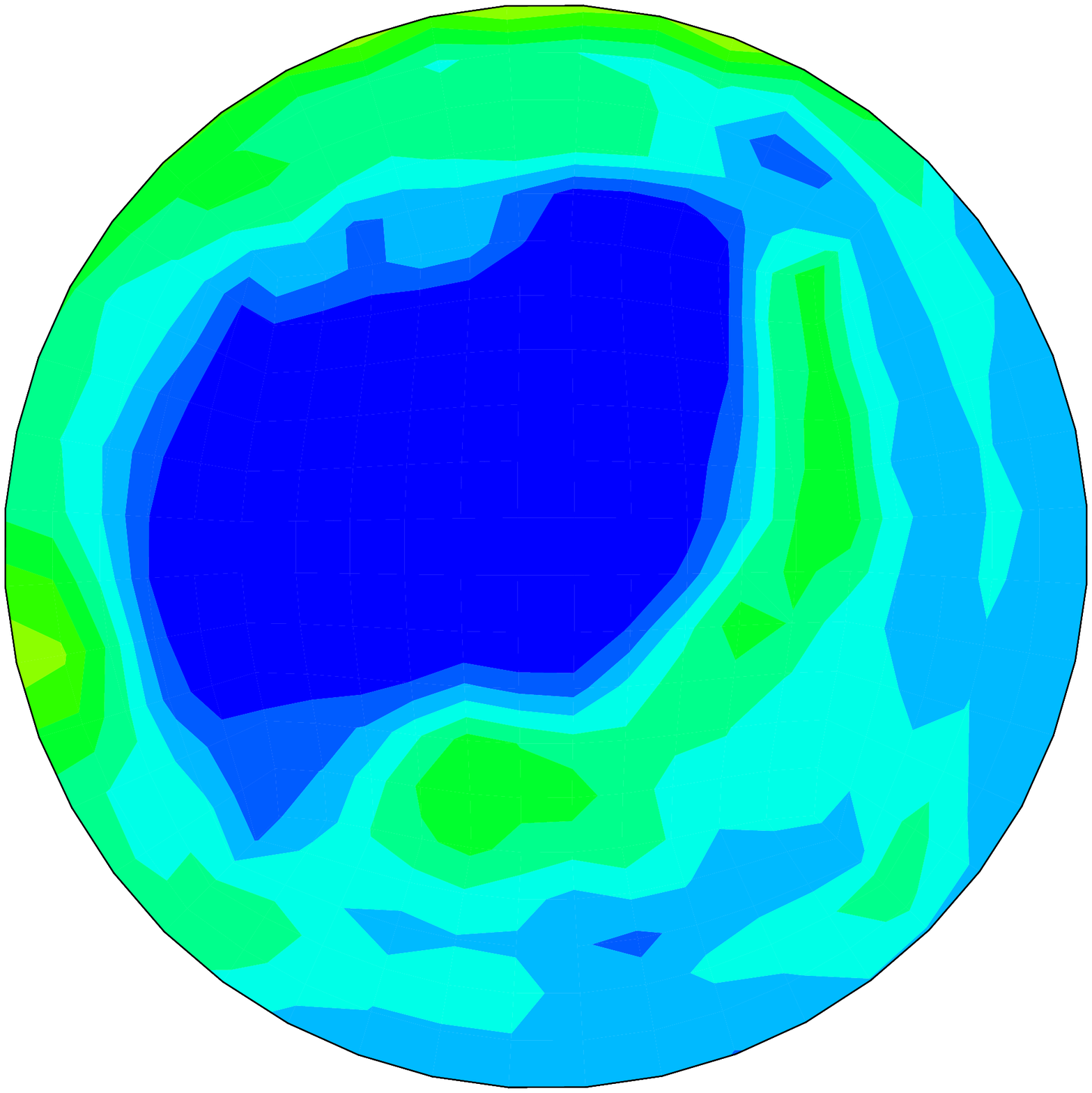}
			\includegraphics[height=2.5cm]{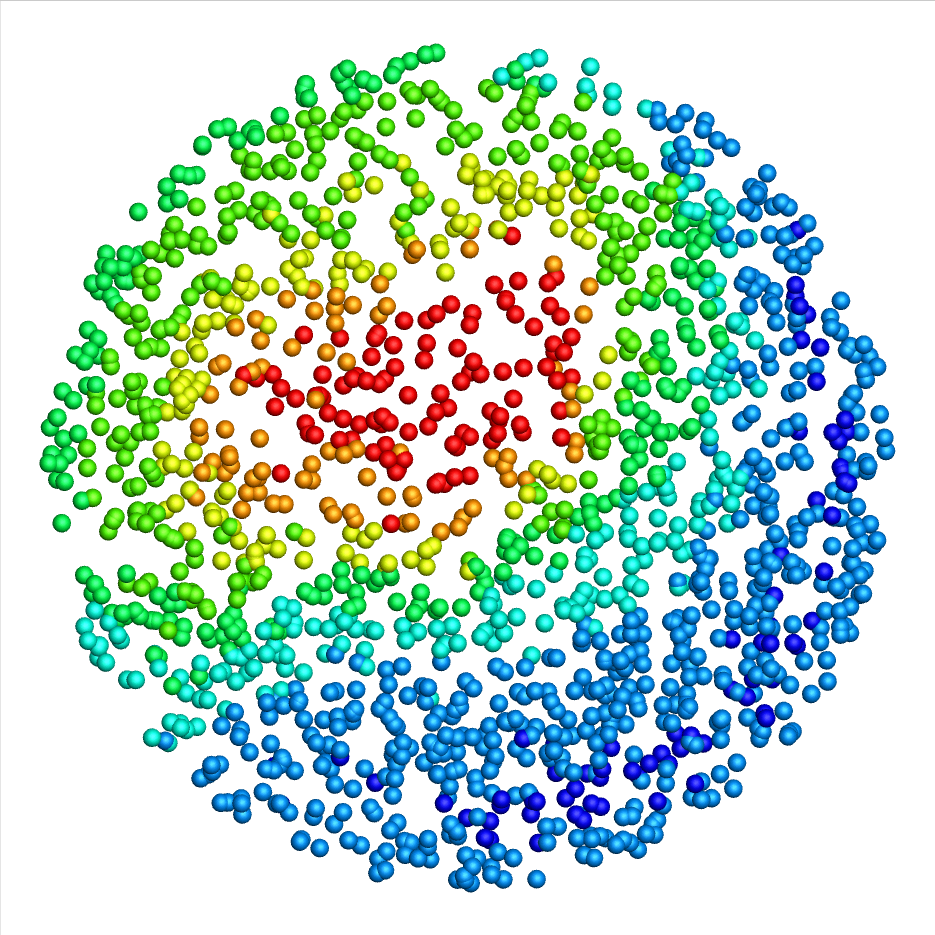}
			\includegraphics[height=2.5cm]{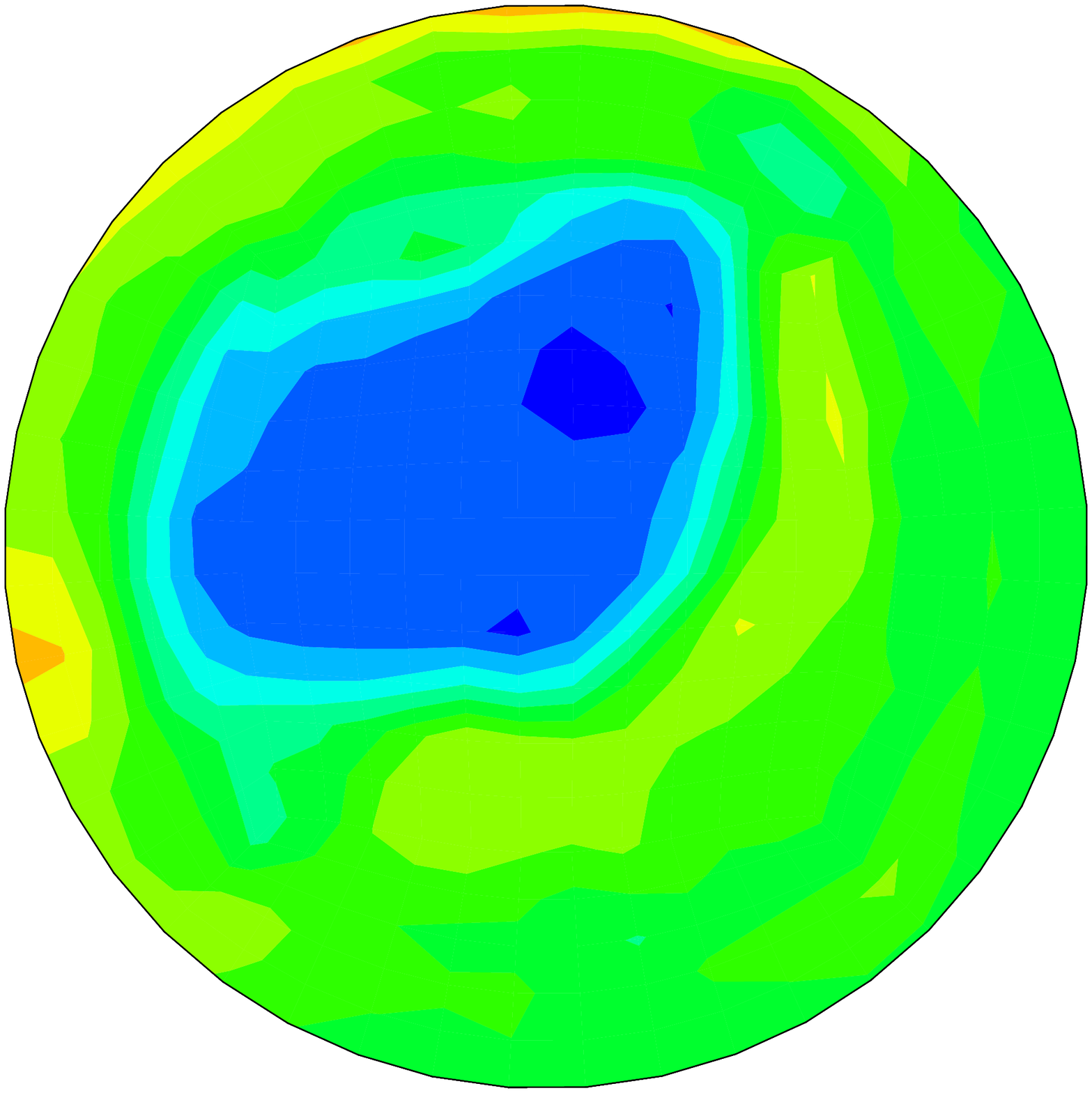}
		\end{minipage}
	}
	\quad
	\subfigure{
		\centering	
		\begin{minipage}[c]{.15\textwidth}
			\centering
			\includegraphics[height=2.5cm]{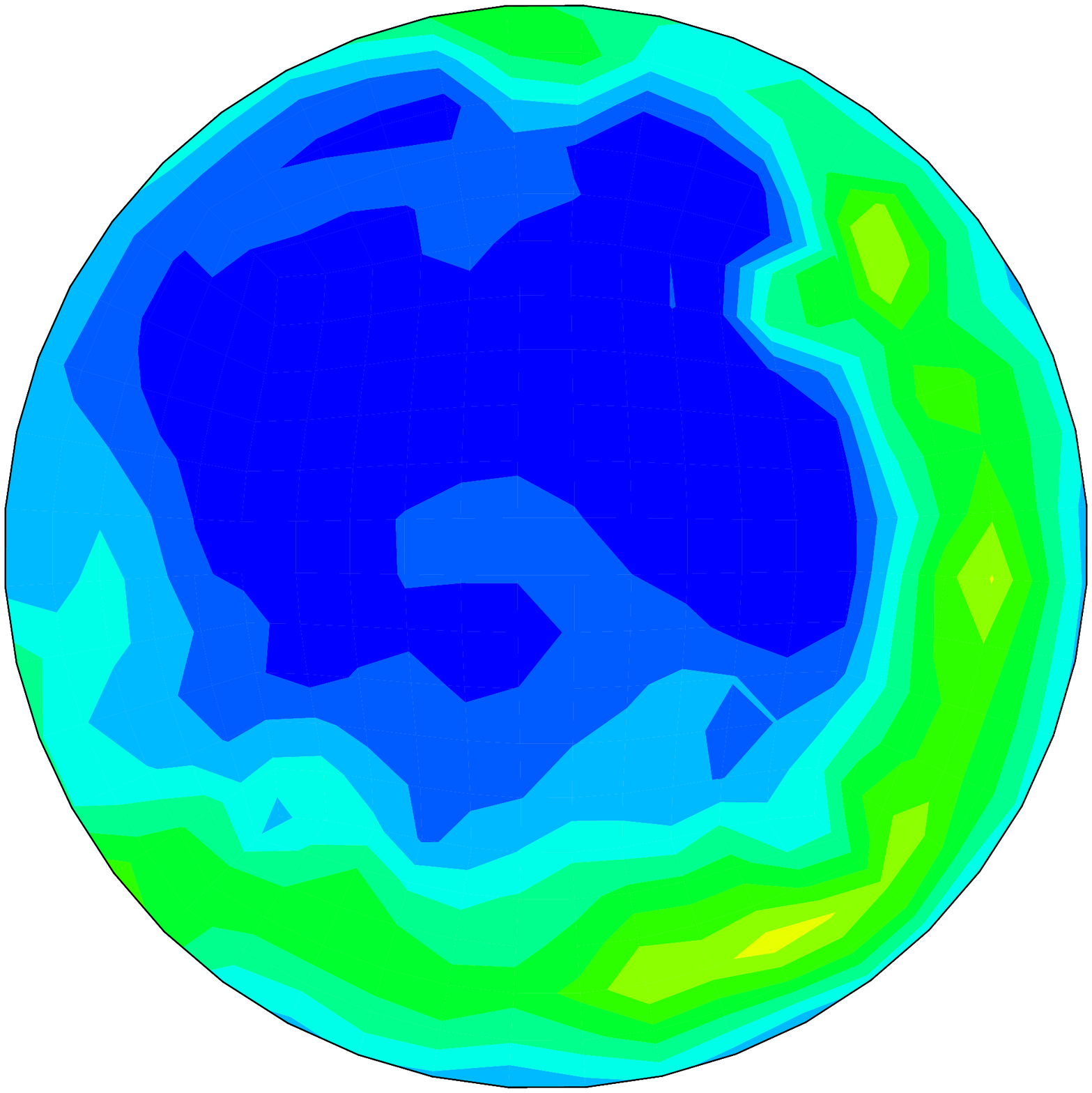}
			\includegraphics[height=2.5cm]{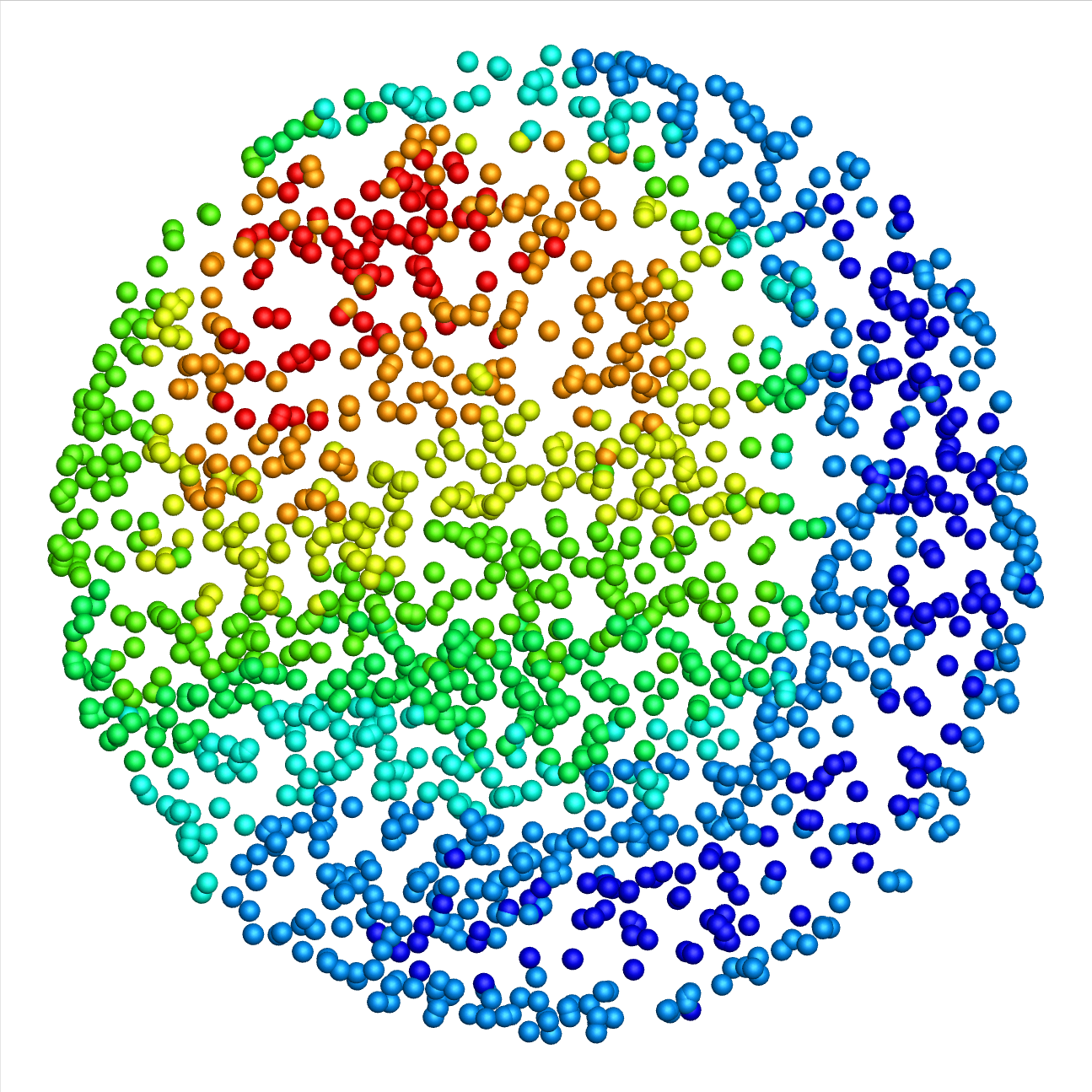}
			\includegraphics[height=2.5cm]{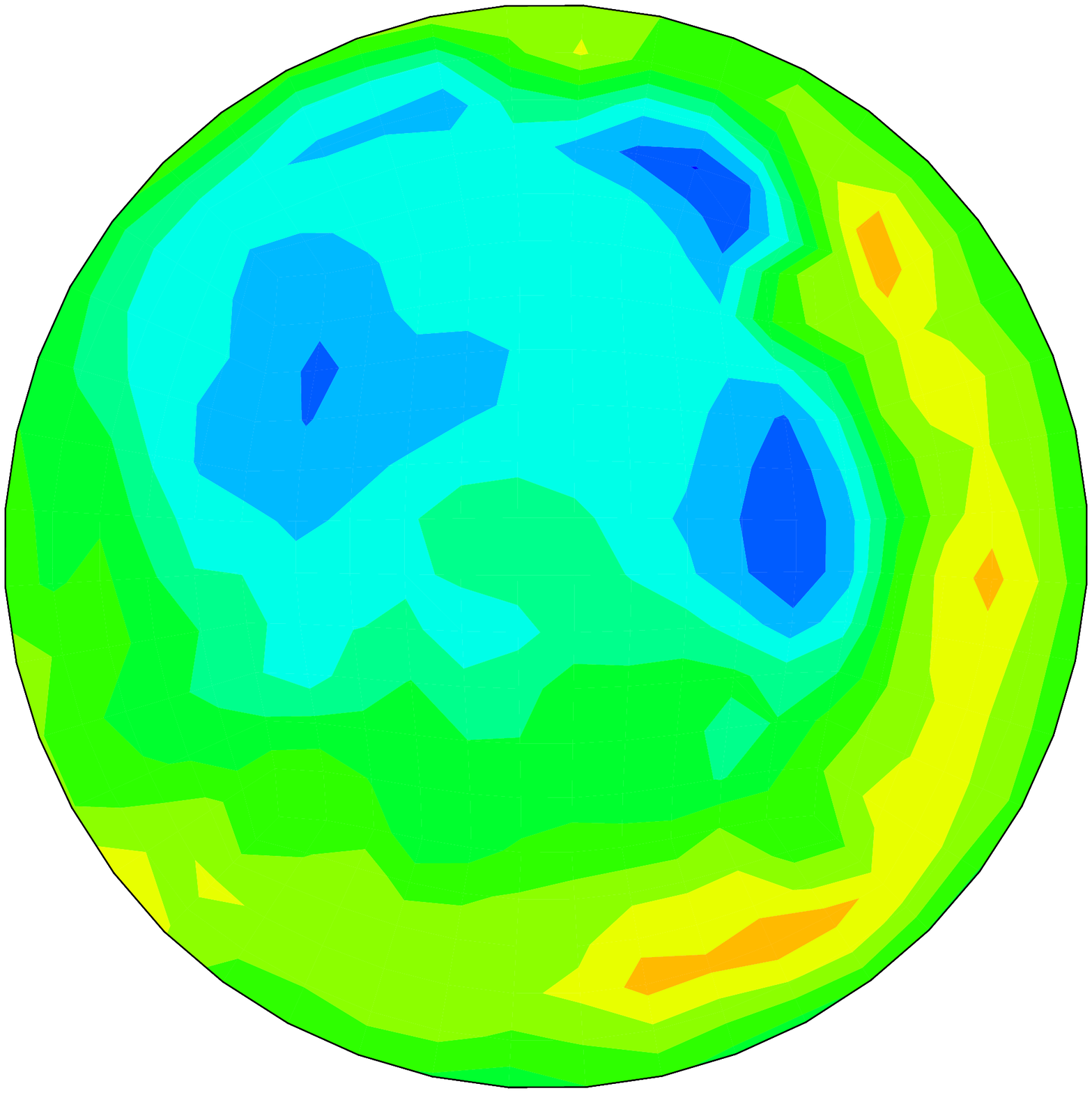}
		\end{minipage}		
	}
	\quad
	\subfigure{
		\centering
		\begin{minipage}[c]{.15\textwidth}
			\centering
			\includegraphics[height=2.5cm]{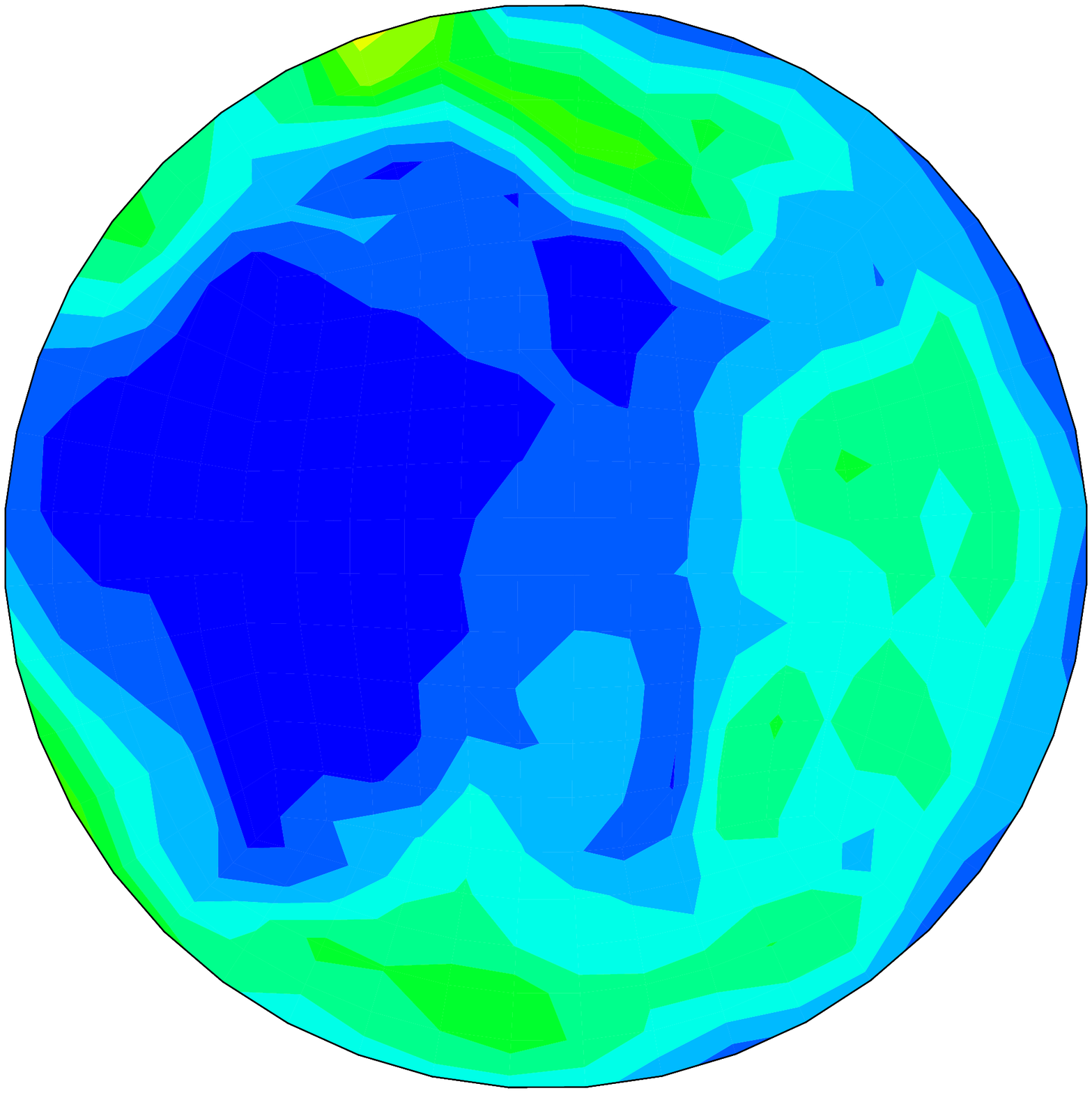}
			\includegraphics[height=2.5cm]{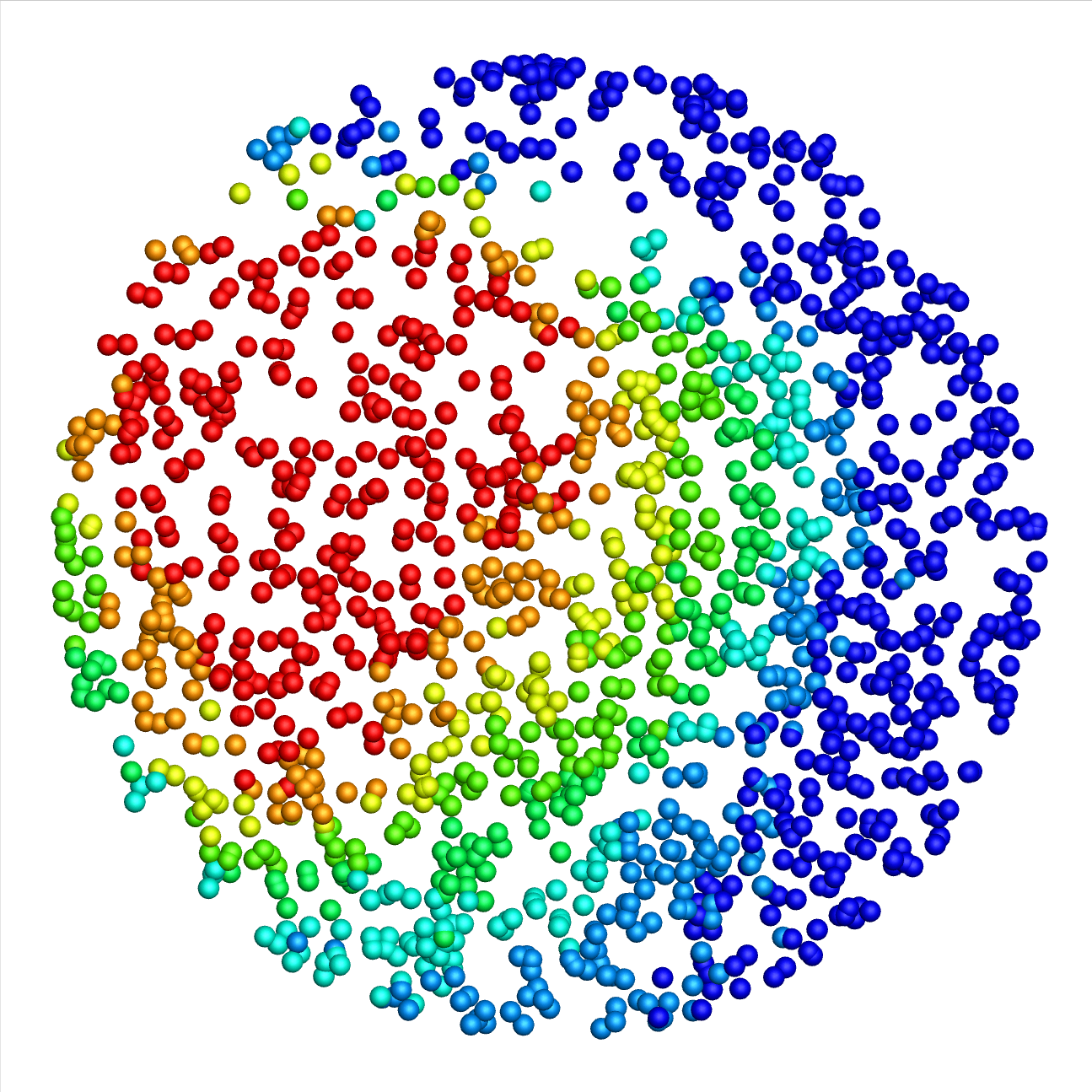}
			\includegraphics[height=2.5cm]{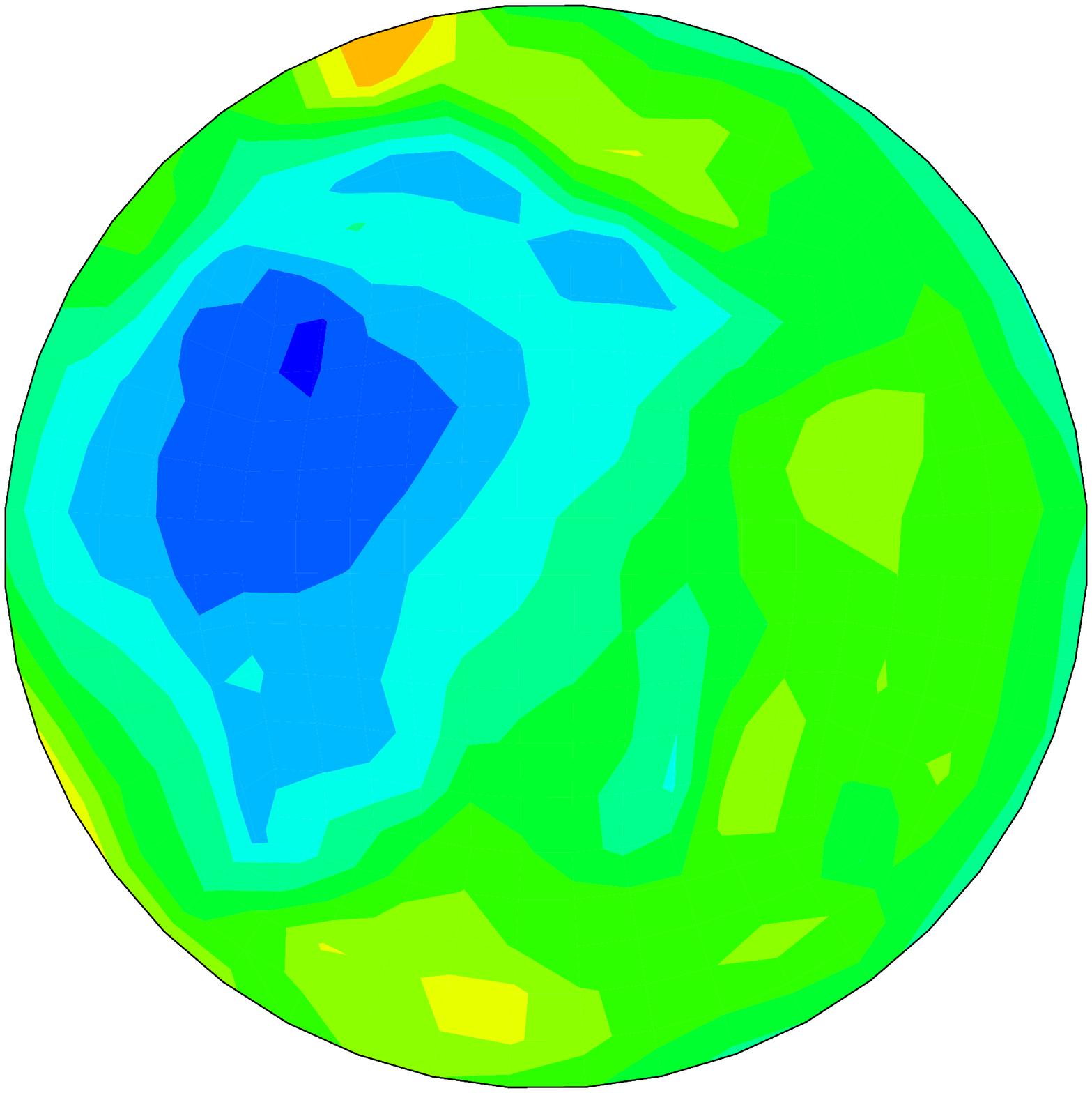}	
		\end{minipage}		
	}
	\quad
	\subfigure{
		\centering
		\begin{minipage}[c]{.15\textwidth}
			\centering
			\includegraphics[height=2.5cm]{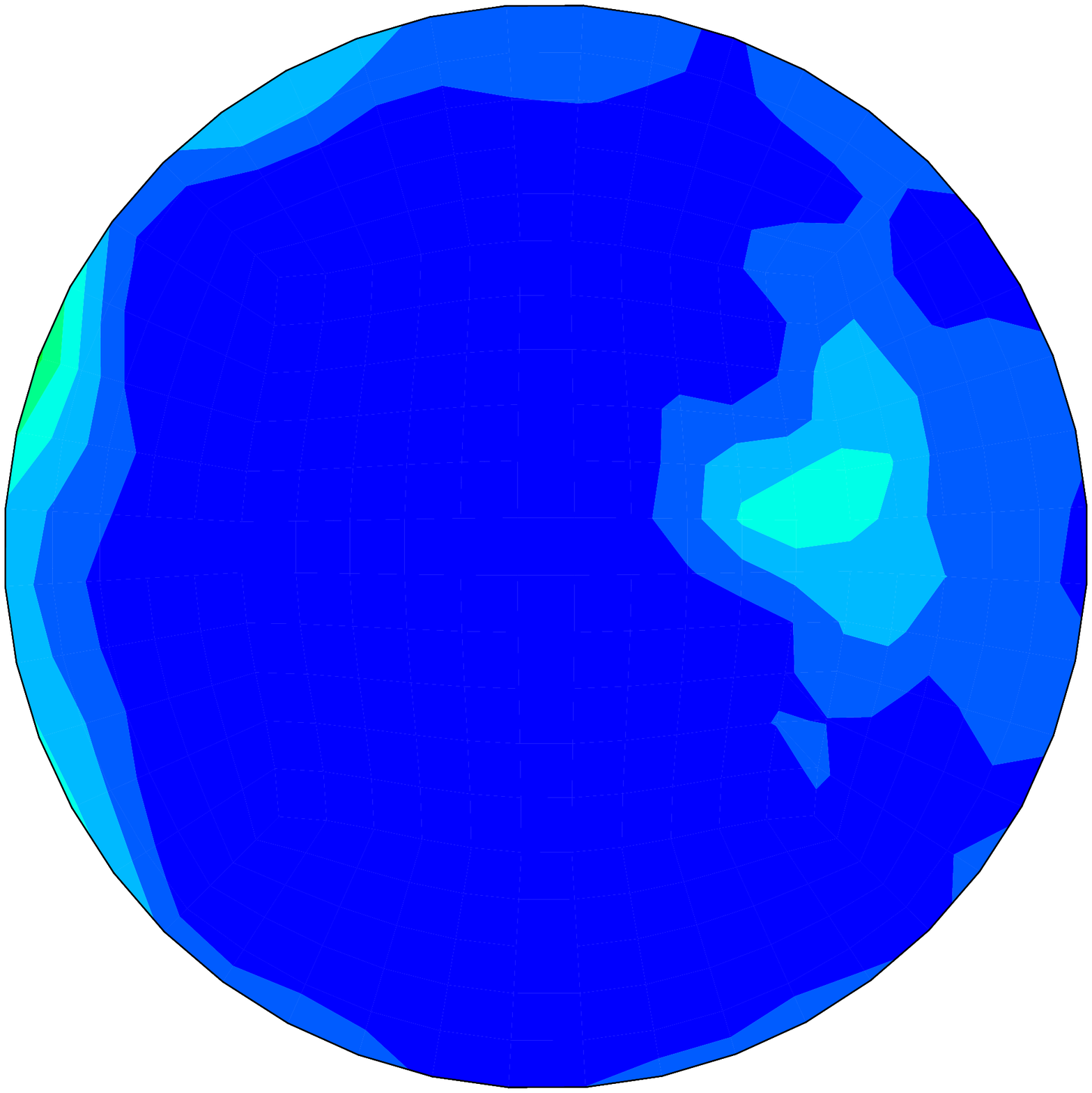}
			\includegraphics[height=2.5cm]{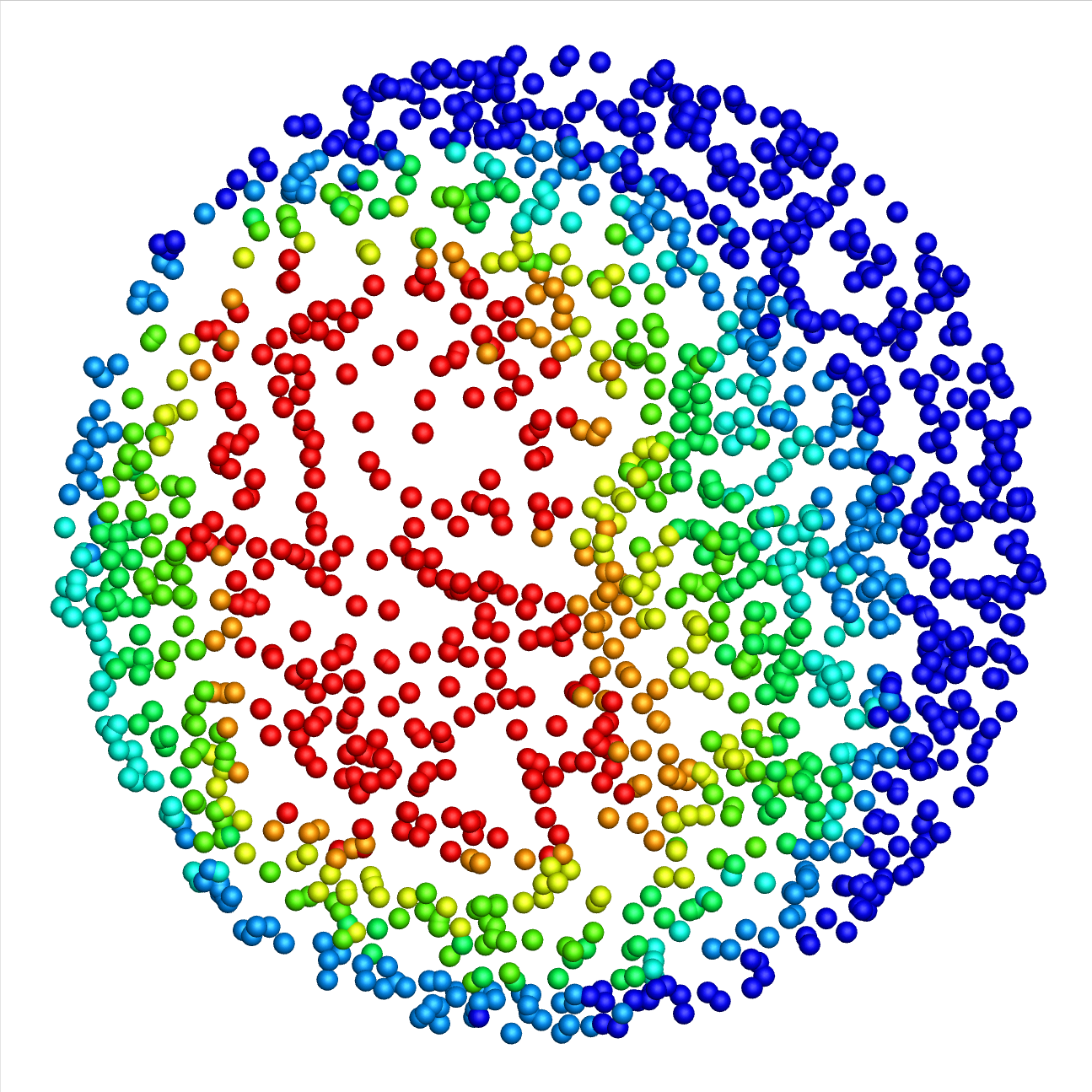}
			\includegraphics[height=2.5cm]{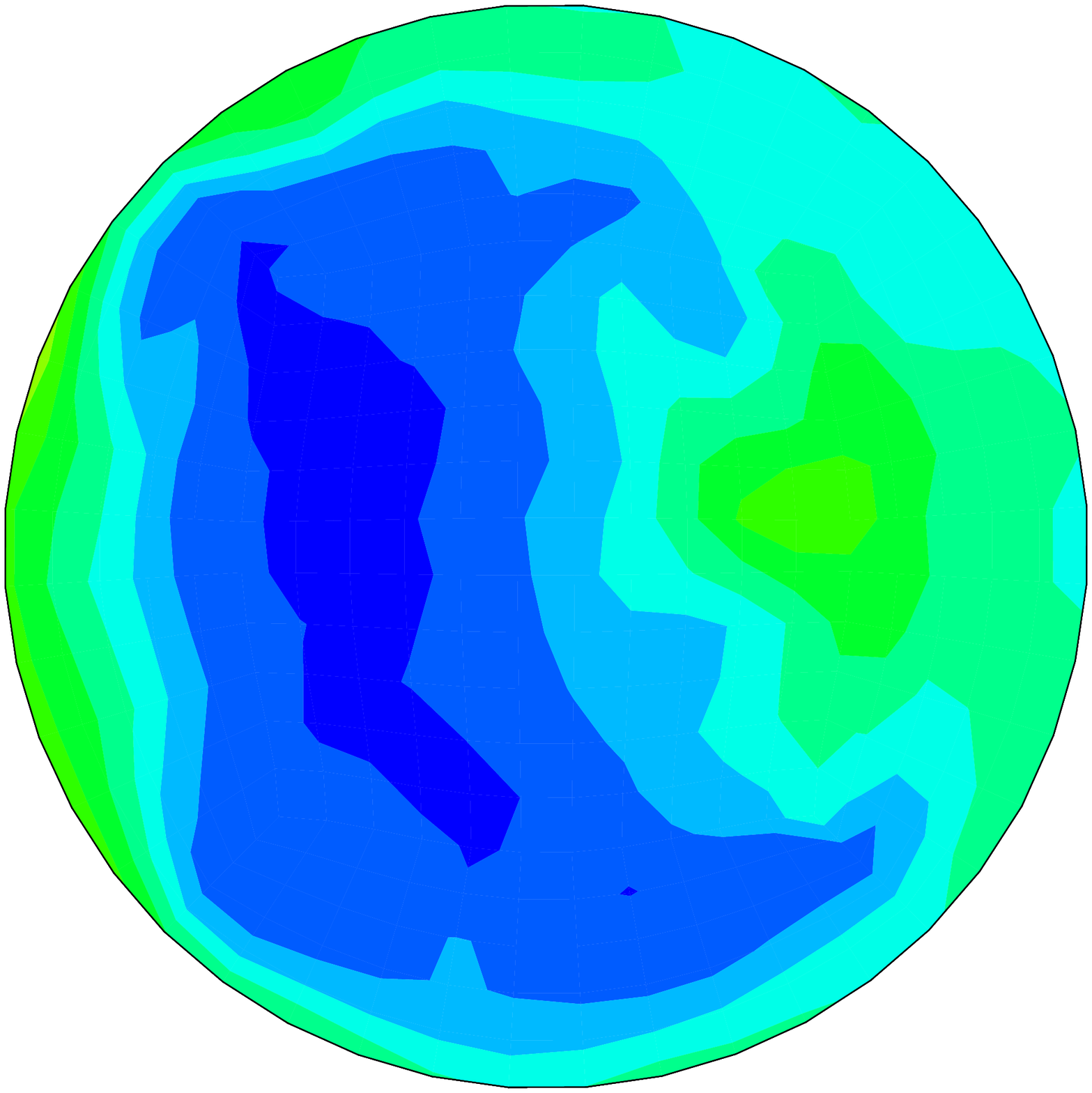}	
		\end{minipage}		
	}
	\quad
	\subfigure{
		\centering
		\begin{minipage}[c]{.15\textwidth}
			\centering
			\includegraphics[height=2.5cm]{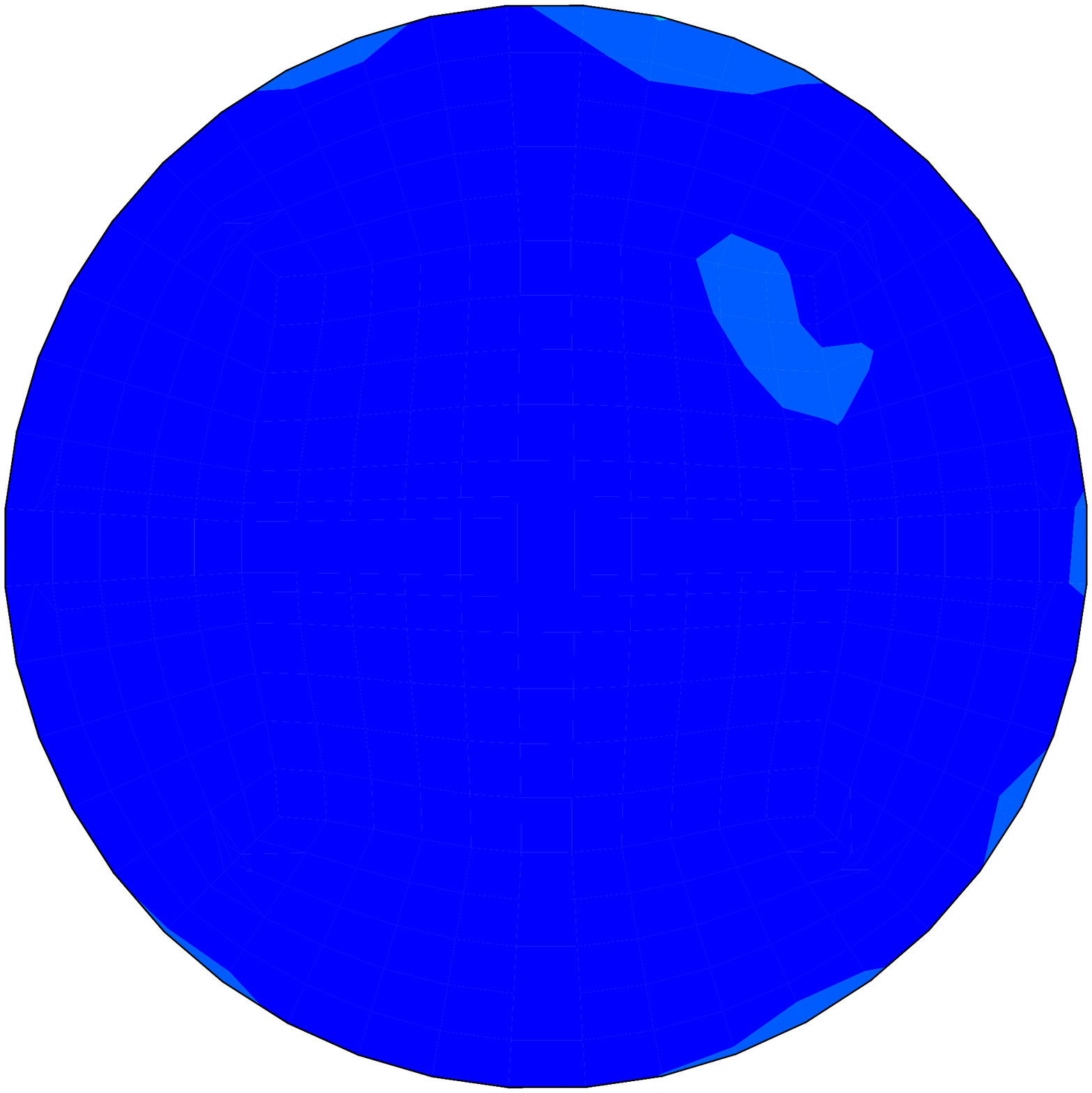}
			\includegraphics[height=2.5cm]{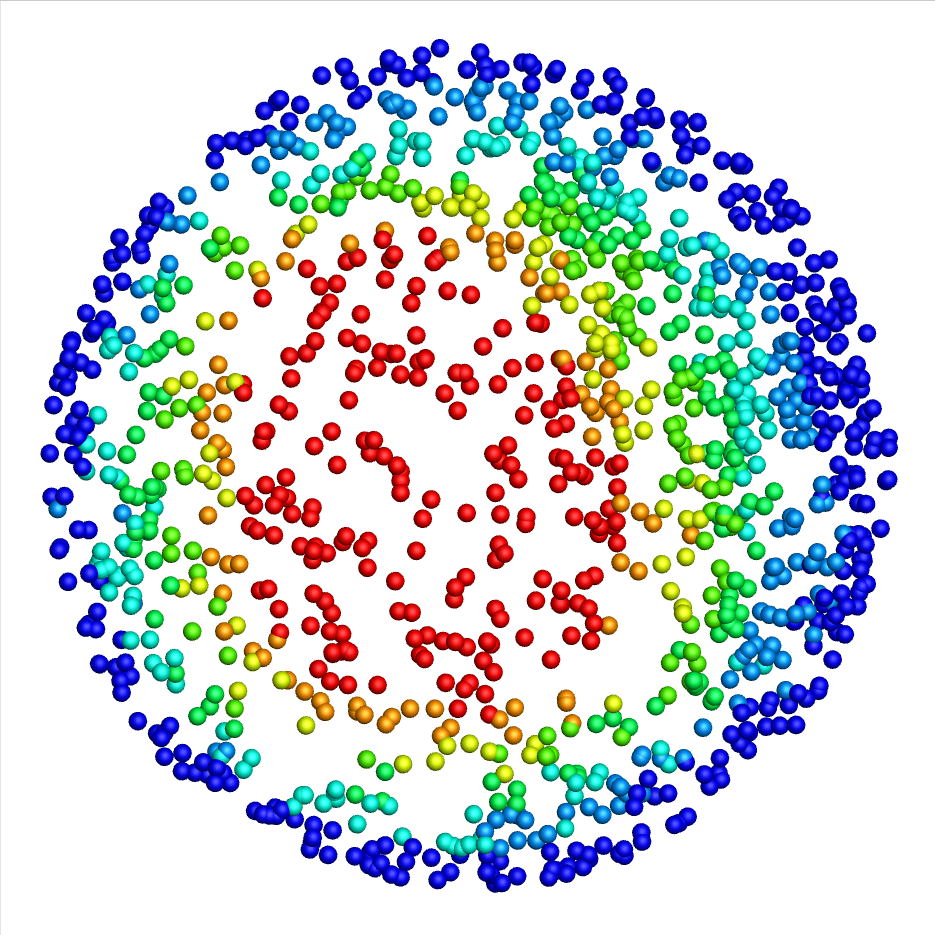}
			\includegraphics[height=2.5cm]{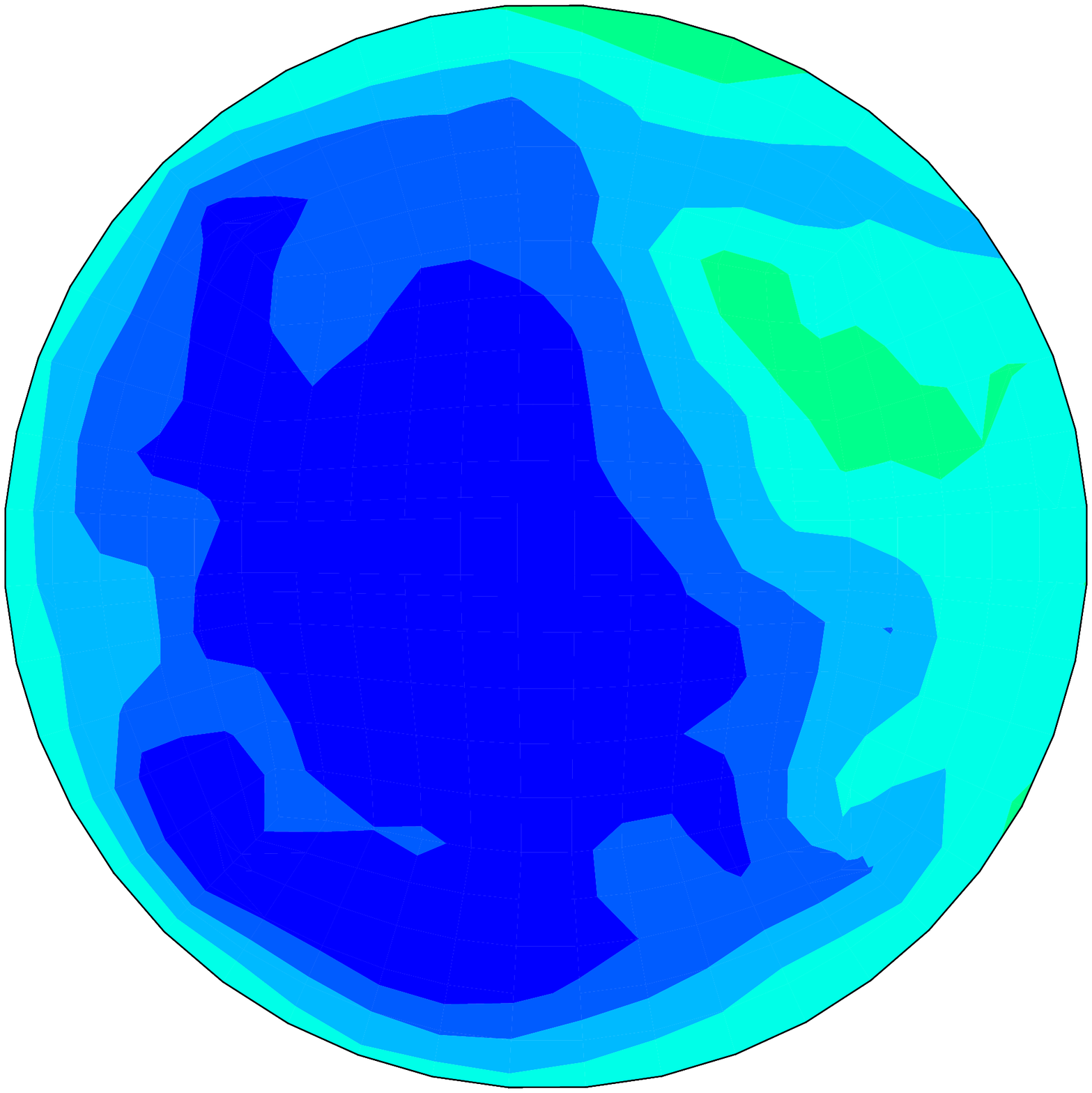}	
		\end{minipage}		
	}		
	\begin{center}
		\footnotesize \hspace{-5.0mm} (a) $0.25m$ \qquad\qquad\quad (b) $0.36m$ \qquad\quad\qquad (c) $0.80m$ \qquad\quad\quad~ (d) $1.63m$ \qquad\quad\quad~ (e) $2.00m$
	\end{center}
	\caption{Solid volume fraction $\epsilon_s$ of wave decomposition and discrete particles at $t=7.4s$ at different heights: (a) $0.25m$, (b) $0.36m$, (c) $0.80m$, (d) $1.63m$, (e) $2.00m$. At each height of (a) $\sim$ (e), three pictures shown from up to down are: \textbf{wave}, \textbf{particle}, \textbf{total}. \textbf{wave} and \textbf{total} are colored by epsilons-legend, and \textbf{particle} is colored
	by vertical velocity-legend. $\epsilon_{s}$ in \textbf{total} is exactly the sum of  \textbf{wave} and \textbf{particle}.}	
	\label{FBHorio 3D epsilon 2D horizontal plane shown}		
\end{figure}

Several instantaneous snapshots of solid volume fraction $\epsilon_s$ at a few times in the interval $7.0s\sim7.4s$ are shown in Figure \ref{FBHorio 3D epsilon shown}. Here the results are shown at the bottom region with height below $0.4m$ on the symmetric plane.
The particle concentrating clusters and diluting bubbles form, move, and vanish dynamically, which may introduce challenges
for the hybrid EE and EL numerical methods in identifying the interface between dilute/dense regions.
For UGKWP, the wave and particle decompositions are automatically distributed according to the local cell's $Kn$,
where particle appears in tracking the non-equilibrium dilute region, and vanishes in the intensive collisional dense region.
At $t=7.4s$, the wave and discrete particle decompositions on the symmetric plane at regions $0\sim0.4m$ and $1.4\sim1.8m$, and
on the horizontal cross-sections at heights, $0.25m$, $0.36m$, $0.80m$, $1.63m$, $2.00m$, are presented in Figure \ref{FBHorio 3D epsilon 2D vertical plane shown} and Figure \ref{FBHorio 3D epsilon 2D horizontal plane shown}, respectively.
The sampled particles, shown in Figure \ref{FBHorio 3D epsilon 2D vertical plane shown} and Figure \ref{FBHorio 3D epsilon 2D horizontal plane shown}, are colored by their vertical velocity. These results clearly show the evolution of solid particle phase through hydrodynamic wave and discrete particle and smooth transition in different regions.
The typical core-annular structures from the time-averaged variables are clearly shown in Figure \ref{FBHorio 3D epsilon and velocity profile},
Figure \ref{FBHorio 3D epsilon 2D vertical plane shown}, and Figure \ref{FBHorio 3D epsilon 2D horizontal plane shown}.
Figure \ref{FBHorio 3D epsilon 2D horizontal plane shown} presents low concentration ($\epsilon_s<0.1$) solid particle region,
with high vertical velocity ($V_s > 1.0m/s$).
Solid particles move upward in the center region, gather and fall down in the near-wall region.
The simulation results validate the GKS-UGKWP method for the study of gas-solid circulating fluidized bed problem.

Another interesting observation is that the experiment shows a very sharp jump in the solid particle vertical velocity around $r/R=0.6$ position and height $1.63m$, with a velocity change from $1.8m/s$ to $-0.8m/s$, as shown in Figure \ref{FBHorio 3D epsilon and velocity profile}(c).
It indicates the highly non-equilibrium transition layer in the solid phase.
In practice, UGKWP is capable of capturing the strong non-equilibrium physics, such as keeping a bimodal distribution in the particle velocity distribution function, such as the verification in the problem of two impinging particle jets \cite{WP-six-gas-particle-yang2021unified}.
In addition, the stratified flow structure, shown in Figure \ref{FBHorio 3D epsilon 2D vertical plane shown}(b), is most likely the flow pattern associated with such a sharp velocity jump observed in the experiment.
With the above consideration, even though the current time-averaged flow distribution shows a smooth transition radially,
UGKWP has the potential to give a complete picture about the underlying non-equilibrium particle transport mechanism.

\section{Conclusion}
In this paper, the gas-solid particle two phase flows, i.e., the turbulent fluidized bed and the circulating fluidized bed,
 are simulated by 3D GKS-UGKWP method.
In both cases, the solid particle flow shows the characteristic flow pattern, such as the bottom dense/middle transition/top dilute regions, the core-annular flow structures in the circulating fluidized bed, and the particle clustering phenomenon, etc.
The solid particle flow is dynamically dominant by particle-particle collisions in the high volume fraction region,
and the particle motion is driven by the gas flow in the dilute particle region.
The complex physics of gas-solid particle two-phase flow brings huge challenges for the development of multi-scale and multi-physics numerical algorithms, and the applications in gas-solid fluidization problem.
UKGWP is a multi-scale solver, which couples the wave and particle formulation in the evolution.
The modeling mechanism in UGKWP is intrinsically consistent with the flow physics in different regimes for the particle phase in the fluidized bed riser.
Specifically, the distribution of wave and particle decompositions in UGKWP is determined by the cell's $Kn$,
reflecting the degree of local non-equilibrium of the state of solid particles.
For a large $Kn$, the solid particle is in a collisionless transport regime and is mainly controlled by gas flow.
The UGKWP will sample discrete particles to capture the non-equilibrium physics.
For a small $Kn$, the solid particle takes intensive collisions and is in near equilibrium regime.
The UGKWP will automatically use analytical wave formulation to capture the particle phase, which reduces the computational cost greatly
due to the absence of particle-tracking.
For an intermediate $Kn$, both wave and particle decompositions contribute to the dynamic evolution of the particle phase.
UGKWP will find the most efficient way through the distribution of wave and particle to capture accurate flow physics and keep
efficient numerical simulation.

Since UGKWP can automatically distribute the wave and particle decompositions based on the local cell's $Kn$,
it is suitable for the fluidized bed simulation with the co-existence of multiple regimes, where
no prior division of dense/dilute regions is needed.
GKS-UGKWP provides a reliable tool for capturing the multi-phase and multi-scale flow physics. The simulation results of both turbulent fluidized bed and circulating fluidized bed
agree well with the experimental measurement.

\section*{Acknowledgements}
The current research is supported by National
Science Foundation of China (No.12172316), Hong Kong research grant council 16208021,
and Department of Science and Technology of Guangdong Province (Grant No.2020B1212030001).

\setcounter{equation}{0}
\renewcommand\theequation{A.\arabic{equation}}
\section*{Appendix A: Parameters in EMMS drag model}
The specific values of model parameters $\left(a,b,c\right)$ to calculate the heterogeneity index $H_D$ in EMMS drag model are listed as below \cite{Gasparticle-subgridmodel-EMMS-drag-lu2011eulerian, Gasparticle-subgridmodel-EMMS-MPPIC-li2012mp},
\begin{table}[h]
	\caption{Parameters in EMMS drag model}
	\vspace{2pt}
	\small
	\centering
	\setlength{\tabcolsep}{3.4mm}{
		\begin{tabular}{lc}\toprule[1.5pt]	
			$\left \{ \begin{array}{lc}
			a=0.8526 - \frac{0.5846}{1+\left(\epsilon_{g}/0.4325\right)^{22.6279}}\\
			c=0
			\end{array}\right.$
			& 0.4 $\leq$ $\epsilon_g$ \textless 0.46 \\
			\midrule[0.5pt]
			$\left \{ \begin{array}{lc}
			a=0.0320 + \frac{0.7399}{1+\left(\epsilon_{g}/0.4912\right)^{54.4265}}\\
			b=0.00225 + \frac{772.0074}{1+10^{66.3224\left(\epsilon_{g}-0.3987\right)}}
			+ \frac{0.02404}{1+10^{53.8948\left(0.5257-\epsilon_{g}\right)}} \\
			c=0.1705 - \frac{0.1731}{1+\left(\epsilon_{g}/0.5020\right)^{37.7091}}
			\end{array}\right.$
			& 0.46 $\leq$ $\epsilon_g$ \textless 0.545 \\
			\midrule[0.5pt]
			$\left \{ \begin{array}{lc}
			a=\left(2124.956-2142.3\epsilon_{g}\right)^{-0.4896}\\
			b=\left(0.8223-0.1293\epsilon_{g}\right)^{13.0310}\\
			c=\frac{\epsilon_{g}-1.0013}{-0.06633+9.1391\left(\epsilon_{g}-1.0013\right)+6.9231\left(\epsilon_{g}-1.0013\right)^2}
			\end{array}\right.$
			& 0.545 $\leq$ $\epsilon_g$ \textless 0.99 \\
			\midrule[0.5pt]
			$\left \{ \begin{array}{lc}
			a=0.4243+\frac{0.8800}{1+exp(-\left(\epsilon_{g}-0.9942\right)/0.00218)}\left(1-\frac{1}{1+exp(-(\epsilon_{g}-0.9989)/0.00003)}\right)\\
			b=0.01661+0.2436 exp\left(-0.5\left(\frac{\epsilon_{g}-0.9985}{0.00191}\right)^2\right)\\
			c=0.0825-0.0574 exp\left(-0.5\left(\frac{\epsilon_{g}-0.9979}{0.00703}\right)^2\right)
			\end{array}\right.$
			& 0.99 $\leq$ $\epsilon_g$ \textless 0.9997 \\
			\midrule[0.5pt]
			$\left \{ \begin{array}{lc}
			a=1\\
			c=0
			\end{array}\right.$
			& 0.9997 $\leq$ $\epsilon_g$ $\leq$ 1.0 \\
			\bottomrule[1.5pt]		
	\end{tabular}}
	\label{pneumatic conveying problem three cases table}
\end{table}

\newpage

\bibliographystyle{plain}%
\bibliography{jixingbib1}
\end{document}